\documentclass[iop,revtx4]{emulateapj}
\usepackage{lscape}
\usepackage{graphicx}
\usepackage{color}
\newcommand{\heiii}{He {\scshape iii}}
\newcommand{\civ}{C {\scshape iv}}

\newcommand{\cii}{C {\scshape ii}}
\newcommand{\ciii}{C {\scshape iii}}
\newcommand{\sii}{Si {\scshape i}}
\newcommand{\siii}{Si {\scshape ii}}
\newcommand{\siiii}{Si {\scshape iii}}
\newcommand{\siiv}{Si {\scshape iv}}
\newcommand{\nv}{N {\scshape v}}
\newcommand{\niti}{N {\scshape i}}
\newcommand{\ovi}{O {\scshape vi}}

\newcommand{\oiv}{O {\scshape iv}}

\newcommand{\heii}{He {\scshape ii}}
\newcommand{\hei}{He {\scshape i}}

\newcommand{\feii}{Fe {\scshape ii}}

\newcommand{\ACS}{\textit{ACS/SBC}}

\newcommand{\STIS} {\textit{STIS}}           

\newcommand{\GHRS}{\textit{GHRS}}
\newcommand{\COS}{\textit{COS}}

\newcommand{\rstar}{R$_*$}
\newcommand{\msun}{M_\odot}

\newcommand{\degs}{$^\circ$}
\newcommand{\kms}{km s$^{-1}$}

\newcommand{\htwo}{H$_2$}

\shorttitle{Hot lines in T Tauri stars}
\shortauthors{Ardila et al.}
\begin{document}
\title{Hot Gas Lines in T Tauri Stars}
\author{David R. Ardila\altaffilmark{1}, Gregory J. Herczeg\altaffilmark{2}, Scott G. Gregory\altaffilmark{3,4}, Laura Ingleby\altaffilmark{5}, Kevin France\altaffilmark{6}, Alexander Brown\altaffilmark{6}, Suzan Edwards\altaffilmark{7}, Christopher Johns-Krull\altaffilmark{8}, Jeffrey L. Linsky\altaffilmark{9}, Hao Yang\altaffilmark{10}, Jeff A. Valenti\altaffilmark{11}, Herv\'{e} Abgrall \altaffilmark{12}, Richard D. Alexander\altaffilmark{13}, Edwin Bergin\altaffilmark{5}, Thomas Bethell\altaffilmark{5}, Joanna M. Brown\altaffilmark{14}, Nuria Calvet\altaffilmark{5}, Catherine Espaillat\altaffilmark{14}, Lynne A. Hillenbrand\altaffilmark{3}, Gaitee Hussain\altaffilmark{15}, Evelyne Roueff\altaffilmark{12}, Eric R Schindhelm \altaffilmark{16}, Frederick M. Walter\altaffilmark{17}}

\altaffiltext{1}{ardila@ipac.caltech.edu; NASA Herschel Science Center, California Institute of Technology, MC 100-22, Pasadena, CA 91125, USA}
\altaffiltext{2}{The Kavli Institute for Astronomy and Astrophysics, Peking University, Beijing 100871, China}
\altaffiltext{3}{Cahill Center for Astronomy and Astrophysics, California Institute of Technology, MC 249-17, Pasadena, CA 91125, USA}
\altaffiltext{4}{School of Physics and Astronomy, University of St Andrews, St Andrews, KY16 9SS, UK}
\altaffiltext{5}{Department of Astronomy, University of Michigan, 830 Dennison Building, 500 Church Street, Ann Arbor, MI 48109}
\altaffiltext{6}{Center for Astrophysics and Space Astronomy, University of Colorado, Boulder, CO 80309-0389, USA}
\altaffiltext{7}{Department of Astronomy, Smith College, Northampton, MA 01063, USA}
\altaffiltext{8}{Department of Physics and Astronomy, Rice University, Houston, TX 77005, USA}
\altaffiltext{9}{JILA, University of Colorado and NIST, 440 UCB Boulder, CO 80309-0440, USA}
\altaffiltext{10}{Institute for Astrophysics, Central China Normal University, Wuhan, China 430079}
\altaffiltext{11}{Space Telescope Science Institute, 3700 San Martin Drive, Baltimore, MD 21218, USA}
\altaffiltext{12}{LUTH and UMR 8102 du CNRS, Observatoire de Paris, Section de Meudon, Place J. Janssen, F-92195 Meudon, France}
\altaffiltext{13}{Department of Physics and Astronomy, University of Leicester, University Road, Leicester LE1 7RH, UK}
\altaffiltext{14}{Harvard-Smithsonian Center for Astrophysics, 60 Garden St. MS 78, Cambridge, MA 02138, USA}
\altaffiltext{15}{ESO, Karl-Schwarzschild-Strasse 2, D-85748 Garching bei M\''{u}nchen, Germany}
\altaffiltext{16}{Southwest Research Institute, Department of Space Studies, Boulder, CO 80303, USA}
\altaffiltext{17}{Department of Physics and Astronomy, Stony Brook University, Stony Brook, NY 11794-3800, USA}

\begin{abstract}
For Classical T Tauri Stars (CTTSs), the resonance doublets of \nv, \siiv, and \civ, as well as the \heii\ 1640 \AA\ line, trace hot gas flows and act as diagnostics of the accretion process. In this paper we assemble a large high-resolution, high-sensitivity dataset  of these lines in CTTSs and Weak T Tauri Stars (WTTSs). The sample comprises 35 stars: one Herbig Ae star, 28 CTTSs, and 6 WTTSs.  We find that the \civ, \siiv, and \nv\ lines in CTTSs all have similar shapes.  We decompose the \civ\ and \heii\ lines into broad and narrow Gaussian components (BC \& NC).  The most common (50 \%) \civ\ line morphology in CTTSs is that of a low-velocity NC together with a redshifted BC. For CTTSs, a strong BC is the result of the accretion process. The contribution fraction of the NC to the \civ\ line flux in CTTSs increases with accretion rate, from $\sim$20\% to up to $\sim$80\%.  The velocity centroids of the BCs and NCs are such that V$_{BC}\gtrsim4$ V$_{NC}$, consistent with the predictions of the accretion shock model, in at most 12 out of 22 CTTSs.  We do not find evidence of the post-shock becoming buried in the stellar photosphere due to the pressure of the accretion flow. The \heii\ CTTSs lines are generally symmetric and narrow, with FWHM and redshifts comparable to those of WTTSs. They are less redshifted than the CTTSs \civ\ lines, by $\sim$10 \kms.   The amount of flux in the BC of the \heii\ line is small compared to that of the \civ\ line,  and we show that this is consistent with models of the pre-shock column emission.  Overall, the observations are consistent with the presence of multiple accretion columns with different densities or with accretion models that predict a slow-moving, low-density region in the periphery of the accretion column. For HN Tau A and RW Aur A, most of the \civ\ line is blueshifted suggesting that the \civ\ emission is produced by shocks within outflow jets.  In our sample, the Herbig Ae star DX Cha is the only object for which we find a P-Cygni profile in the \civ\ line, which argues for the presence of a hot (10$^5$ K) wind. For the overall sample, the \siiv\ and \nv\ line luminosities are correlated with the \civ\ line luminosities, although the relationship between  \siiv\ and \civ\ shows large scatter about a linear relationship and suggests that TW Hya, V4046 Sgr, AA Tau, DF Tau, GM Aur, and V1190 Sco are silicon-poor, while CV Cha, DX Cha, RU Lup, and RW Aur may be silicon-rich.   
\end{abstract}








\section{Introduction}

Classical T Tauri stars (CTTSs) are low-mass, young stellar objects surrounded by an accretion disk. They provide us with a laboratory to study the interaction between stars, magnetic fields, and accretion disks. In addition to optical and longer wavelength excesses, this interaction is responsible for a strong ultraviolet (UV) excess, Ly-$\alpha$ emission and soft X-ray excesses, all of which have a significant impact on the disk evolution, the rate of planet formation, and the circumstellar environment.

The observed long rotational periods, the large widths of the Balmer lines, and the presence of optical and UV excesses of CTTSs are naturally explained by the magnetospheric accretion paradigm (e.g. \citealt{uchida1984}, \citealt{koenigl1991}, \citealt{shu1994}). According to this paradigm, the gas disk is truncated some distance from the star ($\sim$5 \rstar, \citealt{mey97}) by the pressure of the stellar magnetosphere. Gas from the disk slides down the stellar gravitational potential along the magnetic field lines, reaching speeds comparable to the free-fall velocity ($\sim$300 \kms). These velocities are much larger than the local $\sim$20 \kms\ sound speed (but less than the Alfven speed of $\sim$500 \kms\  implied by a 2 kG magnetic field -- \citealt{joh07}).  The density of the accretion stream depends on the accretion rate and the filling factor but it is typically of the order of $\sim10^{12}\ cm^{-3}$ (pre-shock, see \citealt{cal98}). The supersonic flow, confined by the magnetic field, produces a strong shock upon reaching the star, which converts most of the kinetic energy of the gas into thermal energy (e.g. \citealt{lamzin1995}). 

The gas reaches temperatures of the order of a million degrees at the shock surface and cools radiatively until it merges with the stellar photosphere. Part of the cooling radiation will heat the stellar surface, resulting in a hot spot, observed spectroscopically as an excess continuum (the ``veiling''). Cooling radiation emitted away from the star will illuminate gas before the shock surface, producing a radiative precursor of warm (T$\sim10^4$ K), ionized gas (\citealt{cal98}). In this paper we will use the term ``accretion shock region'' as shorthand for the region that includes the pre-shock, the shock surface, the post-shock column, and the heated photosphere.

Because the accretion column should be in pressure equilibrium with the stellar photosphere, \citet{drake2005} suggested that for typical accretion rates ($\sim10^{-8} \msun$/yr) the post-shock region would be buried, in the sense that the shortest escape paths for post-shock photons would go through a significant column of photospheric gas. \citet{sacco2010} have argued that the burying effects appear at accretion rates as small as a few times $\sim10^{-10} \msun$/yr and that the absorption of the X-rays by the stellar photosphere may explain the one to two orders of magnitude discrepancy between the accretion rates calculated from X-ray line emission and those calculated from optical veiling or near-UV excesses, and the lack of dense (n$_e \gtrsim 10^{11}$ cm$^{-3}$) X-ray emitting plasma in objects such as T Tau \citep{gudel2007}. 

Time-dependent models of the accretion column by \cite{sacco2008} predict that emission from a single, homogeneous, magnetically dominated post-shock column should be quasi-periodic, on timescales of $\sim$400 sec, because plasma instabilities can collapse the column. Such periodicity has not been observed \citep{drake2009, gunther2010}, perhaps suggesting that accretion streams are inhomogeneous, or that there are multiple, uncorrelated accretion columns. 

Observations during the last two decades have resulted in spectroscopic and photometric evidence for the presence of accretion hot spots or rings with filling factors of up to a few percent (see review by \citealt{bou07}).  The surface topology of the magnetic field (e.g. \citealt{gregory2006, mohanty2008}) and/or the misalignment between the rotational and magnetic axes result in a rotationally modulated surface flux \citep{johns1995, argiroffi2011, arg2012}, with a small filling factor \citep{cal98, valenti2004}.  The spots appear and disappear over timescales of days  \citep{rucinski2008} to years \citep{bouvier1993}. Analysis of the possible magnetic field configurations  indicates that although CTTSs have very complex surface magnetic fields, the portion of the field that carries gas from the inner disk to the star, is well ordered globally \citep{cmj1999b, adams2012}. 


In this paper we use the strong emission lines of ionized metals in order to probe the characteristics of the accretion shock region.  In particular, we are interested in understanding the role that accretion has in shaping these lines, where the lines originate, and what the lines reveal about the geometry of the accretion process. Our long-term goal is to clarify the UV evolution of young stars and its impact on the surrounding accretion disk.

We analyze the resonance doublets of \nv \ ($\lambda \lambda$ 1238.82, 1242.80 \AA), \siiv\  ($\lambda \lambda$ 1393.76, 1402.77 \AA), and \civ\ ($\lambda \lambda$ 1548.19, 1550.77 \AA), as well as the \heii\ ($\lambda$ 1640.47 \AA) line. If they are produced by collisional excitation in a low-density medium, their presence suggests a high temperature ($\sim10^5$ K, assuming collisional ionization equilibrium) or a photoionized environment. In solar-type main sequence stars, these "hot" lines are formed in the transition region, the narrow region between the chromosphere and the corona, and they are sometimes called transition-region lines.

The \civ\ resonance doublet lines are among the strongest lines in the UV spectra of CTTSs \citep{ard02}. Using {\it International Ultraviolet Explorer} (IUE) data, \citet{jvl00} showed that the surface flux in the \civ\ resonant lines can be as much as an order of magnitude larger than the largest flux observed in Weak T Tauri stars (WTTSs), main sequence dwarfs, or RS CVn stars. They also showed that the high surface flux in the \civ\ lines of CTTSs is uncorrelated with measures of stellar activity but it is strongly correlated with accretion rate, for accretion rates from $10^{-8} \msun$/yr to $10^{-5} \msun$/yr. The strong correlations among accretion rate, \civ\ flux and Far Ultraviolet (FUV) luminosity have been confirmed by \cite{ingleby2011a} and \cite{yang2012} using \ACS\ and \STIS\ data. Those results further suggest a causal relationship between the accretion process and the hot line flux. 

Previous surveys of the UV emission lines in low-resolution spectra of T Tauri stars include the analysis of IUE spectra \citep{valenti2000, jvl00} and GHRS, ACS, and STIS spectra \citep{yang2012}. Prior analysis of high resolution observations of the hot gas lines in T Tauri stars had been published for some objects (BP Tau, DF Tau, DG Tau, DR Tau, EG Cha, EP Cha, GM Aur, RU Lup, RW Aur A, RY Tau, T Tau, TW Hya, TWA 5, V4046 Sgr, and HBC 388, see  \citealt{lamzin2000a, lamzin2000c, errico2000, errico2001, ard02, her02, lamzin2004, her05, her06, gunther2008, ingleby2011a}).  In addition, analysis of the \civ\ lines for the brown dwarf 2M1207 has been published by \cite{france2010}. The observations show that the \civ\  lines in CTTSs have asymmetric shapes and wings that often extend to  $\pm$ 400 \kms\ from the line rest velocity. The emission lines are mostly centered or redshifted, although some stars occasionally present strong blueshifted emission (e.g.  DG Tau, DR Tau, RY Tau). Doublet ratios are not always 2:1 (e.g. DG Tau, DR Tau, RU Lup, and RW Aur A). Overall, the UV spectra of CTTSs also show large numbers of narrow \htwo\ emission lines, and CO bands in absorption and emission \citep{france2011, schindhelm2012}, produced primarily by Ly-$\alpha$ fluorescence.  In most observations published so far, the \siiv\  line is strongly contaminated by \htwo\ lines, and the \nv\ 1243 \AA\ line is absorbed by circumstellar or interstellar N I. High spectral resolution observations are crucial to fully exploit the diagnostic power of the UV observations as the  \htwo\ emission lines can be kinematically separated from the hot lines only when the resolution is high enough. In addition, we will show here that the hot gas lines have multiple kinematic components that can only be analyzed in high resolution spectra.  

Both the correlation between accretion rate and \civ\ surface flux and the presence of redshifted hot gas line profiles in some stars, are consistent with formation in a high-latitude accretion flow. However, according to \cite{lam03, lamzin2003} \civ\ line formation in the accretion shock region should result in double-peaked line profiles, which are generally not observed. In this context \cite{gunther2008} explored the shape of the hot gas lines (primarily \ovi\ and \civ) in a sample of 7 stars. They considered formation in the accretion shock, in an outflow, in the surface of the disk, in an equatorial boundary layer, and in the stellar transition region, and concluded that no single explanation or region can be responsible for all the observed line characteristics. In particular, they concluded that the shape of the redshifted lines was incompatible with models of magnetospheric accretion. 


With the primary goal of providing a unified description of the hot gas lines and understanding their origin, we have obtained Far (FUV) and Near Ultraviolet (NUV) spectra of a large sample of CTTSs and WTTSs, using the Cosmic Origins Spectrograph ({\it{COS}}) and the Space Telescope Imaging Spectrograph ({\it{STIS}}). Most of the data for this paper comes from the Cycle 17  Hubble Space Telescope ({\it HST}) proposal ``The Disks, Accretion, and Outflows (DAO) of T Tau stars'' (PI G. Herczeg, Prop. ID HST-GO-11616). The DAO program is the largest and most sensitive high resolution spectroscopic survey of young stars in the UV ever undertaken and as such it provides a rich source of information for these objects. The program is described in more detail in \cite{herczeg2013}. We have complemented the DAO data with GTO data from {\it HST} programs 11533 and 12036 (PI J. Green - BP Tau, DF Tau, RU Lup, V4046 Sgr) as well as UV spectra from the literature and from the Mikulski Archive at STScI ({\it MAST}).

As shown below, there is a wide diversity of profiles in all lines for the stars in our sample. The spectra are rich and a single paper cannot do justice to their variety nor to all the physical mechanisms that likely contribute to their formation. Here we take a broad view in an attempt to obtain general statements about accretion in CTTSs.  Details about the sample and the data reduction are presented in Section \ref{observations}. We then analyze the \civ,  \siiv, \heii\ and \nv\ lines. The analysis of the \civ\ lines takes up most of the paper, as this is the strongest line in the set, and the least affected by absorptions or emissions by other species  (Section \ref{civ}). We examine the line shapes, the relationship with accretion rate, and correlations among quantities associated with the lines and other CTTSs parameters. We also obtain from the literature multi-epoch information on line variability (Section \ref{variability}). The other lines play a supporting role in this analysis and are examined in Sections \ref{heii} and \ref{siiv}. Section \ref{discussion} contains a summary of the observational conclusions and a discussion of their implications. The conclusions are in Section \ref{conclusions}.

\section{Observations}
 \label{observations}

Tables \ref{TableAncil1},  \ref{TableAncil2}, and \ref{TableAncil3} list the 35 stars we will be analyzing in this paper and the references for all the ancillary data we consider. Table \ref{TableData} indicates the origin of the data (DAO or some other project), the datasets and slit sizes used for the observations. Details about exposure times will appear in \cite{herczeg2013}.

The data considered here encompass most of the published high resolution {\it HST}\ observations of the \civ\  doublet lines for CTTSs. \citet{ard02} provides references to additional high-resolution \civ\  data for CTTSs obtained with the {\it Goddard High Resolution Spectrograph} (\GHRS). We do not re-analyze those spectra here, but they provide additional context to our paper. Non-DAO \STIS\ data were downloaded from the {\it HST \STIS\ Echelle Spectral Catalog of Stars} (StarCAT, \citealt{ayres2010}).  Non-DAO \COS\ data were downloaded from the  Mikulski Archive for Space Telescopes (MAST) and reduced as described below.

The sample of stars considered here includes objects with spectral types ranging from A7 (the Herbig Ae star DX Cha\footnote{We refer to the overall sample of accreting objects as the ``CTTSs sample," in spite of the inclusion of this Herbig Ae star.}) to M2, although most objects have mid-K spectral types. We assume stellar ages and distances to be $\sim$2 Myr and 140 pc, respectively, for Taurus-Aurigae (see \citealt{Loinard2007} and references therein), $\sim$2 Myr and 150 pc for Lupus I (\citealt{comer2009}), $\sim$5 Myr and 145 pc for Upper-Scorpius (see \citealt{alencar2003} and references therein), $\sim$5 Myr and 114 pc for the $\epsilon$ Chamaeleontis cluster (see \citealt{Lyo2008} and references therein), $\sim$5 Myr and 160 pc for Chamaeleon I (see \citealt{hussain2009} and references therein), $\sim$8 Myr and 97 pc for the $\eta$ Chamaeleontis cluster (\citealt{mamajek1999}),  $\sim$10 Myr and 55 pc for the TW Hydrae association (\citealt{zucker2004}), $\sim$12 Myr and 72 pc for V4046Sgr in the $\beta$ Pictoris moving group (\citealt{torres2006}). The sample includes MP Mus ($\sim$7 Myr and 100 pc, \citealt{Kastner2010} and references therein), not known to be associated with any young region.

The sample includes stars with transition disks (TD)  \citep{espaillat2011} CS Cha, DM Tau, GM Aur, IP Tau, TW Hya, UX Tau A, and V1079 Tau (LkCa 15). Here we take the term ''transition disk" to mean a disk showing infrared evidence of a hole or a gap. As a group, transition disks may have lower accretion rates than other CTTS disks \citep{espaillat2012}. However, for the targets included here the difference between the accretion rates of TDs and those of the the rest of the accreting stars is not significant.  The sample also includes six Weak T Tauri Stars (WTTSs): EG Cha (RECX 1), V396 Aur (LkCa 19), V1068 Tau (LkCa 4), TWA 7, V397 Aur, and V410 Tau. 

Of the 35 stars considered here, 12 are known to be part of binary systems. Dynamical interactions among the binary components  may re-arrange the circumstellar disk or preclude its existence altogether. The presence of an unaccounted companion may result in larger-than-expected accretion diagnostic lines. In addition, large instantaneous radial velocities may be observed in the targeted lines at certain points of the binary orbit.

AK Sco, DX Cha, and V4046 Sgr are spectroscopic binaries with circumbinary and/or circumstellar disks.  CS Cha is a candidate long-period spectroscopic binary, although the characteristics of the companion are unknown \citep{guenther2007}. The accretion rates listed in Table \ref{TableAncil3} are obtained from optical veiling and NUV excesses and represent total accretion for the overall system.  At the end of this paper we will conclude that binarity may be affecting the line centroid determination only in AK Sco. 

For non-spectroscopic binaries, the effect of the binarity may be relevant only if both companions are within the \COS\ or \STIS\ apertures. This is the case for DF Tau, RW Aur A, and the WTTSs EG Cha, V397 Aur, and V410 Tau. For DF Tau and RW Aur A, the primary component dominates the FUV emission (see \citealt{her06, alencar2005}).

Table \ref{TableAncil3}  lists the accretion rate, obtained from literature sources. For some stars, the DAO dataset contains simultaneous NUV and FUV observations, obtained during the same HST visit. \cite{ingleby2013} describe those NUV observations in more detail and calculate accretion rates based on them. In turn, we use those accretion rate determinations here. The uncertainty in the accretion rate is dominated by systematic factors such as the adopted extinction correction and the color of the underlying photosphere. These may result in errors as large as a factor of 10 in the accretion rate. We do not list $v \sin i$ measurements for the sample but typical values for young stars are $\sim$10-20 \kms\ \citep{basri1990}.


\subsection{Data Reduction}
 \label{datared}
 
The \COS\ and \STIS\ DAO data were all taken in time-tagged mode.

\begin{itemize}

\item \COS\ observations: The FUV (1150-1790 \AA ) spectra were recorded using multiple exposures with the G130M grating (1291, 1327 \AA \ settings) and the G160M grating (1577, 1600, 1623 \AA \ settings). These provide a velocity resolution of $\Delta v$ $\sim$ 17 km s$^{-1}$ (R$\sim$17,000) with seven pixels per resolution element \citep{osterman2011, green2012}. 

The maximum exposure time was accumulated in the region of the \siiv\ resonance lines and the exposures ranged from 3 ksec (for most stars with 2 orbit visits) to a maximum of 8 ksec for 
a few stars with 4 orbit visits. Use of multiple grating positions ensured full wavelength coverage and reduced the effects of fixed pattern noise.

All observations were taken using the primary science aperture (PSA), which is a 2.5'' diameter circular aperture. The acquisition observations used the ACQ/SEARCH algorithm followed by ACQ/IMAGE \citep{dixon2011}. The absolute wavelength scale accuracy is $\sim$ 15 \kms\ ($1\sigma$), where the error is dominated by pointing errors. We obtained one dimensional, co-added spectra using the \COS\ calibration pipeline (CALCOS) with alignment and co-addition obtained using the IDL routines described by \cite{danfoth2010}.
 
\item \STIS\ observations: Those targets that are too bright to be observed by \COS\ were observed using the \STIS\ E140M echelle grating, which has a spectral resolution of $\Delta v$ $\sim$ 7 km s$^{-1}$ 
(R$\sim$45,000), over the region 1150 - 1700 \AA\ . All the observations were obtained using the 0.2''$\times$0.2'' ``photometric'' aperture, during two {\it HST} orbits.

The \STIS\ FUV spectra were calibrated and the echelle orders co-added to provide a single spectrum  using the IDL software package developed for the StarCAT project \citep{ayres2010}.
This reduction procedure provides a wavelength-scale accuracy of 1$\sigma$ = 3  km s$^{-1}$. Non-DAO \STIS\ data were taken directly from StarCAT.
\end{itemize}
For a given pointing, errors in the positioning of the target within the aperture result in an offset in the wavelength scale. As indicated above, these are supposed to be 15 \kms\ for \COS\ and 3 \kms\ for \STIS, according to the instrument observing manuals \citep{dixon2011, ely2011}.   In addition, the geometric correction necessary to account for the curvature of the \COS\ FUV detector may make longer wavelength features appear redder than they really are. The offset depends on the exact position of the target on the detector.  In most cases this intra-spectrum wavelength uncertainty is $<$10 \kms, but in a few extreme cases it will introduce a $\sim$15 \kms\  shift from the reddest to the bluest wavelengths in a single \COS\ grating mode (See Figure 4 from \citealt{linsky2012}).

To determine how important the errors due to pointing and calibration are in the \COS\ data, we focus on the \htwo\ lines. We use the line measurements from K. France (personal communication, see also \citealt{france2012}) for P(2)(0-5) at 1398.95 \AA,  R(11)(2-8) at 1555.89 \AA,  R(6)(1-8) at 1556.87 \AA,  and P(5)(4-11) at 1613.72 \AA. \cite{france2012} show that in the case of  DK Tau, ET Cha (RECX 15), HN Tau, IP Tau, RU Lup, RW Aur, and V1079 Tau (LkCa 15), some \htwo\ lines show a redshifted peak and a blueshifted low-level emission, which makes them asymmetric. Ignoring these stars, the four \htwo\ lines are centered at the stellar rest velocity, with a standard deviations of 7.1, 6.5, 6.1, and 12.1 \kms, respectively. The large scatter in the P(5)(4-11) line measurement is partly the result of low signal-to-noise in this region of the spectrum. Unlike the results reported by \citealt{linsky2012}), we do not observe a systematic increase in the line center as a function of wavelength, neither star by star nor in the average of all stars for a given line. This is likely the result of the different acquisition procedure followed here. For the \STIS\ data, the average standard deviation in all the \htwo\ wavelengths is 4.4 \kms.

This shows that the errors in the \COS\ wavelength scale are smaller than the 15 \kms\ reported in the manuals and that no systematic shifts with wavelength are present. For the rest of the paper we will assume that the pointing errors in the \COS\ data result in a velocity uncertainty of 7 \kms\ (the average of the first three \htwo\ lines considered) while  the \STIS\ errors are 5 \kms. We do not correct the spectra for the \htwo\ velocities, as it is not clear how to implement this correction in the case of stars with asymmetric \htwo\ lines.



\section{Introduction to the shape of the lines}
\label{analysis}

Figures \ref{panel_1} to \ref{panel_5} in the Appendix  show the lines that we are discussing in this paper. Each line is plotted in velocity space (\kms) centered on the stellar photospheric rest frame. The ordinate gives flux density in units $10^{-14}$ erg sec$^{-1}$  cm$^{-2}$ \AA$^{-1}$. In the case of doublets, the nominal wavelength of the strongest member of the line is set to zero velocity, and the positions of the strongest \htwo\ lines are marked with dashed lines. Dotted lines mark the positions of other features in the spectra. The plotted spectra have been smoothed by a 5-point median.

The \civ, \siiv, and \nv\ lines are resonance doublets with slightly offset upper levels. For each doublet, both lines should have the same shape when emitted. This redundancy allow us to identify extra spectral features and to distinguish real features from noise. If the lines are emitted from an optically thin or effectively thin plasma, their flux ratio should be 2:1. If the lines are emitted from a medium that has a thermalization depth smaller than its optical depth (i.e. the medium is optically thick, but not effectively thin) the flux ratio will tend to 1.

For a plasma at rest in coronal ionization equilibrium, the peak ion abundance occurs at temperatures of $\log$(T)=5.3 (K) for \nv, $\log$(T)=5.0 (K) for \civ, $\log$(T)=4.9 (K) for \siiv, and $\log$(T)=4.7 (K) for \heii\ \citep{mazzotta1998}. In the context of the magnetospheric accretion paradigm, the velocities of the post-shock gas are high enough that the collisional timescales are longer than the dynamical timescale \citep{ardila2007}. This implies that the post-shock gas lines may trace lower temperature, higher density plasma than in the coronal ionization equilibrium case. The pre-shock plasma is radiatively ionized to temperatures of $\sim 10^4$ K. If the pre- and post-shock regions both contribute to the emission, conclusions derived from the usual differential emission analyses are not valid (e.g. \citealt{brooks2001}). In the case of non-accreting stars, the resonance \nv, \civ, and \siiv\ lines are collisionally excited while the \heii\ line is populated by radiative recombination in the X-ray ionized plasma \citep{zirin1975}.

\begin{figure}
\epsscale{1}
\includegraphics[width=3.4in]{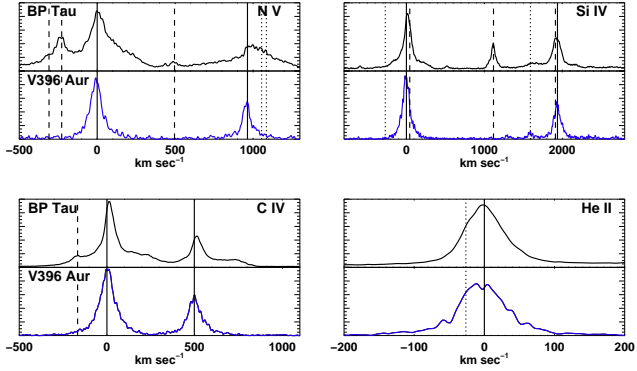}
\caption{Comparison between the CTTS BP Tau (black line) and the WTTS V396 Aur (blue line) for the four emission lines we are analyzing.  The vertical scale is arbitrary. The abscissas are velocities (\kms) in the stellar rest frame. {\bf Solid lines}: Nominal line positions;  {\bf Dashed lines}: Nominal locations of the strongest \htwo \ lines. {\bf N V, dotted lines}: \niti\ (1243.18 \AA, +1055 \kms; 1243.31 \AA, +1085 \kms). {\bf Si IV , dotted lines}: CO A-X (5-0) bandhead (1392.5 \AA, -271 \kms), {O {\scshape iv}} (1401.17 \AA, +1633 \kms). {\bf He II, dotted line}: Location of the secondary \heii\ line. \label{ctts_wtts_comparison}}
\end{figure}

Figure \ref{ctts_wtts_comparison} compares the hot gas lines of BP Tau and the WTTS V396 Aur. These stars provide examples of the main observational points we will make in this paper:

\begin{enumerate}
\item For \nv\ and \civ\ the lines of the CTTSs  have broad wings not present in the WTTSs. This is generally also the case for \siiv, although for BP Tau the lines are weak compared to \htwo. 
\item The \nv\ 1243 \AA\ line of BP Tau (the redder member of the multiplet), and of CTTSs in general, looks truncated when compared to the 1239 \AA\ line (the blue member), in the sense of not having a sharp emission component. This is the result of  \niti\ circumstellar absorption.
\item The \siiv\ lines in CTTSs are strongly affected by \htwo\ lines and weakly affected by CO A-X absorption bands \citep{france2011, schindhelm2012} and \oiv. At the wavelengths of the \siiv\ lines, the dominant emission seen in BP Tau and other CTTSs is primarily due to fluorescent \htwo\ lines. The WTTS do not show \htwo\ lines or CO bands within $\pm400$ \kms\ of the gas lines we study here.
\item The \civ\ BP Tau lines have similar shapes to each other, to the 1239 \nv\ line, and to the \siiv\ lines, when observed.  Generally, the CTTS \civ\ lines are asymmetric to the red and slightly redshifted with respect to the WTTS line. 
\item The \heii\ lines are similar (in shape, width, and velocity centroid)  in WTTSs and CTTSs, and similar to the narrow component of \civ.  
\end{enumerate}

\begin{figure}
\epsscale{1}
\includegraphics[width=3.4in]{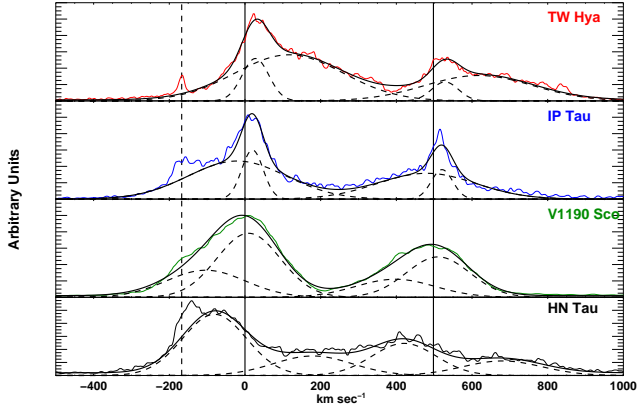}
\caption{The diversity of \civ\ CTTSs profiles. {\bf Dashed lines}: Narrow and broad gaussian components; {\bf Smooth solid lines}: Total model fit. The most common morphology is that of TW Hya, with a lower peak BC redshifted with respect to the NC. Stars like IP Tau have a BC blueshifted with respect to the NC. For V1190 Sco, both components have similar widths. For HN Tau the NC profile is blueshifted. \label{all_ctts}}
\end{figure}

The general statements above belie the remarkable diversity of line profiles in this sample. Some of this diversity is showcased in Figure \ref{all_ctts}. For analysis and interpretation, we focus on the \civ\ doublet lines, as they are the brightest and ``cleanest" of the set, with the others playing a supporting role.  In order to test the predictions of the magnetospheric accretion model, in section \ref{fits} we perform a Gaussian decomposition of the \civ\ profiles, and representative results are shown in Figure \ref{all_ctts}. 

TW Hya has a profile and decomposition similar to those of BP Tau: strong narrow component (NC) plus a redshifted, lower-peak broad component (BC). $\sim$50\% (12/22) of the stars for which a Gaussian decomposition is possible show this kind of profile and $\sim$70\% (21/29) of CTTSs in our sample have redshifted \civ\ peaks.  HN Tau also has a NC plus a redshifted BC, but the former is blueshifted with respect to the stellar rest velocity by 80 \kms. For IP Tau and V1190 Sco, the BC is blueshifted with respect to the NC. The magnetospheric accretion model may explain some of objects with morphologies analogous to TW Hya but blueshifted emission requires extensions to the model or the contributions of other emitting regions besides the accretion funnel \citep{gunther2008}. 

\begin{figure}
\epsscale{1}
\includegraphics[width=3.4in]{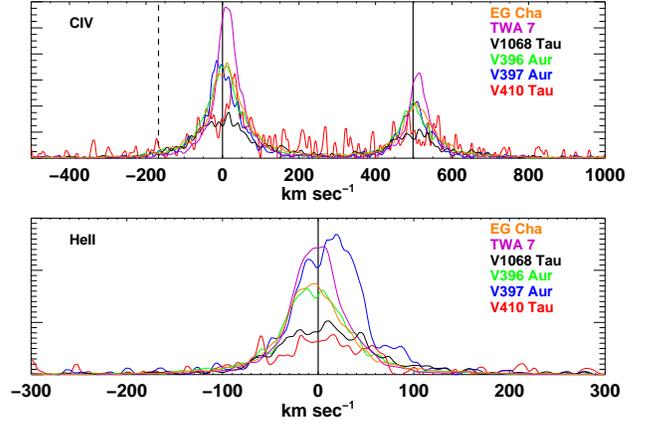}
\caption{The WTTSs spectra in \civ\ and \heii. All of the spectra have been smoothed by a 5-point median. All the spectra have been scaled to have the same mean value in the wing between -150 and -50 \kms. The solid vertical lines mark the rest-velocity positions of the \civ\ and \heii\ lines. {\bf Top}: \civ. The dashed vertical line indicates the location at which the \htwo\ line R(3)1- 8 would be, if present. The large value of the TWA 7 line is an artifact of the scaling procedure, due to the line redshift (25.3 \kms).  {\bf Bottom}: \heii. Except for V1068 Tau and V410 Tau, the WTTSs have similar characteristics (shape, shifts, and FWHM) in \heii\ and \civ. V1068 Tau and V410 Tau appear to be truncated or broadened in both lines. \label{all_wtts}}
\end{figure}

Figure \ref{all_wtts} shows the \civ\ and \heii\ lines in all the WTTSs, scaled to the blue wings of the lines. 
For all except for V1068 Tau (Lk Ca 4) and V410 Tau,  the \civ\ and \heii\ lines appear very similar to each other in shape, width, and shift. They are all fairly symmetric, with velocity maxima within $\sim$20 \kms\ of zero, and FWHM from 60 to 100 \kms. V1068 Tau and V410 Tau have the largest FWHM in both \civ\ and \heii. The lines of these two stars appear either truncated or broadened with respect to the rest of the WTTSs. In the case of V410 Tau, $v \sin i=73$ \kms\ \citep{gleb2005}, which means that rotational broadening is responsible for a significant fraction of the width. However, V1068 Tau has  $v \sin i=26$ \kms. For the rest of the WTTSs, $v \sin i$ ranges from 4 to 20 \kms\ \citep{gleb2005, torres2006}. 

To characterize the line shapes of \heii\ and \civ\ we have defined non-parametric and parametric shape measurements. The non-parametric measurements (Table \ref{ForPaper_NonParametric}) do not make strong assumptions about the line shapes and they provide a intuitive summary description of the line. These are the velocity at maximum flux, the full width at half maximum (FWHM), and the skewness, defined in Section \ref{non_par}. For \civ\ we have also measured the ratio of the 1548 \AA\ to the 1550 \AA\ line, by scaling the line wings to match each other. This provides a measure of the line optical depth. The parametric measurements (Tables \ref{ForPaper_TableFits_CIV} and \ref{ForPaper_HeII}) assume that each \civ\ and \heii\ line is a combination of two Gaussians. 

Table \ref{ForPaper_TableLineFluxes} lists the flux measurements that we will be considering in the following sections. The fluxes are obtained by direct integration of the spectra over the spectral range listed in the table, after subtracting the continuum and interpolating over known blending features (see Section \ref{civ}).


\section{The relationship between accretion rate and C IV luminosity}

As mentioned in the introduction,  \citet{jvl00} showed that the accretion rate is correlated with excess \civ\ luminosity. The excess  \civ\ luminosity is obtained by subtracting the stellar atmosphere contribution to the observed line. They estimated the stellar contribution to be $6 \times 10^{-5}$ L$_\odot$ for a 2R$_\odot$ object. They also showed that the correlation of \civ\ excess luminosity with accretion rate is very sensitive to extinction estimates. More recently, \cite{yang2012} obtain a linear correlation between the \civ\ luminosity (from \STIS\ and \ACS\ low-resolution spectra) and the accretion luminosity (from literature values) for 91 CTTSs. Here we show that our data are consistent with a correlation between accretion rate \civ\ luminosity and explore the role that the lack of simultaneous observations or different extinction estimates play in this relationship.

\begin{figure}
\epsscale{1}
\includegraphics[width=3.4in]{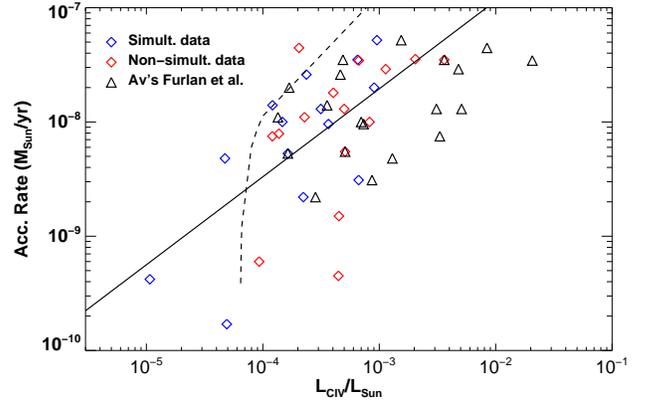}
\caption{Accretion rate vs. \civ\ luminosity.  The blue diamonds correspond to stars with simultaneous determinations of accretion rate and line luminosity. The red diamonds are objects with non-simultaneous determinations of accretion rate. The solid line is the correlation obtained by using all of the diamonds. Black triangles use the extinctions from \cite{furlan2009,furlan2011} to calculate L$_{C IV}$. Errors in L$_{C IV}$/L$_{\odot}$ are $\sim$5-10\%. The dashed line is Equation 2 from \cite{jvl00}, assuming R$_*$= 2R$_\odot$ for all stars. The lowest luminosity blue diamond corresponds to ET Cha (RECX 15) which has R$_*$= 0.9 R$_\odot$ \citep{siess2000}.  \label{lacc_vs_lciv}}
\end{figure}

Figure \ref{lacc_vs_lciv} compares accretion rates (references given in Table \ref{TableAncil1}) with the \civ\ line luminosities.  Blue diamonds correspond to objects with simultaneous determinations of  ${\dot {M}}$ and L$_{C IV}$ when available \citep{ingleby2013} and green diamonds correspond to  ${\dot {M}}$ determinations for the rest of the objects. Note that the accretion rate estimates from \cite{ingleby2013} are derived using the extinction values from \cite{furlan2009,furlan2011}.

Using only the simultaneous values (blue diamonds), we obtain a Pearson product-moment correlation coefficient r=$0.73$ (p-value=0.3\%\footnote{The p-value is the probability of obtaining a value of the test statistic at least as extreme as the observed one, assuming the null hypothesis is true. In this case, it is the probability that the Pearson's r is as large as measured or larger, if the two quantities are uncorrelated. We reject the null hypothesis if p-value$\leq$0.05.}) while with all values we obtain r=$0.61$ (p-value=$<$0.05\%). The difference between using all of the data or only the simultaneous data is not significant for the purposes of the correlation.  Including all data, we obtain $\log \dot {M}= (-5.4\pm0.2)+(0.8\pm0.1)\ \log L_{C IV}/L_{\odot}$, where $\dot {M}$ is given in $M_{\odot}$/yr and the errors indicate 1$\sigma$ values obtained by the bootstrap method. The correlation is plotted in Figure  \ref{lacc_vs_lciv}.

As argued by \cite{jvl00}, the observed relationship between $\log L_{C IV}/L_{\odot}$ and accretion rate is very sensitive to extinction estimates. This is shown in Figure \ref{lacc_vs_lciv}, for which the  \civ\ luminosities indicated by the black triangles were calculated using the extinction estimates from \cite{furlan2009,furlan2011}, for all the targets we have in common with that work. For $\log L_{C IV}/L_{\odot}$ and  $\log \dot {M}$ the correlation is weakened when using the \cite{furlan2009,furlan2011} extinction values:  the value of the Pearson's $r$ is $0.44$, with p-value=5\%.

The extinction values we adopt in this paper come from a variety of sources, but a significant fraction come from \cite{gull98, gullbring2000}.  They argue that the colors of the WTTSs underlying the CTTSs are anomalous for their spectral types, which biases the near-IR extinction estimates. They obtain the extinctions reported here by deriving models of the UV excess in CTTSs. For other stars, we have adopted extinction estimates based on spectroscopic observations of the accretion veiling, when possible.  The \cite{furlan2009,furlan2011} extinction values are obtained by de-reddening the observed near-infrared colors until they match the colors for the target's spectral type. They are significantly larger than the values we adopt in this paper, resulting in larger estimates of the \civ\ luminosity. Differences in the extinction estimates can have a substantial impact in the adopted flux, as a 10\% increase in the value of A$_V$ results in a 30\% increase in the de-reddened \civ\ line flux. 

Figure \ref{lacc_vs_lciv} also shows the relationship between accretion rate and \civ\ luminosity derived by  \citet{jvl00} (their equation 2) assuming that all stars have a radius of 2R$_\odot$. They obtained their relationship based on the accretion rates and extinctions from \cite{hartigan1995}. Both of those quantities are higher, on average, than the ones we adopt here, and so their correlation predicts larger accretion rates. Note that the relationship from \citet{jvl00} is not defined for stars with excess \civ\ surface fluxes smaller than 10$^6$ erg sec$^{-1}$  cm$^{-2}$.  

Overall, there is enough evidence to confirm that for most CTTSs the \civ\ line luminosity is powered primarily by accretion, and we will adopt this hypothesis here. However, the exact relationship between accretion rate and \civ\ luminosity remains uncertain. This is not surprising considering the complexity of the processes that contribute to the line flux, as we show in this work. The dominant uncertainty is the exact value of the extinction, which depends on the assumed stellar colors and on the shape of the extinction law in the UV \citep{jvl00,calvet2004}. 

We do not detect a monotonic decrease in the CTTSs \civ\ luminosity as a function of age, for the range of ages considered here (2 - 10 Myrs). We do not observe a significant difference in the \civ\ luminosities of the TDs as compared with the rest of the sample, consistent with the results from  \cite{ingleby2011a} who found no correlations between FUV luminosities and tracers of dust evolution in the disk.

\section{The C IV line shape}
\label{civ}

The \civ\ lines of CTTSs are generally redshifted, broad (with emission within $\sim$400 \kms\ of the stellar rest velocity), and asymmetric to the red (positive skewness). We will show that none of these characteristics is correlated with the line luminosity or with accretion rate. 

\subsection{Comparing the two  C IV lines}
 \label{comparison}
If optically thin, both \civ\ lines form should have the same shape. Differences between the components can help us discover the presence of extra sources of absorption or emission. To exploit this redundancy, in Figure \ref{all_civ} we plot both members of the \civ\ doublet, with the 1550 \AA\ line scaled to match the 1548 \AA\ one. The scaling is done by matching the line peaks or the line wings from $\sim0$ to $\sim$150 \kms.  We expect this scaling factor to be 2 for optically thin or effectively thin emission. However, as we discuss in Section \ref{shape}, the opacity characteristics of the broad and narrow line components are different, and the overall scaling factor may not be the best predictor of opacity. 

Figure \ref{all_civ} shows that the line wings tend to follow each other closely, at least until about +200 to +300 \kms, when the 1548 \AA\ line start to bump into the 1550 \AA\ one.  The 1548 \AA\ line is usually contaminated by the \htwo\ line R(3)1- 8 (1547.3 \AA), at -167 \kms\ with respect to its rest velocity (e.g. DK Tau).  Figure \ref{all_civ} also reveals examples of extra emission at $\sim$-100 \kms\ which do not appear in the 1550 \AA\ line (AK Sco, DE Tau, DK Tau, DR Tau, HN Tau A, RU Lup, and UX Tau A).   We tentatively identify emission from \feii\ (1547.807, -73.6 \kms\ from the rest velocity of the 1548 \civ\ line), \cii\ (1547.465; -139.8 \kms), and \sii\ (1547.452 \AA, -142.3 \kms; 1547.731 \AA, -88.3 \kms), for these stars, as responsible for the emission to the blue of the \civ\ 1548 \AA\ line.

To measure the \civ\ line flux listed in Table \ref{ForPaper_TableLineFluxes} we subtracted the continuum and interpolated over the \htwo\ R(3)1-8 line and the \sii, \cii, \feii -complex, when present. The resulting spectrum is then integrated  from -400 \kms\ to 900 \kms\ of the 1548 \civ\ line. We also detect the CO A-X (0-0) absorption band at 1544.4 \AA\  (-730 \kms\ from the  1548 \AA\ line, see \citealt{france2011, mcjunkin2013}) in a significant fraction of the sample. The wing of the CO absorption may extend to the blue edge of the 1548 \civ\ line. However its impact in the overall \civ\ flux is negligible.

Note that the red wings of each \civ\ line for DX Cha, RW Aur A, DF Tau, and RU Lup do not follow each other, and AK Sco and CS Cha have extra emission features near the 1550 \AA.

For DX Cha we will argue in section \ref{discussion} that the peculiar shape of the \civ\ lines can be explained by the existence of a hot wind.  

\begin{figure}
\epsscale{1}
\includegraphics[width=3.4in]{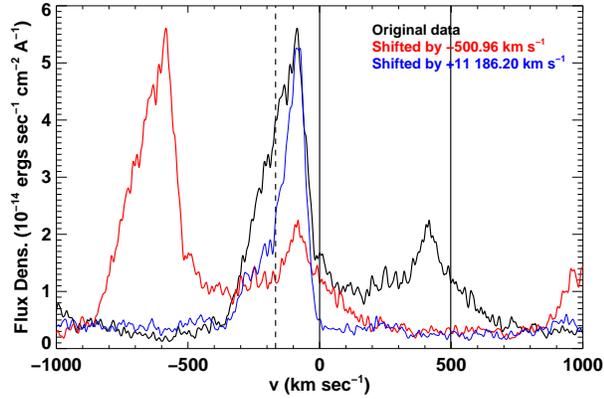}
\caption{\htwo\ contamination of the \civ\ line in RW Aur.  The plot shows that the \civ\ and \htwo\ emissions are blueshifted and redshifted, respectively, by almost 100 \kms. The result is that most of the \civ\ 1548 \AA\ line emission is covered by the redshifted \htwo.  {\bf Vertical solid lines}: nominal positions of the \civ\ doublet lines;  {\bf Vertical dashed line}: nominal R(3) 1-8 \htwo\ line position. {\bf Black trace}: observed spectrum in the \civ\ region; {\bf Red trace}: the \civ\ spectrum blueshifted by 500.96 \kms. With this shift, the nominal position of the 1550 \AA\ line should match the nominal position of the 1548 \AA\ line. {\bf Blue trace}: The R(3)1-7 \htwo\ line (at 1489.57 \AA), but redshifted to the nominal position of the R(3)1-8 \htwo\ line (at 1547.34 \AA). \label{rw_aur}}
\end{figure}

The strange appearance of  \civ\ lines of RW Aur A is due to a bipolar outflow (See Figure \ref{rw_aur}).  \cite{france2012} show that the \htwo\ lines from RW Aur A are redshifted by $\sim$100 \kms\ in at least two progressions ([v',J']=[1,4] and [1,7]), as they originate in the receding part of the outflow. In the observations we present here, the \civ\ doublet lines are blueshifted by $\sim$-100 \kms, as can be seen from the position of the 1550 \civ\ line. This points to an origin in the approaching part of the outflow. The net result of these two outflows is that the blueshifted 1548 \civ\ line is buried under the redshifted \htwo\ emission. RW Aur A is the only unequivocal example of this coincidence in the current dataset. For HN Tau A the \htwo\ R(3) 1-8 line is also redshifted, although the redshift (+30 \kms) is within the $2\sigma$ of the error introduced by the pointing uncertainty (see also \citealt{france2012}). Therefore, HN Tau A may be another object for which we are observing two sides of the outflow, although the velocities of the blueshifted and redshifted sides do not match as well as in RW Aur.

For DF Tau, the apparent red wing in the 1550 \AA\ line is an artifact of the line scaling, because for this star the ratio between the two \civ\ lines is almost 3, indicating either extra absorption or emission in one of the components.  Unlike the case in RW Aur A, redshifted \htwo\ emission is not responsible for the extra emission, as the R(3) 1-8 line is observed at -167 \kms. Other observations show that the ratio between the two lines remains high among different epochs (Section \ref{variability}). The origin of this extra emission in the 1548 \AA\ line or absorption in the 1550 \AA\ one remains unexplained in this work. 

For RU Lup, the extra ``bump" 200 \kms\ to the red of the 1548 \AA\ \civ\ line is also present in other epoch observations of the system and remains unexplained in this work. AK Sco and CS Cha also present anomalous profiles, but of a different kind. AK Sco shows a broad emission -250 \kms\ from the 1550 \AA\ line, not present in the 1548 \AA\ line. CS Cha also shows and extra emission in the 1550 \AA\ line at $\sim$-100 \kms\ and $\sim$+200 \kms. 

\subsection{Non-parametric description}
 \label{non_par}

\subsubsection{The velocity of the peak emission as a function of luminosity and accretion rate}

The velocity of peak emission for each line is defined as the mean velocity of the top 5\% of the flux  between -100 and +100 \kms\ (for the 1548 \civ\ line) or between +400 and +600 \kms\ (for the 1550 \AA\ line). This is calculated from spectra that have been smoothed by 5-point median. The velocity of peak listed in Table \ref{ForPaper_NonParametric} is the average of both lines.

The \civ\ line is centered or redshifted for 21 out of 29 CTTSs, uncorrelated with line luminosity or accretion rate  (Figure \ref{vel_max} - Top).  The most significant exceptions are:  DK Tau which shows two emission peaks, one blueshifted and the other redshifted, present in both lines; HN Tau A, for which both \civ\ lines have an emission peak at $\sim$-80 \kms\ from their rest velocity\footnote{We note that for HN Tau A, the radial velocity (4.6$\pm$0.6 \kms), derived by \citet{nguyen2012}, which we use here, deviates significantly from the velocity of the surrounding molecular cloud (21 \kms; \citealt{kenyon1995}).  Using the cloud velocity would shift the \civ\ lines to the blue even more. \citealt{france2012} finds that the \htwo\ lines have maxima at $\sim$ 20 \kms from the rest velocity, although the line asymmetry suggest components at lower velocity.}; RW Aur A, for which the peak of the 1550 \AA\ line is at -86 \kms. In the case of AK Sco there is extra emission in the 1548 \AA\ line that may be due to \sii\ but there is also extra emission $\sim$-250 \kms\ from the 1550 \AA\ line. It is unclear therefore, whether the \civ\ lines are centered or blueshifted, although both the \siiv\ and the \nv\ lines (which follow the \civ\ lines) are well centered. Those four stars are among the higher accretors in the sample and have jets seen in the high-velocity components of forbidden lines \citep{hartigan1995}. Four other CTTSs have slightly blueshifted peaks, but with velocities $>-4$ \kms.

The average velocity at maximum flux for the \civ\ CTTSs lines is  $\overline{V_{Max}}=18\pm4$ \kms\ (Table \ref{velocities}. In this paper the uncertainties refer to uncertainties in the calculated mean values), not including the four stars with strong blueshifts (DK Tau, HN Tau A, and RW Aur A for which outflowing material is contributing to the profile, nor AK Sco, for which the two \civ\ lines are different from each other). We have argued that the velocity uncertainties for this \COS\ dataset are $\sim$7 \kms. A one-sample Kolmogorov -- Smirnov (KS) test confirms that the probability of the observed distribution being normal centered at zero, with $1\sigma =7$ \kms\ is negligibly small. The conclusion is the same if we take $1\sigma =15$ \kms, the nominal wavelength error for \COS. In other words, the overall redshift of the CTTSs \civ\ lines is significant. 


The WTTSs are also redshifted, with $\overline{V_{Max}}=11\pm4$ \kms. A two-sided KS test comparing the CTTSs and WTTSs $V_{Max}$ distribution gives p-value=0.6, which indicates that they are consistent with the null hypothesis. In other words, although the mean velocity of the WTTSs is $\sim2\sigma$ less that the mean CTTS velocity, the observations are consistent with the two quantities having the same distribution. \cite{linsky2012} have shown that for rotation periods such as those observed in T Tauri stars, the \siiv\ and \civ\ lines in low-mass dwarfs present redshifts of $\sim$7 \kms, the result of gas flows produced by magnetic heating. Given the scatter in their sample, our redshifts in WTTSs are consistent with theirs. 


 
\subsubsection{The line width as a function of luminosity and accretion rate}

Overall, most of the CTTSs \civ\ lines show detectable emission within $\pm$400 \kms\ of the nominal velocity. The FWHM is measured using the 1550 \AA\ line of the smoothed \civ\ spectrum. The smoothing is the same used to measure the velocity of maximum flux. 

\begin{figure}
\epsscale{1}
\includegraphics[width=6in]{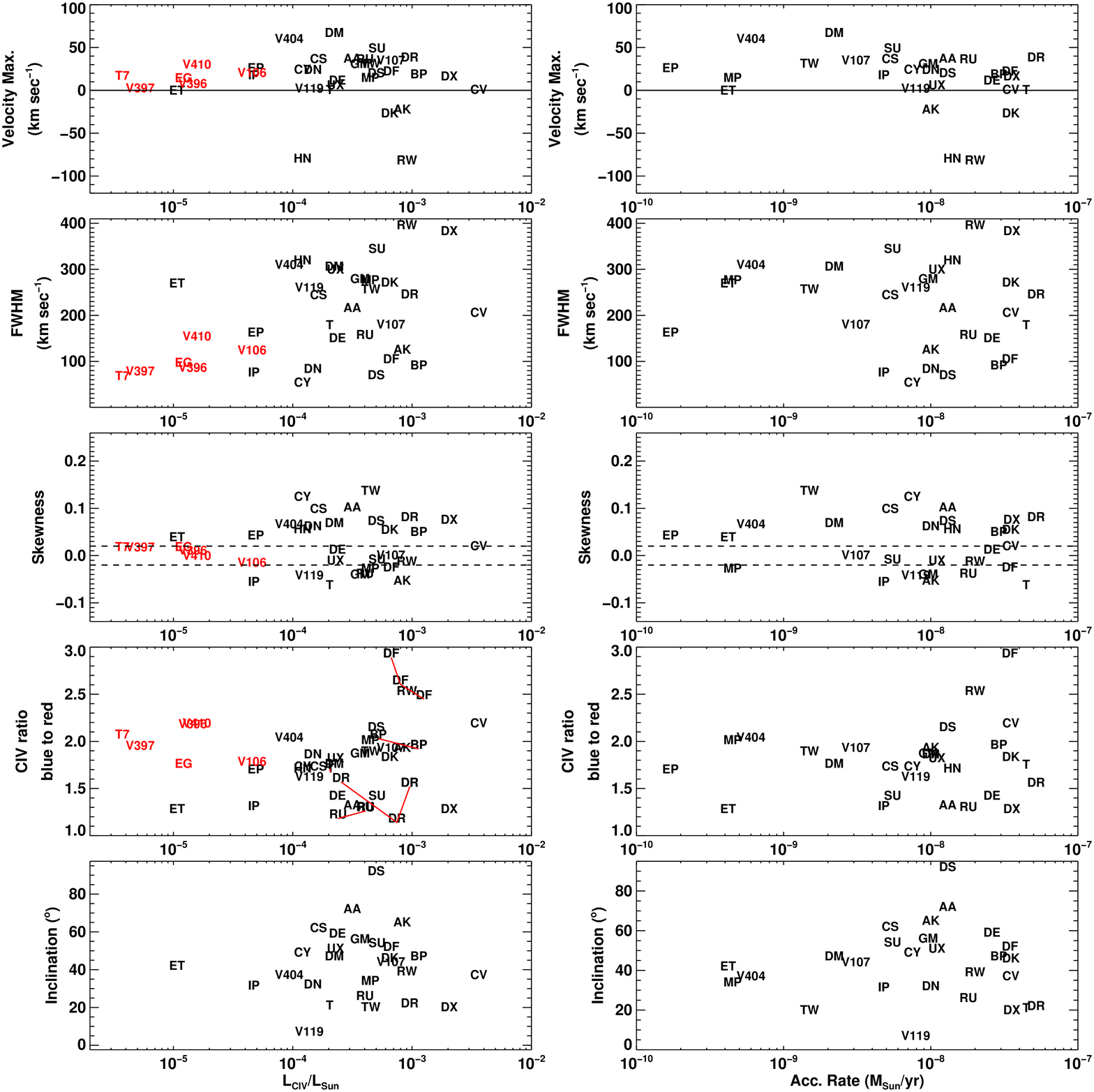}
\caption{Measurements of the line shape. An abbreviated name is used for each star. WTTSs are plotted in red. To avoid crowding the plots, error bars are not shown. {\bf First row}: Velocity at the maximum flux (measurement errors: $\sim$2 \kms; systematic error: $\sim$3 \kms for \STIS\ data, $\sim$15 \kms for \COS). {\bf Second row}: FWHM (errors: $\sim$5 \kms). {\bf Third row}: Skewness (errors: $\sim$0.005). For the skewness, the dashed lines indicate $\pm0.02$, the region within which we consider the profiles to be symmetric. {\bf Fourth row:} Scale factor from the 1550 \AA\ to the 1548 \AA\ \civ\ line. Errors are $\sim$0.1 - 0.3. We also indicate the variability path followed by BP Tau, DF Tau, DR Tau, T Tau N, and RU Lup.;  {\bf Fifth row:} Inclination, with zero meaning ``face-on." When the inclination angle errors are given in the literature, they are $\sim10 \deg$.   \label{vel_max}}
\end{figure}

Figure \ref{vel_max} (Second row) shows that, as a group, the \civ\ WTTSs lines are narrower ($\overline{FWHM}=90\pm10$ \kms) than the CTTSs ones ($\overline{FWHM}=210\pm20$ \kms).  The p-value of the KS test comparing the two samples is 0.02, implying that the difference is significant. 

The WTTS also have a smaller FWHM range (from 60 to 145 \kms) than the CTTSs (from 45 to 387 \kms). Note that the FWHM of the WTTSs is comparable to that of the CTTSs in some stars, the result of the strong narrow component dominating the CTTS lines (see for example the \civ\ lines for DS Tau in Figure \ref{panel_2}). The FWHM clearly fails to capture most relevant information regarding the line as it does not take into account the multi-component nature of the lines.

The observed FWHM is uncorrelated with line luminosity or accretion rate. The plots in Figure \ref{vel_max} (Second row) suggest that the FWHM scatter increases with accretion rate, but this apparent increase is not statistically significant.
 
\subsubsection{The skewness as a function of luminosity and accretion rate}

In Figure \ref{vel_max} (Third row) we plot skewness versus \civ\ luminosity. The skewness compares the velocity of the peak to the mean velocity  of the profile. It is defined as $(V_{Max}-\overline{V})/\Delta V$, where $V_{Max}$ is the velocity at maximum flux and $\overline{V}$ is the flux-weighted mean velocity over an interval $\Delta V$ centered on the maximum velocity.  To calculate the skewness of the \civ\ line, we subtracted the continuum and interpolated over the \htwo\ lines present in the \civ\ intervals. For the 1548 and 1550 \AA\ lines, $\overline{V}$ is measured within $\pm$250 km s$^{-1}$ and $\pm$150 km s$^{-1}$ from the maximum, respectively. To make the values comparable, we normalize to $\Delta V=250$ \kms\ in each line. The 1548 \AA\ and 1550 \AA\ values are then averaged.  

Qualitatively, values of skewness within $\sim \pm0.02$ indicate a symmetric line. Positive values indicate a line extending to the red. The absolute value of the skewness serves as a quantitative measure of asymmetry. 70\% (20/29) CTTS have skewness greater or equal to zero, with 52\% (15/29) having positive ($>0.02$) skewness.

All the WTTS have skewness values consistent with symmetric lines.  A KS test comparing both populations shows that the difference in asymmetry is significant (p-value=0.002).

As with FWHM, the skewness is uncorrelated with line luminosity or accretion rate.


\subsubsection{The line scaling as a function of luminosity and accretion rate}
  \label{shape}
  
 The fourth row panels of Figure \ref{vel_max} compare the ratios of the 1548 \AA\ to the 1550 \AA\ \civ\ lines as a function of $L_{C IV}$ (left) and $\dot{M}$ (right).   These are not flux ratios, but the scaling factors used to match both line profiles in Figure \ref{all_civ}. As mentioned before (Section \ref{comparison}), the scaling factors between the \civ\ lines are indicative of the line's optical depth compared to the thermalization depth (Table \ref{ForPaper_TableLineFluxes}). 
 
 All WTTSs and over half of the CTTSs (19/29 - 66\%; 6/7 - 86\% of TD, 12/22 - 55\% of the non-TD) have ratios $>$1.65, consistent within the errors with thin or effectively thin emission. The difference in line opacity between stars with TDs and those without is not significant. For the rest of CTTSs, 10/29 (34\%) have ratios which are consistent with small absorption mean free paths.  
 
We do not observe a correlation between this measure of opacity and either line luminosity or accretion rate. The apparent increase in scatter at high accretion rates is not significant, according to a two-sided KS test (p-value=0.6). Furthermore, objects with \civ\ ratios $<1.7$ are found for all luminosities and accretion rates. 

As indicated in Section \ref{comparison}, DF Tau presents a particular case for which the scaling factor is significantly $>2$, an impossible value unless there is extra emission in the 1548 \AA\ line or extra absorption in the 1550 \AA\ one.

 \subsubsection{The role of the inclination in the measured luminosity and accretion rate}
 \label{inc_and_rate}
Geometric explanations are often invoked in the literature to explain the shapes of these emission lines.  For example, \cite{lamzin2004} suggest that the observed accretion in TW Hya occurs at low stellar latitudes, and that the lack of separate line components coming from the pre- and post-shock in other stars may be due to equatorial layer accretion.  

Figure \ref{vel_max} (Bottom) shows the relationship between inclination and line luminosity or accretion rate. We do not observe particularly larger or smaller fluxes at high or low inclinations. According to the KS test, the distributions of $L_{C IV}$ and  $\dot{M}$ are the same between objects with  $i>45^\circ$ and those with $i\leq45^\circ$. Also, the distribution of inclinations for objects with high $\dot{M}$ or high $L_{C IV}$  is statistically the same that for objects with low $\dot{M}$ or low $L_{C IV}$. In particular, the apparently empty region in the panel at high inclinations and low accretion rates is not significant and does not provide evidence that the disk or the accretion flow are obscuring the accretion diagnostics. Furthermore, we do not find any significant correlations between the line ratio, $V_{Max}$, FWHM, or skewness, and inclination. 

If the \civ\ lines originate in a localized accretion stream one would expect to detect more stars at low inclinations than at high inclinations: for face-on systems the accretion stream will always be visible, while for systems almost edge-on this is not the case, if the region below the disk is blocked from view. For a random distribution of accretion spot positions in the stellar hemisphere, we expect to see an accretion spot in 58\% of stars with $0^\circ<i<45^\circ$ and in 42\% of stars with $45^\circ<i<90^\circ$ (ignoring disk flaring). In our sample, we have 37\% of targets with inclinations larger than 45$^\circ$, and the standard deviation of this count is 10\%. Therefore, the observed difference in the number of stars with high and low inclinations is not significant. Over a hundred CTTSs with known inclinations are needed before we can distinguish 42\% from 58\% at the 3$\sigma$ level. The current inclination dataset is not complete enough to reveal geometric information about the \civ\ lines.  
 
On the other hand, one could assume that the \civ\ UV lines are not emitted from a particular place but covers the whole star. For a given star, the \civ\ luminosity should then decrease linearly, by a factor of two, as the inclination increases from 0$^\circ$ to 90$^\circ$.  We do not observe this effect either, indicating, at least, that the observed scatter is dominated by intrinsic differences in the objects and not by geometry.




\subsubsection{Conclusions from the non-parametric analysis}

In conclusion, WTTS and CTTS \civ\ lines have comparable velocities at maximum flux, but the CTTS lines are generally broader and more asymmetric. In the case of CTTSs, neither the velocity of maximum flux, the FWHM, the skewness of the line, nor the ratio between the two \civ\ lines are correlated with \civ\ luminosity or accretion rate. The right column panels from Figure \ref{vel_max} do show increased scatter in these quantities as the accretion rate increases, suggesting that objects with large accretion rates have more diverse line shapes. However, the differences in the distributions (of the velocity, FWHM, skewness) with high and low accretion rates are not significant.  More observations of stars with accretion rates $\leq4 \times 10^{-9} \msun$/yr are needed. Among the three pairs of shape quantities ($V_{Max}$ vs. FWHM, FWHM vs. Skewness,   $V_{Max}$ vs. Skewness) there are no significant correlations. 

The scaling of the 1548 \AA\ line to the 1550 \AA\ one should be 2 if the lines are emitted from a thin or effectively thin medium. This is the case in all the WTTSs and in about 70\% of CTTSs. This measure of the opacity is not correlated with accretion rate or line luminosity. However, in Section \ref{onset} we show that the NC of the line is correlated with accretion rate.

There are no correlations between line shape parameters and inclination or between inclination and accretion rate or line luminosity, but we conclude that the data are not complete enough for inclination to be a strong descriptor in the sample. 
 
\subsection{C IV Gaussian decomposition}
\label{fits}

Different regions of the T Tauri system may be contributing to the  \civ\ lines. In the context of the magnetospheric accretion paradigm, the pre- and post-shock regions should be the dominant sources of the observed line emission. Shocks in an outflow, hot winds, and the stellar atmosphere may also contribute to the emission. To examine the line kinematics of the different regions we decompose each \civ\ line in one or two Gaussian functions.  We are not asserting that the mechanism giving origin to the lines produces Gaussian shapes, although turbulent flows will do so \citep{gunther2008}. Representative decompositions are shown in Figure \ref{all_ctts}. 

The primary goal of the Gaussian decomposition is to obtain widths and centroids for the main line components. According to the magnetospheric accretion model, post-shock emission lines should have small velocity centroid distribution about the stellar rest frame velocity, and if no turbulence is present, the lines should be narrow. Pre-shock emission lines, which likely originate in a larger volume upstream from the accretion flow \citep{cal98}, should have larger velocity centroid distribution in the stellar rest frame and broader lines. 

Table \ref{ForPaper_TableFits_CIV} presents the results of fitting one or two Gaussians to each of the \civ\ lines, after subtracting the continuum and interpolating over the \htwo\ line R(3)1-8, and the \sii, \cii, \feii -complex, when present. For each target, we assume that the \civ\ lines are always separated by 500.96 \kms, and that they have the same shape. This results in a 4-parameter fit when fitting one Gaussian to each line: Height for the 1548 \AA\ line; height for the 1550 \AA\ line; width $\sigma$s; centroid velocity for the 1548 \AA\ line. When fitting two Gaussians to each line, we have an 8-parameter fit: For the 1548 \AA\  line, the heights of the broad (A$_{BC}$) and narrow (A$_{NC}$) components; analogous parameters for the 1550 \AA\ line; $\sigma$ for the broad component; $\sigma$ for the narrow component; centroid velocities for the broad (v$_{BC}$) and narrow (v$_{NC}$) components of the 1548 \AA\  line.

Table \ref{ForPaper_TableFits_CIV}  also indicates whether the data were taken with \COS\ (possible systematic wavelength error assumed to be 7 \kms) or \STIS\ (possible systematic wavelength error of 3 \kms). We also list the parameters derived for the multi-epoch observations of BP Tau, DF Tau, DR Tau, RU Lup, and T Tau N that we will consider in Section \ref{variability}. 

\cite{wood1997} analyzed the \civ\ lines of 12 stars with spectral types F5 to M0. They found that the observed line profiles could be better fit with both a narrow and a broad Gaussian component, than with a single Gaussian component. For the type of stars included in their sample (dwarfs, giants, spectroscopic binaries, and the Sun), the \civ\ line is a transition-region line, as it is in the WTTSs. Because of this, we fitted both NCs and BCs to all the WTTSs, except V410 Tau. For V410 Tau, each \civ\ line can be well fitted with only one NC, although this may just be the result of the low S/N in the spectrum.

When comparing the two \civ\ lines in section \ref{comparison}, we mentioned that AK Sco, CS Cha, DX Cha, DF Tau, RU Lup, and RW Aur A were objects for which the line doublet members had different shapes and/or extra unidentified emission in one of the doublet members. We do not perform the Gaussian decomposition for RW Aur A or DX Cha.  For AK Sco, we list the parameters derived from the Gaussian fits (Table \ref{ForPaper_TableFits_CIV}), but we do not use these results when exploring correlations. For the rest, we interpolate the profiles over the apparent the extra emission.

\begin{figure}
\epsscale{1}
\includegraphics[width=3.4in]{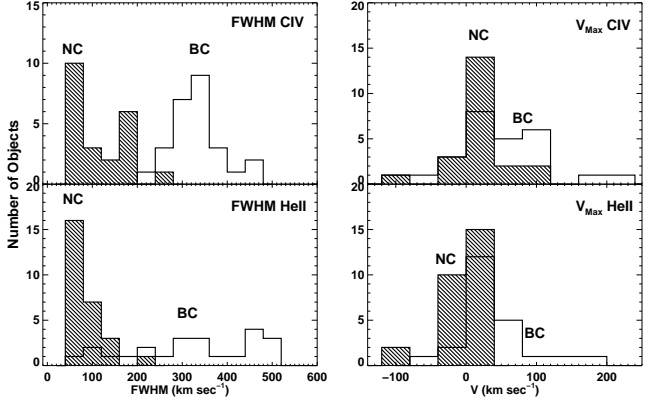}
\caption{Histogram of line components for CTTSs. {\bf Left Column}: Full widths at half maxima for \civ\ and \heii. {\bf Right Column}: Velocity at maximum flux for \civ\ and \heii. {\bf Clear histogram}: BC; {\bf Diagonal hatch histogram}: NC\label{histogram_vel}}
\end{figure}

For 4 objects (CY Tau, DM Tau, ET Cha, and UX Tau A) we decompose the \civ\ lines in only one broad Gaussian component. For the rest of the CTTSs, the decomposition requires both a narrow and a broad Gaussian components, and Figure  \ref{histogram_vel} shows the distributions of velocity centroids and FWHM.  Average velocity and FWHM values are given in Table \ref{velocities}. Typical full widths at half maxima of CTTS NCs range from 50 to 240 \kms, with an average of 130 \kms, while BCs widths range from 140 to a 470 \kms, with an average of 350 \kms. The velocity centroids range from -100 \kms\ to 200 \kms.  The BC velocity is larger than the NC velocity in 70\% of the CTTS sample, and the distribution of BC velocities tend to be more positive ($\overline{V_{BC}}\sim40$ \kms) than that of the NC velocities ($\overline{V_{NC}}\sim30$ \kms), giving some of the profiles the characteristic ``skewed to the red" shape. 


\subsubsection{The optical depth as a function of accretion}
\label{onset}

\begin{figure}
\epsscale{1}
\includegraphics[width=3.4in]{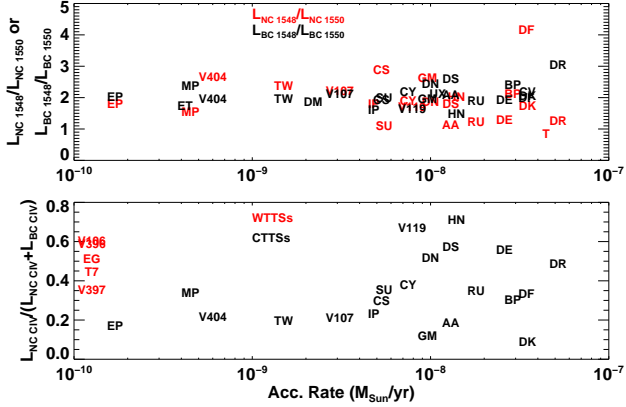}
\caption{The \civ\ line shape as a function of accretion rate. Typical errors in the ordinate axes are $\sim$10\%. {\bf Top}: The ratio of the luminosity 1548 \AA\ to 1550 \AA. Red labels indicate the ratio of the luminosities of the BCs. Black labels are the ratio of the luminosities of the NCs. {\bf Bottom}: The fraction of the line luminosity in the NC. Black labels are for CTTSs, while red are for WTTSs. For CTTSs, the NC contributes about 20\% of the luminosity at low accretion rates, and up to 80\% at high accretion rates.\label{opacity}}
\end{figure}

Figure  \ref{opacity} (Top) shows the ratio of the height of the NC in the 1548 \AA\  line to the NC in the 1550 \AA\  \civ\ line as a function of accretion rate, as well as the ratio of the heights of the BCs. Most observations are grouped around 2 although there is a lot of scatter, particularly at high accretion rates.  The plot reveals that the NC of DF Tau is anomalous ($>2$), while the BC has an optically thin ratio. Individual Gaussian components of other objects (CS Cha, DR Tau) are also anomalous. Furthermore, there is a population of objects with NC ratios close to 1: AA Tau, DE Tau, DR Tau, RU Lup,, SU Aur, and T Tau.  However, the difference in the distribution of BC ratios and NC ratios is marginal according to the KS test (p-value=6\%), and the correlation of the ratio of NC luminosities  with accretion rate is not significant (Pearson's r=-0.37, p-value=10\%).  We conclude that for most objects, both the NC and BC ratios are close to 2, although there is considerable scatter, and that there are some peculiar objects at accretion rates $>4 \times 10^{-9}$ M$_\odot$/yr.



The bottom panel of Figure \ref{opacity} shows that the contribution of the NC to the overall profile increases with accretion rate. For low accretion rates ($<4 \times 10^{-9}$ M$_\odot$/yr), the average NC contribution to the luminosity is $\sim$20\% while for high accretion rates it is  $\sim$40\% on average.  While the Pearson's r=0.4 (p-value=6\%) suggest that this correlation is not significant, this statistical test assumes that the quantities being compared are sampled from a bivariate Gaussian distribution. This is likely not the correct assumption for the heterogeneous sample of CTTSs being considered here. A better correlation test uses the Kendall rank correlation statistic, which considers only (non-parametric) rank orderings between the data \citep{feigelson2012}. The Kendall's $\tau$=0.30 (p-value=0.05) is at the threshold of what we consider significant.

If the increase in accretion rate is due to larger accreting area or larger density, and both the narrow and broad components are emitted from regions that are optically thin or effectively thin, no correlation should be observed. This is because the rate of collisional excitation will change linearly with density and both lines will increase at the same rate. The observed correlation implies that the region responsible for the BC may be becoming optically thick at high accretion rates.

As in main-sequence stars, the WTTS line shapes (blue labels in Figure \ref {opacity}) are characterized by a strong NC and a weak BC \citep{wood1997, linsky2012} and tend to have stronger NC contributions to the total flux than low accretion rate CTTSs. The luminosity in NCs and BCs, in both WTTSs and CTTSs increases with total \civ\ luminosity (not shown). 

 The bottom panel of Figure \ref{opacity} raises the issue of the transition from WTTS to CTTS as a function of accretion rate. The WTTSs generally have strong NC, while low accretion rate objects have very weak NC, compared to BC.  Does the accretion process suppress the NC present in the WTTSs or does it enhance the BC? We believe the latter to be true. The lowest accretion rate object shown in Figure \ref{opacity} (Bottom) is EP Cha (RECX 11). For this star the total \civ\ luminosity is $\sim$6 times larger than for most of the WTTSs, with the exception of V1068 Tau. However, the NC luminosity of EP Cha  is $4\times10^{-6}$ L$_\odot$, similar to that of a low-luminosity WTTSs. In other words, most of the extra \civ\ luminosity that distinguishes this CTTS from the WTTSs is due to the generation of the BC. This suggest that the accretion process generates first a BC, with the NC becoming increasingly important at larger accretion rates.  

From this it follows that all accreting stars should show a BC. In average, the flux in the BC grows at a slower rate than the flux in the NC as the accretion rate increases, but it is not clear why some stars develop a strong NC (like DS Tau) while others do not (like GM Aur). These statements may not be valid outside the mass or accretion rate range of the sample considered here. For example, \cite{france2010} present UV spectra of the brown dwarf 2M1207, from which they derive an accretion rate of $10^{-10}$ M$_\odot$/yr, comparable to that of EP Cha. This value is derived using the calibration between \civ\ flux and accretion rate from \cite{jvl00}. On the other hand \cite{herczeg2009} derives an accretion rate of $10^{-12}$ M$_\odot$/yr based on the Balmer excess emission. At any rate, each \civ\ line can be fit with only one, very narrow, Gaussian component with FWHM=36.3$\pm$2.3 \kms.\footnote{In their published analysis,  \cite{france2010}  fit the \civ\ lines of 2M1207 with two Gaussian components each. However, that analysis is based on an early reduction of the COS data. A new re-processing and new Gaussian decomposition shows that the \civ\ line can be fit with one Gaussian component (K. France, personal communication).} This is then a case of a low-mass accreting young object without a BC. The very narrow \civ\ lines may be the result of the smaller gas infall velocity in 2M1207 ($\sim$ 200 \kms, using  the stellar parameters from \citealt{riaz2007}) compared to the sample presented here ($\sim$300 \kms). The smaller gas infall velocity will result in lower turbulence broadening.

\subsubsection{The kinematic predictions of the magnetospheric accretion model}
 \label{centroids}
For the magnetospheric accretion model, the gas speed in the accretion flow before the shock should reach velocities $\sim$300 \kms\ for typical stellar parameters \citep{cal98}, although the interplay between line-of-sight and the complex magnetospheric structure may result in line-of-sight velocities smaller than this.

If it originates primarily in the accretion shock region, the \civ\ line emission (as well as the \siiv, \nv, and \heii\ lines) comes from plasma very close to the stellar surface and the gas in this region should be moving away from the observer. Therefore, we expect that the velocity of the flow should be positive (v$_{BC}>0$, v$_{NC}>0$). If the dominant emission in the broad and narrow components comes only from the pre- and post-shock  regions, respectively, then v$_{BC}>$v$_{NC}$. The velocity of the post-shock gas decreases after the shock surface, and depending on the origin of our observational diagnostic, we may observe velocities ranging from 4$\times$ less than the pre-shock gas velocity to zero \citep{lamzin1995, lam03,lamzin2003}. 

\begin{figure}
\epsscale{1}
\includegraphics[width=3.4in]{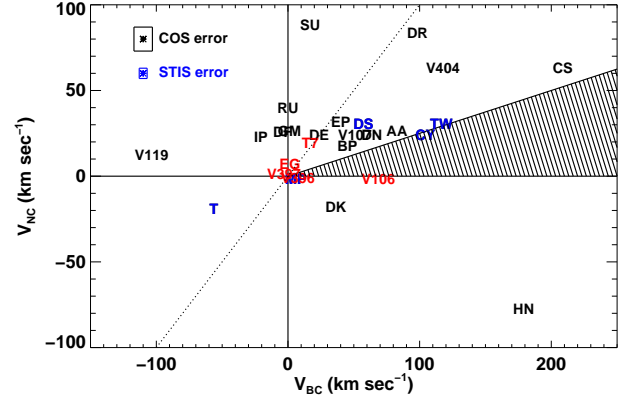}
\caption{Narrow vs. the broad velocity components for the Gaussian decomposition of the \civ\ lines. The plot is divided in quadrants (solid horizontal and vertical lines). {\bf blue labels} indicate targets observed with STIS, while {\bf black labels} are for targets observed with COS. {\bf Red labels} are for WTTSs (all observed with COS). The errors in the velocities are dominated by systematic errors, shown as boxes in the upper left of the diagram. Errors in the wavelength scale will move the points parallel to the dotted line. If the line emission is dominated by the magnetospheric shock, the data points should reside within the hatched region (v$_{NC}<$ v$_{BC}/4$). More generally, when taking into account the wavelength errors, all points below the dotted line in the upper right-hand quadrant may reside within the hatched region  \label{vbc_vs_vnc}}
\end{figure}

 Observationally, v$_{BC}$ and v$_{NC}$ are uncorrelated with accretion rate or line luminosity.  Figure \ref{vbc_vs_vnc} compares the velocity centroids of the broad and narrow components, for 22 CTTSs and 5 WTTSs. Black labels indicate CTTSs observed with \COS, while orange labels are CTTSs observed with \STIS. The plot also shows, in red, the velocity centroids of the WTTSs, all observed with \COS.  The errors in these velocities are dominated by the systematic wavelength scale error, illustrated by the boxes on the upper-left corner of Figure \ref{vbc_vs_vnc}. We have argued in Section \ref{non_par} that the errors in the \COS\  dataset are $\sim$7 \kms. Errors in the wavelength scale cause the data points to move parallel to the dotted line. 
 
We focus first on the upper right quadrant of Figure \ref{vbc_vs_vnc}, those objects with both v$_{BC}>0$ and v$_{NC}>0$. The hatched region of the plot corresponds to v$_{BC} \geq 4$ v$_{NC}$. As we can see, only TW Hya and CY Tau reside fully within this region.  In addition, errors in the wavelength calibration may explain why some objects (AA Tau, BP Tau, CS Cha, DE Tau, DN Tau, DS Tau, EP Cha, MP Mus, V1079 Tau, and V4046 Sgr) reside away from the hatched region. Note that while the rotation of the star may contribute to  v$_{BC}$ and v$_{NC}$, typical values of $v \sin i$ for CTTSs are 10 to 20 \kms, and so the velocity pair would only move to higher or lower velocities by up to this amount, parallel to the dotted line.

The other 10 objects  are ``anomalous," either because one or both of the velocities is negative and/or v$_{BC}\lesssim$v$_{NC}$, beyond what could be explained by pointing errors. Stellar rotation alone will not bring these objects to the hatched region. As a group, they all have accretion rates larger than $4\times 10^{-9} \msun$/yr, and they make up half of the objects with accretion rates this large or larger.  From the point of view of the magnetospheric accretion paradigm, they present a considerable explanatory challenge.

In turn, these anomalous objects come in two groups: those for which both components are positive or close to zero but v$_{BC}<$v$_{NC}$ (DF Tau, perhaps DR Tau, GM Aur, RU Lup, SU Aur), and those objects for which one or both of the components are negative (DK Tau, HN Tau, IP Tau, T Tau N, V1190 Sco). Emission from the latter group may have contributions from outflows or winds.



\subsection{The accretion process in time: multi-epoch information in C IV}
 \label{variability}

Line flux variability in CTTSs occurs on all time scales, from minutes to years. It is therefore relevant to ask how does the line shape change in time and what is the impact of these changes on the general statements we have made. 

\begin{figure}
\epsscale{1}
\center
\includegraphics[width=6in]{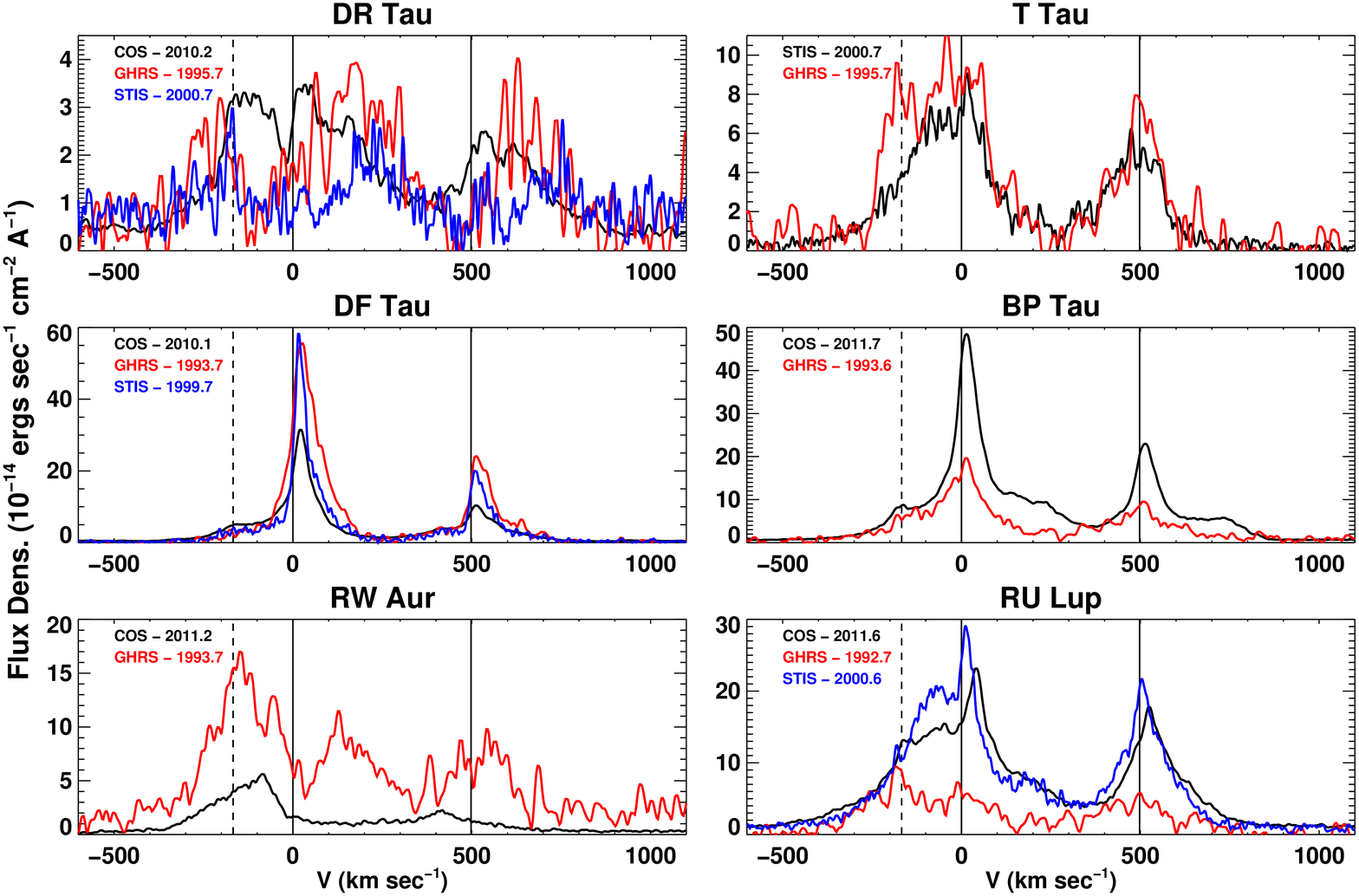}
\caption{CTTSs with multi-epoch, high-resolution observations of \civ.  Nominal line centroids are indicated by a vertical solid line and the position of the R(3)1- 8 \htwo\ line is indicated by a dashed line. {\bf Black}: Observations from this paper. {\bf Red}: GHRS observations from  \cite{ard02}. The systematic error in the GHRS wavelength scale is 20 \kms. For DR Tau there are two GHRS observations available in the literature, one from 1993 and the other from 1995. To avoid crowding the figure we plot here only the one from 1995. {\bf Blue}: For DR Tau, this observation belongs to HST program GO 8206 (PI Calvet).  For DF Tau and RU Lup, the spectra are from \cite{her05, her06}. \label{one_panel_multiepoch}}
\end{figure}

High-resolution multi-epoch observations of the \civ\ lines are available for a subset of our objects, plotted in Figure \ref{one_panel_multiepoch}. The spectra were obtained from \citet{ard02} (GHRS observations: BP Tau, DF Tau, DR Tau, RU Lup, T Tau, and RW Aur), \citet{her05} (\STIS: RU Lup), \citet{her06} (\STIS: DF Tau) and the MAST archive (HST-GO 8206; \STIS: DR Tau). The epoch of the observations is indicated in the figure. Note that the apertures of the three spectrographs are different. The GHRS observations were performed with the Large Science Aperture (2"$\times$ 2" before 1994, 1.74"$\times$1.74" after 1994), the Primary Science Aperture for \COS\ is a circle 2.5" in diameter, and the \STIS\ observations where obtained with the apertures 0.2"$\times$ 0.06" (for T Tau) or 0.2"$\times$0.2" (for the other objects). 

For DR Tau the DAO observations show double-peaked \civ\ emission as well as strong emission in the blue wing of the 1548 \AA\  line. We have identified the additional blue wing emission as a mixture of \htwo, \sii, \cii, and \feii. The parameters that we have listed for DR Tau in Table \ref{ForPaper_TableFits_CIV} provide a good fit to the overall profile, but not to the double-peaked \civ\ emission.  In the \GHRS\ observation (red, from 1995) the extra  \sii, \cii, and \feii\ emissions are not present, leaving only a low S/N \htwo\ line. The low-velocity peak of the \civ\ line observed in the DAO data is not present in the GHRS data. In addition to this 1995 observation, \cite{ard02}  describes a 1993 observation (not shown in Figure \ref{one_panel_multiepoch}), which shows blueshifted emission at --250 \kms\ present in both line components. The \STIS\ observations from HST program GO 8206 (PI Calvet) show a strong  \htwo\ line. The centroids of the \civ\ lines are either redshifted by $\sim$200 \kms\ or the centers (within $\pm$150 \kms\ of the rest velocity) are being absorbed. If redshifted, this is the largest redshift observed in the sample, although a comparable shift is seen in  v$_{BC}$ for CS Cha (Figure \ref{panel_3}).  We may be observing extended \civ\ emission that is not seen in the narrow-slit \STIS\ observations \citep{schneider2013}, \civ\ absorption from a turbulent saturated wind or the disappearing of the accretion spot behind the stellar limb as the star rotates.  

T Tau shows a change in the \htwo\ emission and a small decrease in the strength on the NC.  \cite{walter2003, saucedo2003} show that the \htwo\ emission around T Tau is extended over angular scales comparable to the GHRS aperture, and the smaller \STIS\ flux is due to the smaller aperture.

For DF Tau, the ratio between the two line members remain anomalously high, $\sim$3 in all epochs. For BP Tau, a decrease in the flux is accompanied by a decrease in both components, although the NC decreases more strongly. Changes over time in the velocity centroids are not significant. 

For RW Aur A the difference between GHRS and \STIS\ DAO observations is also dramatic. As we have shown, the DAO observations can be explained by assuming that we are observing two sides of a bipolar outflow. The GHRS flux is larger, and dominated by three peaks, the bluemost of which is likely the R(3)1-8 \htwo\ line. The other two do not match each other in velocity and so they cannot both be \civ. \cite{errico2000} have suggested that the early GHRS observations may be affected by \feii\ absorption. We do not find evidence for  \feii\ absorption within the \civ\ lines for any other star, nor for the DAO observations of RW Aur, and so we discard this possibility. Based on the spectroscopic and photometric variability, \cite{gahm1999} have suggested that a brown dwarf secondary is present in the system, although its role in the dynamics of the primary is uncertain. The \COS\ aperture is larger, suggesting that the changes are due to true varaibility.
 
 Dramatic changes are also seen in RU Lup, as noted by \cite{her06} and \cite{france2012}. The \civ\ line, which is almost absent in the \GHRS\ observations is 4$\times$ stronger in the \STIS\ observations. The DAO observations presented here are similar to the latter, although the strength of the extra \sii, \cii, and \feii\ emissions is also variable. The excess emission in the blue wing of the 1548 \AA\ line described in Section \ref{comparison} is also present in the \STIS\  observations. 
 
For all of the multi-epoch CTTSs observations except RW Aur A, Figure \ref{vel_max} (Fourth row) shows the path that the line ratio (the optical depth indicator) follows as a function of line luminosity. The changes in the value of the line ratio are not correlated with the changes in the \civ\ line luminosity, as we observed before.

  
\subsection{Conclusions from the parametric analysis}

Most CTTSs \civ\ line profiles can be decomposed into narrow ($\overline{FWHM}\sim130$ \kms) and broad ($\overline{FWHM}\sim 350$ \kms) components, with the BC redshifted with respect to the NC in 70\% of the CTTSs sample. The fractional contribution to the flux in the NC increases (from $\sim$20\% to 40\% on average) and may become more optically thick with increasing accretion rate. Strong narrow components will be present in objects of high accretion rate, but high accretion rate by itself does not guarantee that the \civ\ lines will have strong NCs.



The component velocities in about 12 out of 23 CTTSs are roughly consistent with predictions of the magnetospheric accretion model, in the sense that   v$_{BC}>0$, v$_{NC}>0$ and v$_{BC}\gtrsim 4v_{NC}$. For most of the 12 the NC velocity seems too large or the BC velocity too small, compared to predictions, although this may be the result of errors in the wavelength calibration. For 11 of the CTTSs the kinematic characteristics of the \civ\ line cannot be explained by the magnetospheric accretion model: these objects (AK Sco, DF Tau, DK Tau, DR Tau, GM Aur, HN Tau A,  IP Tau, RU Lup, SU Aur, T Tau N, V1190 Sco) have BC velocities smaller than their NC velocities, or one of the velocity components is negative. An examination of the systematic pointing errors which produce offsets in the wavelength scale lead us to conclude that these do not impact the conclusions significantly.

Multi-epoch observations reveal significant changes in morphology in all the lines, from one epoch to the next. The line velocity centroids remain relatively constant in low accretion rate objects such as DF Tau and BP Tau, but the overall line appearance change significantly for high accretion rate CTTSs, like DR Tau.  In the case of RU Lup, the observed variability of the line may be consistent with the accretion spot in \civ\ coming in and out of view.


\section{The He II line }
\label{heii}
The 1640 \AA\ \heii\ line is the ``Helium $\alpha$" line, analogous to neutral hydrogen line H$_\alpha$, corresponding to a hydrogenic de-excitation from level 3 to level 2 \citep{brown1984}. In principle, the line is a blend of 1640.33  \AA , 1640.34 \AA , 1640.37 \AA , 1640.39 \AA , 1640.47 \AA , 1640.49 \AA ,  and 1640.53 \AA. Of these, 1640.47 \AA\ ($Log(gf)=0.39884$) should dominate the emission, followed by 1640.33 \AA\ ($Log(gf)=0.14359$), as the recombination coefficient is largest to 3d $^1D$  \citep{Oster1989}.  \heii\ is strongly correlated with \civ, implying, as with \siiv, that they are powered by the same process \citep{jvl00, ingleby2011a, gdc2012, yang2012}.


For a sample of 31 CTTSs \cite{beristain2001} modeled the \heii\ 4686 \AA\ line with a single Gaussian function. Those lines are narrow, with an average FWHM of 52 \kms\ and somewhat redshifted, with an average centroid of 10 \kms. Based on the high excitation energy of the line (40.8 eV), and the redshift in the velocity centroid,  \cite{beristain2001}  argued that the \heii\ optical emission originated from the post-shock gas.  Below we find that the centroid shifts in the \heii\ UV line are comparable with the average values reported by \cite{beristain2001} for the \heii\ optical line.  

If the \heii\ 1640 \AA\ line originates in the accretion post-shock we expect it to be redshifted, similar in shape to the NC of the \civ\ line, although with lower velocity centroids as it will be emitted from a cooler, slower region of the post shock. Below we show that,  indeed, the shape is similar to the NC \civ\ line, and the velocity centroid is smaller. 


In Figure \ref{all_civ_heii} we compare the 1550 \AA\ \civ\ line to the \heii\ line, scaled to the same maximum flux. The \heii\ lines are similar to the NCs of the \civ\ lines, if present, although the latter appears slightly redshifted with respect to \heii. Most \heii\ lines can be described as having a strong narrow core and a low-level broad component. Significant emission is present within $\pm$200 \kms\ of the nominal wavelength. The exceptions to this description are HN Tau A and RW Aur A, which are blueshifted and present a strong BC, and DX Cha, for which no \heii\ is observed in the background of a strong continuum. 

\cite{gdc2012} state that \heii\ is observed only in a subset of the stars that show \civ\ emission. In their sample of ACS and IUE low-resolution spectra, 15 stars show \heii\ out of the 20 stars that present \civ\ emission.  We do not confirm this statement, as we observe the \heii\ line in all of the stars in our sample. The difference between detection rates is likely due to the differing spectral resolutions and sensitivities between our sample (R$>$10000) and theirs (R$\sim$40 at 1640 \AA). 

It is well-known that the \civ\ and \heii\ luminosities are correlated to each other (e.g. \citealt{ingleby2011a,yang2012}). The  \civ-to-\heii\ luminosity ratio measured here is significantly larger for CTTSs ($3.5\pm0.4$)  than for WTTSs ($1.3\pm0.2$). \cite{ingleby2011a} noted that the \civ-to-\heii\ luminosity ratio is close to one in field stars, but larger in CTTSs.  The fact that we measure a ratio close to unity also in WTTSs supports \cite{ingleby2011a}'s assertion that the \civ-to-\heii\ luminosity ratio is controlled by accretion, and not by the underlying stellar chromosphere \citep{alexander2005}.  
 
For the purposes of a deeper analysis we first use the non-parametric measurements (the velocity of maximum flux, the skewness and the FWHM, Table \ref{ForPaper_NonParametric}) to describe the line.  We also fit one or two Gaussians to the lines (Table \ref{ForPaper_HeII}). 

\subsection{The He II line shape: non-parametric measurements}

\begin{figure}
\epsscale{1}
\includegraphics[width=3.4in]{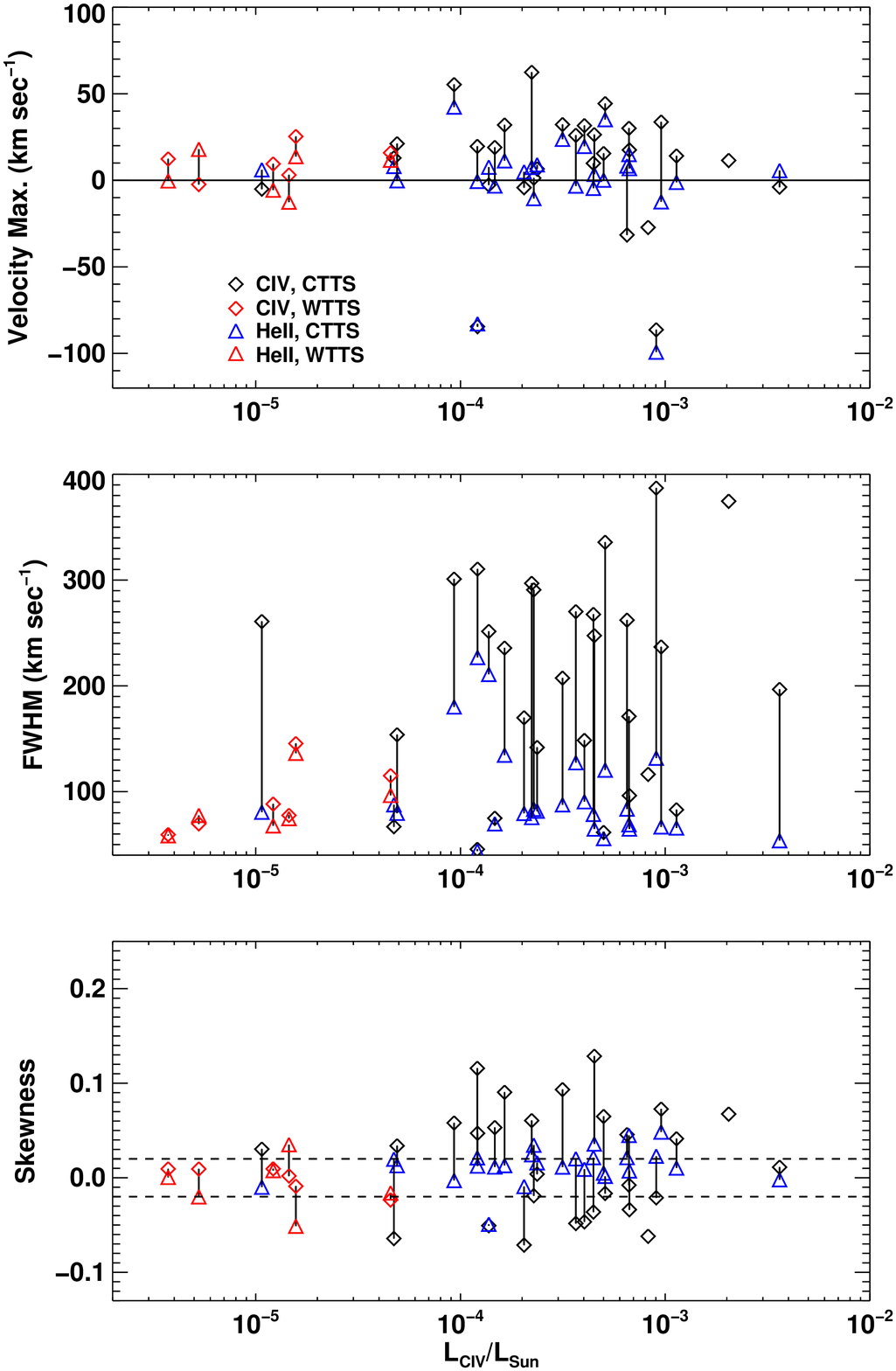}
\caption{Shape characteristics for \heii\ compared with \civ. {\bf Diamonds} indicate the \civ\ values,  {\bf blue} triangles the CTTSs \heii\ values, {\bf red} symbols the WTTSs. Values for the same star are joined by a segment.\label{vel_he_c}}
\end{figure}

We find that, except for HN Tau A and RW Aur A, the CTTSs \heii\ lines are well-centered or slightly redshifted: the average $V_{Max}$ ignoring these two stars is $7\pm3$ \kms\ (Figure \ref{vel_he_c}). As with \civ, we conclude that the error on the \COS\ wavelength scale for \heii\ should be smaller than the nominal value. This is because only 7 out of 23 objects (again ignoring HN Tau A and RW Aur A) have $V_{Max}<0$, a situation expected to occur with negligible probability if the wavelength errors are normally distributed around zero \kms. The difference in redshift between CTTSs and WTTSs is not significant.

The \civ\ lines are redder than the \heii\ lines. For the CTTSs $V_{Max\ C IV}-V_{Max\ He II}=11\pm4$ \kms.  For WTTSs, the difference is $7 \pm 6$ \kms\ (Table \ref{velocities}).

 The FWHM for \heii\ CTTSs range range from $\sim$50 to 400 \kms, with an average of 96$\pm$9 \kms, which is significantly narrower than in \civ. On the other hand, the FWHM of \heii\ lines of CTTSs and WTTSs are consistent with being drawn from the same sample, according to the KS test.

The average skewness for the CTTS  \heii\ sample is $0.01\pm0.01$, whereas the skewness of the WTTS sample is $-0.01\pm0.01$: the difference between the two populations is not significant. The \heii\ lines are significantly more symmetric (as measured by the absolute skewness) than the \civ\ lines.

As with \civ, neither the velocity shift, FWHM, or skewness are correlated with line luminosity, accretion rate or inclination.

 \subsection{The He II line shape: Gaussian decomposition}
 
 Most \heii\ lines require a narrow and a broad Gaussian component to fit the core and the wings of the line, respectively. The average of the ratio of the \heii\ NC line luminosity to the total \heii\ line luminosity is 0.6 and uncorrelated with accretion rate, and it is the same in CTTSs and WTTSs. This large contribution of the NC gives the lines their sharp, peaked appearance. As shown in Figure \ref{histogram_vel} the BC of the \heii\ lines span a much broader range than those of \civ, and their distributions are significantly different. 

Overall, this Gaussian decomposition confirms the conclusions from the non-parametric analysis. The average value of v$_{NC}$ for \heii\ is the same for the CTTS and WTTS (V$_{CTTS\ NC\ He II}$ - V$_{WTTS\ NC\ He II}$=$2\pm3$ \kms), the \civ\ CTTS line is redshifted with respect to the \heii\ line (V$_{CTTS\ NC\ CIV}$ - V$_{CTTS\ NC\ He II}$=$20\pm6$  \kms), and for WTTSs the velocities of \heii\ and \civ\ are the same (V$_{WTTS\ NC\ CIV}$ - V$_{WTTS\ NC\ He II}$=$-2\pm5$  \kms). 

Based both on the non-parametric analysis and the Gaussian decomposition we conclude that the \heii\ line is comparable in terms of redshift and FWHM in WTTSs and in CTTSs. The line is blueshifted with respect to \civ\  CTTSs but has the same velocity shift  as a \civ\ WTTS line, and as the NC of the CTTSs in \civ.  We discuss this further in Section \ref{discussion}.

\section{Si IV and N V: Anomalous abundances}
\label{siiv}
Analyses as detailed as those performed before are not possible for \siiv\ and \nv: the lines are weaker and the extra emissions and absorptions due to other species make a detailed study of the line shape unreliable.  Here we provide a high-level description of the lines and compare their fluxes to that of \civ.

\subsection{Si IV description}

The two \siiv\ lines are separated by 1938 \kms\ (see Figures \ref{panel_1} to \ref{panel_5}). The wavelength of the \siiv\ doublet members coincides with the bright \htwo\ lines R(0) 0-5 (1393.7 \AA, -9 \kms\ from the 1394 \AA\  line), R(1) 0-5 (43 \kms\ from the 1394 \AA\  line), and P(3) 0-5 (1402.6 \AA, -26 \kms\ from the 1403 \AA\  line). The strong narrow line between the two \siiv\ lines is \htwo\ P(2) 0-5 (1399.0 \AA, 1117 \kms\ away from the 1394 \AA\ \siiv\ line). Additional \htwo\ lines  (R(2) 0-5, 1395.3 \AA; P(1) 0-5, 1396.3 \AA; R(11) 2-5, 1399.3 \AA) are observed between the doublet members of DF Tau and TW Hya. For this paper, the \htwo\ lines are considered contaminants in the spectra and we will ignore them. A detailed study of their characteristics has appeared in \cite{france2012}. The panels in the appendix also show the {O {\scshape iv}} line at 1401.16 \AA, sometimes observed as a narrow emission line blueward of the 1403 \AA\ \siiv\ line (see for example DE Tau). We have also indicated the position of the CO 5-0 bandhead \citep{france2011}.  In the case of WTTSs, no \htwo\ lines are observed in the \siiv\ region and narrow well-centered \siiv\ lines characterize the emission. 

To the extent that the \siiv\ lines can be seen under the \htwo\ emission, their shape is similar to that of the \civ\ lines, as was suggested in \cite{ard02} (see for example IP Tau in Fig. \ref{panel_3}). A notable exception is the WTTS EG Cha, for which the \siiv\ lines have broad wings that extend beyond $\pm$500 \kms, not present in \civ. Because the observations were taken in TIMETAG mode, we have time-resolved spectra of the \siiv\ region that shows sharp increase in the count rate ($\sim10\times$ in 300 secs) followed by a slow return to quiescence over 1000 secs. These observations indicate that the WTTS EG Cha was caught during a stellar flare. The broad line wings are observed only during the flare. A detailed study of this flare event is in preparation.


\subsection{N V description}

The two \nv\ lines are separated by 964 \kms. Figures \ref{panel_1} to \ref{panel_5}  also indicate the positions of the \htwo\ lines R(11) 2-2 (1237.54 \AA),  P(8) 1-2 (1237.88 \AA), and P(11) 1-5 (1240.87 \AA). In addition to the \htwo\ lines, we note the presence of \niti\ absorptions  \citep{her05} at 1243.18 \AA\ and 1243.31 \AA\ (1055 \kms\ and 1085 \kms\ from the 1239 \AA\ line). 

The two \niti\ lines are clearly seen in the 1243 \nv\ member (the red line) of  EP Cha (Figure \ref{panel_4}). \niti\ absorptions are observed in most CTTSs (Exceptions: AA Tau -- \niti\ in emission, DN Tau, EG Cha; Uncertain: AK Sco, CV Cha, CY Tau).  Some objects (e.g. RU Lup) show a clear wind signature in \niti, with a wide blueshifted absorption which in this object absorbs most of the \nv\ line.  

The 1239 \AA\ line of \nv\ (the blue line, not affected by extra absorption) and the 1550 \AA\ line of \civ\ (the red line, not affected by \htwo) have  comparable wing extensions, velocity centroids, and overall shapes. As for \civ\ and \siiv, the doublet lines of \nv\ should be in a 2:1 ratio if effectively thin although the presence of the \niti\ line makes estimating the ratio impossible. For CTTSs there are other unidentified absorption sources in vicinity of the 1243 \AA\ \nv\ line. These can be seen most easily in the spectra of DE Tau ( --32 \kms), TW Hya (--87 \kms, --32 \kms), GM Aur, and V4046 Sgr (50 \kms) with respect to the rest velocity of the 1243 \nv\ line. It is unclear if these are \niti\ absorption features in a highly structured gas flow or absorptions from a different species. 


 For the WTTSs both \nv\ lines are copies of each other, and copies of the WTTS \siiv, \civ, and \heii\ lines. Therefore, the extra absorption in the CTTSs are due to the "classical" T Tauri Star phenomena: they may represent absorption in a velocity-structured wind, or in a disk atmosphere. In particular, the \niti\ feature  is likley due to absorption in the CTTSs outflow or disk. We note here the similar excitation \niti\ absorption lines are observed at 1492.62 \AA\ and 1494.67 \AA\ for all stars that show \niti\ absorption in the \nv\ region. A detailed study of the wind signatures in the sample will appear in a future paper.

\subsection{Flux measurements}
 \label{abun}
\begin{figure}
\epsscale{1}
\includegraphics[width=3.4in]{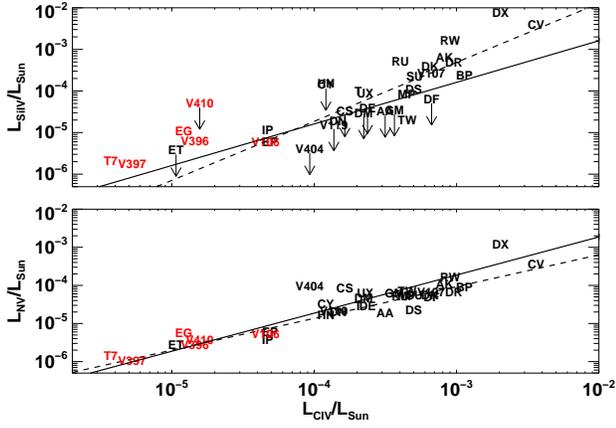}
\caption{ \siiv\ luminosity and \nv\ luminosity vs. \civ\ luminosity.  Errors are $\sim$10-30\% for the \siiv\ and \nv\ measurements and $\sim$1-5\% for \civ, as indicated in Table \ref{ForPaper_TableLineFluxes}. {\bf Red labels} are for WTTS. Downward arrows indicate upper limits. In both plots, the dashed line is the linear fit to the data, ignoring the non-detections and the WTTSs and the solid line corresponds to L$_{Si IV}=0.111 L_{C IV}$ ({\bf Top}) or L$_{N V}=0.183  L_{C IV}$ ({\bf Bottom}). \label{siiv_civ_correlation}}
\end{figure}

Figure \ref{siiv_civ_correlation} shows the relationship between the \civ\ luminosity and the \siiv\ and \nv\ luminosities. 

To measure the flux in the \siiv\ lines (Table \ref{ForPaper_TableLineFluxes}), we have integrated each of them between --400 and 400 \kms, interpolating over the contaminating \htwo\ lines. The \siiv\ lines are much broader than the \htwo\ ones, and at these resolutions can be separated from them, at least in the cases in which the \siiv\ lines are actually observed. Assuming the \htwo\ lines are optically thin, the emission in R(0) 0-5 should be 2$\times$ weaker than P(2) 0-5 and R(1) 0-5 should be 1.4$\times$ weaker than P(3) 0-5, as explained in \cite{ard02}.  In 11 objects no \siiv\ line is seen under the \htwo\ emission. For these stars, we give the 3$\sigma$ upper limit to the flux, over the same velocity range we used to measure the \civ\ line in the same star. 

Table \ref{ForPaper_TableLineFluxes} also lists the flux in the \nv\ lines. Note that the flux in the 1243 \AA\ line is the observed flux, and it has not been corrected for \niti\ absorption or by any of the other absorptions mentioned above.


Figure \ref{siiv_civ_correlation} shows that the \civ\ and \siiv\ luminosities of CTTSs are correlated ($\log L_{Si IV}/L_\odot = (0.9\pm0.6) + (1.4\pm0.2) \log L_{C IV}/L_\odot$, ignoring the non-detections and the WTTSs). To understand the nature of this correlation, we perform the following calculation. We assume that each line has a contribution from two regions, a pre- and a post-shock, and that the ratio of the contributions between the two regions is the same in \siiv\ and in \civ.  In other words, ${L_{Si IV pre} \over L_{Si IV post}}={L_{C IV pre} \over L_{C IV post}}$. This is likely appropriate if all the emission regions contributing to the line are optically thin. If this is the case, the total luminosity in each line can be shown to be proportional to the post-shock luminosity, and we have:

 $${{L_{Si IV}} \over {L_{C IV}}}= {L_{Si IV post} \over L_{C IV post}} \simeq{N_{Si IV}\over N_{C IV}}{Ab_{Si IV}\over Ab_{C IV}}{C_{Si IV}\over C_{C IV}}{\lambda_{C IV}\over \lambda_{Si IV}}$$
 $$ = 0.111 $$ 

further assuming that the post-shock gas is an optically thin plasma in collisional equilibrium and the emission measure (EM) is the same for both lines over the emitting region. $N_x$ is the fraction of the species in that ionization state, at that temperature, $Ab_x$ is the number abundance of the element and $C_x$ is the collisional excitation rate. We assume a Si/C number abundance ratio of 0.13 for the present day Sun \citep{grevesse2007}.  To calculate the collision rate we use the analytical approximation  \citep{burg1992, dere1997}:
 
$$C_x\propto{1\over T^{0.5}} \Upsilon \exp(-{h\nu \over {kT}}) $$ 
 
\noindent where $\Upsilon$ is the thermally averaged collision strength. We use Chianti V. 7.0 \citep{dere1997, landi2012} to calculate this quantity and assume that the \siiv\ and the \civ\ emission come from a plasma at $Log (T) = 4.9$ (K) and $Log (T) = 5.0$ (K), respectively. 

The relationship L$_{Si IV}$ = 0.111 L$_{C IV}$ is indicated with a solid line in Figure \ref{siiv_civ_correlation} (Top). A model that does not assume the same EM for \siiv\ and \civ\ will move the solid line up or down (modestly: a 30\% larger emission measure in  \siiv\  than in \civ\ moves the solid line up by 0.1 dex), but will not change the slope.  

To perform the same comparison between \civ\ and \nv\ requires a correction from the observed values in the latter, because we only fully observe the 1239 \AA\ doublet member, as the 1243 \AA\ doublet member is generally absorbed by \niti.  Therefore, we assume that the ratio of the flux in the 1239 \AA\ to the 1243 \AA\ \nv\ lines is the same as in the \civ\ lines and calculate the total flux that would be observed in both \nv\ doublet lines in the absence of \niti. This is \nv\ flux plotted against \civ\ in  Figure \ref{siiv_civ_correlation} (Bottom). For \nv\ the observed relationship is $\log L_{N V}/L_\odot = (-1.5\pm0.4) + (0.8\pm0.1) \log L_{C IV}/L_\odot$, ignoring the WTTSs. The expected relationship between the luminosities of both lines is L$_{N V}=0.183  L_{C IV}$, using N/C abundance of 0.25, and maximum ionization temperature of $Log (T) = 5.3$ (K). 

In the case of \siiv\ the observed linear fit (dashed line in Figure \ref{siiv_civ_correlation}) is not consistent with our simple model (solid line, Figure \ref{siiv_civ_correlation}). For the current assumptions, V4046 Sgr and TW Hya show a flux deficit in \siiv\ (or an excess in \civ), while objects such as CV Cha, CY Tau, DX Cha, RU Lup, and RW Aur present a flux excess in \siiv\ (or a deficit in \civ).  In average, the WTTSs are above the line predicted by the model. For \nv\ vs. \civ\ this simple model succeeds in explaining the observed linear correlation.  

Based on the absence of \siiii\ lines as well as the weak \siii\ lines  in TW Hya,  \cite{her02} proposed that for this star the silicon has been locked in grains in the disk and does not participate in the accretion. Consistent with this idea, \cite{kastner2002} and \cite{stelzer2004} found depletion of Fe and O in TW Hya. For V4046 Sgr, \cite{gunther2006} found a high Ne/O ratio, similar to that which was found in TW Hya and suggesting that a similar process may be at work. 

More generally, \cite{drake2005b} argues that the  abundance of refractory species reflects the evolutionary status of the circumstellar disk. We see no evidence of this in our sample. The distribution of accretion rates for stars above and below the model line for \siiv/\civ\ is not significantly different. We find that 80\%$\pm$10\% of the non-TD CTTSs are above the Si/C model line, compared to 60\%$\pm$20\% of the TD stars. This difference is not significant either. It is, however, noteworthy that our analysis suggests that most CTTSs are \siiv-rich.

Four out of the 5 WTTSs for which we measure the \siiv\ luminosity lie above the model line. This is reminiscent of the First Ionization Potential (FIP) effect seen in the Sun (e.g. \citealt{mohan2000}) in which elements in the upper solar atmosphere (the transition region and the corona) show anomalous abundances when compared to the lower atmosphere. Upper atmosphere elements with low FIPs (FIP $<$ 10eV, like silicon) show abundance excesses by 4$\times$, on average, while high FIP elements (FIP $>$ 10 eV, like carbon and nitrogen) show the same abundances between the photosphere and the corona. Possible explanations for the effect include the gravitational settling of neutrals in the chromospheric plateau \citep{vau1985} and/or diffusion of neutrals driven by electromagnetic forces \citep{hen1998, laming2004}. The fact that the luminosity in \nv\ for WTTSs follow the expected relationship  with \civ, is consistent with the \siiv\ excess flux in WTTSs being due to the FIP effect \citep{wood2011}.

However, abundance studies of active, non-accreting, low-mass stars based on X-ray observations, find a mass-dependent inverse FIP-effect abundance pattern in Fe/Ne (see \citealt{gudel2007b, testa2010} for reviews). If this applies to WTTSs, we would expect to see a silicon deficit in them, as compared to carbon. If we were to move the model (solid) line from the top panel of Figure \ref{siiv_civ_correlation} upwards, in order to make the WTTSs silicon-poor, most CTTSs would become silicon-poor, suggesting that the disk grain evolution observed in TW Hya starts at younger ages.  This is speculative and requires further study. We note here that  \cite{telleschi2007} find a mass-dependent inverse FIP-effect abundance pattern also in CTTSs.   
 
These statements are only intended to identify candidates for further study. What we can tell from our observations is that there is considerable scatter of CTTSs to both sides of the linear relation between $L_{Si IV}$ and $L_{C IV}$. EM analyses for each star are necessary before significant conclusions can be drawn from this sample regarding abundances. These are possible with the DAO dataset, but they are beyond the scope of this paper.

 


\section{Discussion }
\label{discussion}

In this paper we are primarily interested in what the hot gas lines are telling us about the region they are emitted from. In particular, we want to know whether the observed profiles are consistent with emission in an accretion shock, a stellar transition region, or other volumes within the system, such as shocks in the stellar outflow or hot winds. We conclude that extensions of the accretion shock model involving inhomogeneous or multiple columns, or emission far from the accretion spot are necessary to account for the observations.

The observational description reveals a remarkable diversity of line shapes, and interpretations about them are necessarily qualified with exceptions. However, there are a few general statements that describe the systems: 

\begin{enumerate}
\item  The \civ, \nv, and \siiv\ lines generally have the same shape as each other, while the \heii\ line tends to be narrower and symmetric. The 1548 \AA\ and 1550 \AA\ \civ\ lines are generally similar to each other, except for a scale factor (Exceptions:  Different red wings: DF Tau, DX Cha, RU Lup, RW Aur A; Extra emission close to the 1550 \AA\ line: AK Sco, CS Cha) 
\item On average, for CTTSs the \civ\ lines are redshifted by  $\sim$20 \kms\ from the stellar rest velocity. The \heii\ lines for CTTSs are redshifted by  $\sim$10 \kms.  The \civ\ lines in WTTS may be redshifted from the stellar rest velocity (by $\sim$11 \kms\ at the 2$\sigma$ level). 
\item For CTTSs, the \civ\  lines are broader (FWHM$\sim$200 \kms) than for WTTSs (FWHM$\sim$90 \kms). On average the \heii\ lines have the same widths in CTTSs and WTTSs (FWHM$\sim$100 \kms). The CTTSs \civ\ lines are skewed to the red more than the \heii\ lines, and more than \civ\ lines in WTTSs.   
\item For \civ\ the NC contributes about 20\% of the total line flux at low accretion rates ($<4 \times 10^{-9}$ M$_\odot$/yr). At high accretion rates the NC contribution to the flux becomes comparable to and larger than the BC contribution for some stars. The multi-epoch data also suggest that increases in  \civ\ luminosity are accompanied by an increase in the strength of the NC. For \heii, the NC contributes about 60\% of the total line flux at all accretion rates.
\item The \civ\ line luminosity is stronger than the \heii\ line luminosity by a factor of 3 to 4 in CTTSs, while the luminosities are comparable to each other in WTTSs. 
\item For CTTSs that show BC and NC in \civ, both components are redshifted (exceptions: DF Tau, DK Tau, HN Tau A, IP Tau, MP Mus, T Tau N, V1190 Sco) and the BC is redshifted with respect to the NC (exceptions: DF Tau, GM Aur, IP Tau, RU Lup, SU Aur, T Tau, V1190 Sco). For 12 out of 22 objects (AA Tau, BP Tau, CS Cha, CY Tau, DE Tau, DN Tau, DS Tau, EP Cha, MP Mus, TW Hya, V1079 Tau, V4046 Sgr), the velocities of the narrow and broad components are consistent with a magnetospheric origin, within the instrumental velocity errors.  For the rest (DF Tau, DK Tau, DR Tau, GM Aur, HN Tau A, IP Tau, RU Lup, SU Aur, T Tau, V1190 Sco) other explanations are necessary.
\item There are no significant correlations between the \civ\ luminosity or accretion rate and the velocity at peak flux, FWHM, skewness, inclination, 1548-to-1550 line ratio, or velocity of the Gaussian components. 
\end{enumerate}

\subsection{Emission from the accretion column?}
\label{model}
 
If the lines originate primarily in an accretion shock, one can make at least four predictions: (1) the \civ\ and \heii\ line profiles should be redshifted, with the latter having smaller velocities; (2) the line emission should be well localized on the stellar surface; (3) if increases in the accretion rate are  due to density increases, the post-shock gas emission will be quenched, due to burying of the post-shock column; (4) the velocity of the narrow and broad components of the \civ\ lines should be related as  V$_{BC}\gtrsim4$ V$_{NC}$.
 
Indeed, we observe \civ\ and \heii\ redshifted, with the former redshifted more than the latter. This, however, is not a very constraining prediction. \cite{linsky2012} have shown that \civ\ lines in dwarfs are redshifted by an amount correlated with their rotation period and \cite{ayres1983} have shown that the \civ\ lines in late-type giants are redshifted with respect to the \heii\ lines by variable amounts of up to $\sim20$ \kms. So, for non-CTTSs gas flows in the upper stellar atmosphere may result in line shifts comparable to those we observe in our sample. 

The predictions regarding the localization of the emitting region in the stellar surface are difficult to test with this dataset, because, as we have argued in section \ref{inc_and_rate}, the sample is not large enough for the inclination to serve as a discriminator of the line origin. On the other hand, there is a large body of evidence (see the references in the introduction) suggesting that the accretion continuum and the (optical) line emission are well localized on the stellar surface. No unequivocal rotational modulation of the lines studied here has been reported in the literature, but this may be because the available dataset does not provide enough rotational coverage. The change in line flux observed among the different RU Lup epochs (Figure \ref{one_panel_multiepoch}) may result from rotational modulation.

If we assume that the NC of \civ\ is due to post-shock emission, we would naively expect its importance to diminish for high accretion rates as the post-shock is buried in the photosphere \citep{drake2005}. However, we observe the opposite, with the NC flux increasing with respect to the BC flux as the accretion rate increases. One possible explanation is that increased accretion produces larger accretion areas on the star rather than
much higher densities in the accretion column. This has been shown for BP Tau by \cite{ardila2000}. Larger areas provide more escape paths for the photons emitted from a buried column. Area coverages as small as 0.1\% of the stellar surface of a 2R$_\odot$ star have a radius of $\sim$10$^5$ km, which is larger than the deepest likely burials \citep{sacco2010}.

A complementary insight comes from X-ray observations. Models that assume a uniform density accretion column also predict quenching of the X-ray flux, due to absorption of X-rays in the stellar layers, for accretion rates as small as a few times $10^{-10} \msun$/yr  \citep{sacco2010}.  However, models by \cite{romanova2004} indicate that the accretion column is likely non-uniform in density and the denser core may be surrounded by  a slower-moving, lower-density region.  \cite{sacco2010} argue that most of the observed X-ray post-shock flux should come from this low-density region. In addition, post-shock columns are believed to be unstable to density perturbations, and should collapse on timescales of minutes \citep{sacco2008}.  The lack of observed periodicities in the X-ray fluxes indicate that multiple incoherent columns should be present.  \cite{orlando2010} also conclude that the presence of multiple columns with different densities is necessary to explain the low accretion rates derived from X-ray observations.

In addition, models of low-resolution CTTSs spectra from the near-UV to the infrared indicate that the observed accretion continuum is consistent with the presence of multiple accretion spots \citep{ingleby2013}.

In summary, we observe that the \civ\ flux does not decrease with accretion, which suggests that if burying is occurring, it does not affect the observed flux substantially. This may be because the aspect ratio of the accretion spots is such that the post-shock radiation can escape without interacting with the photosphere, and/or that the flux we observe is emitted from unburied low-density edges of the accretion column, and/or that multiple columns with different densities and buried by different amounts are the source of the emission.

\subsubsection{\heii\ emission in the pre-shock gas}
\label{he_model}
As we have shown, the BC of \heii\ is weak compared to its NC, in contrast with the BC for \civ. For a BC produced in the pre-shock gas, this implies that the pre-shock emits more strongly in \civ\ than in \heii. Observationally, for CTTSs the ratio between luminosity in the BC of the \civ\ to the \heii\ line ranges from 2.5 to 10.4, with a median of 5.6. The ranges are comparable if instead of using the whole sample we use only those CTTSs with redshifted profiles. Are these values consistent with a pre-shock origin?

The pre-shock region is heated and ionized by radiation from the post-shock gas. The model described next confirms that the \heii\ pre-shock gas contribution to the line is produced by recombination of \heiii\ to \heii, while the pre-shock gas contribution to \civ\ 1550 \AA\ line is the result of collisions from the ground state. The \heii\ 1640 \AA\ line is emitted from a smaller region in the pre-shock, closer to the star, than the \civ\ 1550 \AA\ lines, because the energy required to ionize \heii\ to \heiii\ is 54.4 eV, while the energy to ionize \ciii\ to \civ\ is 47.9 eV. 

We use Cloudy version 07.02.02 \citep{fer98} to simulate the pre-shock structure and calculate the ratio between the \civ\ lines at 1550 \AA\ and the \heii\ line at 1640 \AA.  We illuminate the pre-shock gas with a 4000 K stellar photosphere, a shock continuum with the same energy as contained in half of the incoming flux, and half of the cooling energy from the post-shock gas. For the purposes of this model the post-shock cooling radiation is calculated by solving the mass, momentum, and energy conservation equations as described in \cite{cal98}, to derive a temperature and density structure and using Chianti V. 7.0  \citep{dere1997, landi2012} to calculate the emissivity of the plasma at each point in the post-shock. This procedure assumes that the post-shock gas is an optically thin plasma in collisional equilibrium.  These models are parametrized by incoming gas velocities and densities. A typical density for an incoming accretion flow with $\dot {M}=10^{-8} \msun$/yr, covering 1\% of the stellar surface area, is $5 \times 10^{12}$ cm$^{-3}$  \citep{cal98}. 

We find that the \civ\ pre-shock gas emission is always larger than \heii\ emission by factors ranging from 2 (at $10^{14}$ cm$^{-3}$) to 6 (at $10^{10}$ cm$^{-3}$), for incoming pre-shock gas velocities of 300 \kms. This increase with density implies that the pre-shock emission is beginning to become optically thick at high densities, consistent with Figure \ref{opacity} (bottom). Higher velocities result in larger \civ\ emission with respect to \heii: for incoming velocities of 400 \kms and densitites of $10^{10}$ cm$^{-3}$ the pre-shock emits 10 times more flux in \civ\ 1550 \AA, than in \heii\ 1640 \AA. In these simple models, the \civ\ post-shock emission as a fraction of the total emission varies from 0.7  (at $10^{10}$ cm$^{-3}$) to $\sim$1 (at $10^{14}$ cm$^{-3}$).    

These are only illustrative models but they suggest that the observed values are within the range of what is produced by the accretion shock region. It is noteworthy that low pre-shock gas densities  ($10^{10}$ cm$^{-3}$) are required to explain the median \civ/\heii\ ratio.  This again suggests that the accretion spots are very large, or that the observed emission comes from low-density regions in the accretion columns. 

\subsubsection{Kinematic predictions of the accretion shock model}

A crucial prediction of the accretion model is that V$_{BC}\gtrsim4$ V$_{NC}$.  Strictly, the observations are consistent with this prediction for only two objects: TW Hya and CY Tau. However, given the size of the \COS\ and \STIS\ pointing errors, half of the CTTSs sample may comply with the accretion model predictions (see Figure \ref{vbc_vs_vnc}). The rest of the objects are anomalous, as the standard magnetospheric accretion model has trouble explaining stars for which the BC velocity is significantly smaller than the NC velocity, or those stars in which the velocity of one or both of the components is significantly negative. Possible explanations for these anomalous objects include the target having a significant extra radial velocity due to the presence of a close companion, multiple columns being responsible for the emission, regions far from the accretion shock surface or parts of an an outflow contributing to the emission, and winds or outflows dominating the emission. We examine the first three possibilities in this section. We discuss outflows in Section \ref{outflow}. 

For binaries, the relative velocity between close binary components will result in shifts in the velocity centroid of the \civ\ line, especially if one component dominates the accretion or if a circumbinary disk is present \citep{arty1996}. For example, AK Sco is a well know spectroscopic binary in which the radial velocity of the stellar components can reach 100 \kms\ with respect to the system's center-of-mass \citep{alencar2003}. If the characteristics of the accretion stream to each component were different (different accretion rates, different pre- and post-shock emission contributions among the components, etc), this would result in shifted velocity centroids. We have not considered AK Sco in the general description because the Gaussian decomposition is problematic. Regarding the other spectroscopic binaries in the sample, V4046 Sgr and CS Cha have velocity contrasts between the BC and NC close to what is expected from the accretion paradigm. Orbital modulation of X-rays has been detected in V4046 Sgr \citep{argiroffi2012}, which may be responsible for the large NC velocity, compared to the predictions of the accretion shock model. The \civ\ profiles of DX Cha are slightly redshfited, but they cannot be decomposed into velocity components and this object may belong to a different class altogether, as the only Herbig Ae star in the sample. The other systems for which multiplicity may be relevant are DF Tau and RW Aur A. DF Tau has v$_{BC} - $v$_{NC}\sim -27$ \kms,  but the component separation of 12 AU is too large to induce these velocity shifts. In the case of RW Aur A, we do not decompose the \civ\ lines into Gaussian components, and argue that the strong blueshift is the result of outflow emission. \cite{gahm1999} have suggested that the system is accompanied by a brown dwarf companion. According to their observations, the companion produces radial velocity variations smaller than 10 \kms. In summary, with the possible exceptions of AK Sco and V4046 Sgr, binarity does not play a significant role in altering the values of V$_{BC}$ and V$_{NC}$ for any of the CTTSs we have termed anomalous.

Models of the hot gas line shapes have been performed by \cite{lam03, lamzin2003} and \cite{lamzin2004}, among others. For a broad range of geometries, those models predict redshifted double-peaked line profiles. \cite{lam03} modeled the \civ\ emission assuming plane-parallel geometry with the pre- and post-shock emission lines thermally broadened, while \cite{lamzin2003} consider emission from an accretion ring at high stellar latitudes, with gas flow falling perpendicular to the stellar surface. Both sets of models recover two emission kinematic components, one from the pre-shock and one from the post-shock. The peak separation between the pre- and post-shock contributions depends on the velocity of the incoming flow. With the possible exceptions of DK Tau and DR Tau, these double-peaked profiles are not observed.

The failure of those models in predicting observed line profiles, led \cite{lamzin2003} to argue that the incoming flow cannot be perpendicular to the stellar surface, and that a substantial tangential component must be present in the gas velocity.  In models of the UV spectrum of TW Hya, \cite{lamzin2004} argue that to explain the \civ\ line shape, the accretion flow must fall at a very low stellar latitude but in a direction almost parallel to the stellar surface, in such a way that we are able to observe the accretion streams from both sides of the disk through the inner disk hole. However,  \cite{donati2011} showed that the magnetic topology in TW Hya is such that the accretion streams have to be located at high stellar latitudes. We have shown here that the \civ\ line shape in TW Hya is the most common one ($\sim$50\%) in our sample and therefore whatever process is responsible for it must be fairly general.  A more general explanation that the one from \cite{lamzin2004} is required to understand the line shapes.

\cite{gunther2008} considered hot gas observations of 7 CTTSs (RU Lup, T Tau, DF Tau, V4046 Sgr, TWA 5, GM Aur, and TW Hya). Their model of the post-shock contribution to the \ovi\ profiles of TW Hya, including only thermal broadening, results in a line skewed to the red, but very narrow compared to the observations. A turbulent velocity of 150 \kms\ is necessary to obtain widths comparable to those observed. However, this turbulence results in a very symmetric line, with significant emission to the blue of the line, which is not observed. Based on these analyses, \cite{gunther2008} concluded that the \ovi\ emission in those stars that show redshifted profiles (DF Tau, V4046 Sgr, TWA 5, GM Aur, and TW Hya) is incompatible with current models of magnetospheric accretion.

The larger sample that we present here provides some insights into these issues. Focusing only on those objects for which the velocities are such that they could in principle be produced in an accretion shock (the wedge with 12 CTTSs in the upper right quadrant of Figure \ref{vbc_vs_vnc}, below the dotted line, including MP Mus but excluding DR Tau), we conclude that the lack of observed double peaked profiles is due to the small difference between velocity components and to the fact that both components are very broad (perhaps as a result of turbulence in the flow). Notwithstanding the conclusions from \cite{lamzin2004} and \cite{gunther2008}, the ratio between the velocity components for the \civ\ profile from TW Hya is perfectly consistent with magnetospheric accretion. 

The small difference between velocity components is due to either too large  V$_{NC}$ values or too small V$_{BC}$ values, compared with expected infall speeds. For TW Hya, for example, the models by \cite{gunther2007} predict an infall velocity of 525 \kms, and we observe V$_{BC}=116$ \kms. This difference requires an angle between the line of sight and the accretion column of $\sim$77 degrees. For most of the rest of the stars likely to come from an accretion shock, similarly large angles are implied: AA Tau: 76$^\circ$, BP Tau: 82$^\circ$,  CY Tau: 68$^\circ$, CS Cha: 49$^\circ$, DE Tau: 80$^\circ$, DN Tau: 75$^\circ$, DS Tau: 83$^\circ$, EP Cha: 85$^\circ$, MP Mus: 89$^\circ$, V1079 Tau: 83$^\circ$, V4046 Sgr: 76$^\circ$ (assuming literature values for stellar masses and radii).

These values are calculated using the observed BC velocity and the predicted free-fall velocity. Observationally, they represent a flux-weighted average of the velocities along the line of sight. It is surprising that most are close to 90 degrees, indicating that the average column is seen sideways and that no emission is observed from the top of the accretion column.  However, observations of red-wing absorption in the \hei\ 1.1$\mu$m line \citep{fischer2008} indicate that the accretion flow is slower than the free-fall velolocities by $\sim$50\%. In addition, in the models by \cite{romanova2004, romanova2011} mentioned before, the periphery of the accretion column is moving more slowly, by factors of $\sim$2, than the column core. If the BC emission is not coming from the fast free-falling core of the pre-shock gas flow but from the slower edges, the calculated angle will be smaller. 

So far, this exploration of the expected relationship between the broad and narrow velocity components assumes that the pre- and post-shock flows share the same line-of-sight angle. There are two situations in which this may not be the case. If some of the post-shock ionizing radiation reaches regions of the accretion flow in which the line-of-sight to the observer is different than for the accretion spot, we may end up with BC velocities that are unrelated to the NC velocities. After all, even close to the stellar surface the magnetic field twists and curves \citep{gregory2008, mohanty2008} and photoionized regions may be produced in flow moving in different directions. The Cloudy models we develop in section \ref{he_model} create fully ionized post-shock columns with sizes ranging from 10000 km for densities $\sim 10^{12}$ cm$^{-3}$ to $\sim$R$_\odot$, for densities $\sim 10^{10}$ cm$^{-3}$ \citep{cal98}. In other words, for low densities the ionizing photons may reach far from the stellar surface. Without more detailed models including at least some notional information regarding the configuration of the magnetosphere, it is not possible to say if this concept is relevant, but it may offer an explanation for objects  with either small or negative BC velocities, but positive NC velocities (upper half of Figure \ref{vbc_vs_vnc}), such as DF Tau, DR Tau, GM Aur, IP Tau, RU Lup, SU Aur, and V1190 Sco. 

It is also possible that we are observing multiple columns for which the ratios of pre- to post-shock emission are not the same in all columns, resulting in a situation in which we observe the pre-shock of one column but the post-shock of another with a different orientation. This may occur, for example, if the post-shock is occulted by the stellar limb, or buried, or if the columns have different optical depths. In this case, we will see pre- and post-shock velocities that are essentially unrelated to each other. This effect may explain objects for which both velocities are positive, but the BC is small compared to the NC (e.g., DF Tau, DR Tau, GM Aur, maybe IP Tau, RU Lup, SU Aur).



\subsubsection{A contribution from the stellar transition region?}

An alternative hypothesis to the line origin in an accretion shock is that some of the observed flux originates in the stellar transition region outside of the accretion spot. Models by \cite{cranmer2008, cranmer2009} indicate that accretion energy may contribute to the powering of the corona. Furthermore, based on iron and helium line observations in the optical of five CTTSs, \cite{petrov2011} suggest that an area of enhanced chromospheric emission, more extended than the hot accretion spot,  is produced by the accretion process. In addition, observations by \cite{brickhouse2010} suggest the existence of a larger region than the accretion spot as the source of a third X-ray component (after the corona and the accretion spot itself). 

The idea of an atmospheric contribution to the observed lines in CTTSs is not new \citep{herbig1970}, although its limitations were the inspiration for the magnetospheric accretion paradigm. \cite{cram1979} and \cite{calvet1984} showed that a dense chromosphere cannot reproduce the strength of the observed H$_\alpha$ line in CTTSs, and \cite{batalha1993} showed that chromospheric-based models are unable to reproduce the veiling or the size of the Balmer jump in CTTSs. On the other hand, magnetospheric accretion models are able to reproduce the hydrogen-line fluxes and shapes (e.g. \citealt{muzerolle1998}, \citealt{kurosawa2006}).

If the lines are primarily emitted from the transition region, then a model would predict that: (1) the line profiles should have comparable redshifts in WTTSs and CTTSs; (2) the emission should not be localized to a small area of the stellar surface; and (3) the \heii\ and \civ\ lines should have similar shapes. Our observations do not confirm any of the predictions of a transition region origin for the lines: the \civ\ line redshifts are different in CTTS and WTTSs, the \heii\ and \civ\ lines have different shapes, and other observations show that the accretion indicators are localized in a small area of the stellar surface.  

However, we observe that the NCs of the \heii\ lines have comparable widths in CTTSs and in WTTSs, and the same widths for the NC of \civ. Within the picture of an accretion shock column, it is surprising that the WTTSs line widths, formed in the upper stellar atmosphere, should be the same as the CTTSs NCs line widths, formed in the turbulent post-shock gas. The velocity differential between high and low density regions in the accretion column will result in large amounts of turbulence, which tends to produce broad lines \citep{gunther2008}. It may be that at least some of the flux in the NCs in CTTSs comes from the stellar transition region, while the BCs come from the pre-shock gas. This would only apply to objects in which both Gaussian components have positive velocities as there is no evidence of blueshifted \civ\ or \heii\ profiles in the atmospheres of young active stars.

Could the stellar transition region respond to accretion by producing enough \civ\ or \heii\ emission to contribute to the observed lines? For \heii\ a detailed model of the heating process would have to show that the X-ray emission from the accretion spot in CTTSs is enough to increase the \heii\ line luminosity in the stellar transition region by approximately one order of magnitude from the WTTS values (\citealt{yang2012} and this work). In the Sun, between 30\% (quiet regions) and 60\% (active regions) of the 1640 \AA\ \heii\ flux comes from ionization by soft X-rays followed by radiative recombination. The rest is due to collisional or radiative excitation of ground-level \heii\  \citep{hartmann1979, kohl1977}. In CTTSs, the accretion spot acts as a source of soft  X-rays \citep{kastner2002, stelzer2004, gunther2007}, although the overall X-ray emission is dominated by hot plasma produced by enhanced magnetic activity. Even the corona may increase its X-ray emission as a response to accretion events \citep{dupree2012}. Typical observed values of L$_{He II}\sim10^{30}$ erg/sec are comparable to X-ray luminosities between 0.2 and 10 keV \citep{ingleby2011a,gdc2012}. However, based on solar models, \cite{hartmann1979} concluded L$_X$=50 L$_{He II}$ in the 0.25 keV band. This suggests that the observed amount of X-ray flux in young stars is small compared to what would be required to produce the observed \heii\ line, and perhaps other mechanisms besides radiative recombination may be at play. On the other hand, the models of the coronal heating by \cite{cranmer2008} show that it is plausible to assume that the accretion energy is sufficient to drive CTTS stellar winds and coronal X-ray emission. 

The \civ\  resonance doublet observed in stellar atmospheres is the result of collisional excitation from \civ\ followed by radiative de-excitation \citep{golub2009}, and an increase ranging from one to two orders of magnitude from WTTSs values would be required to match the surface flux \citep{jvl00} or luminosity (Figure \ref{vel_max}) observed in CTTSs. This would require a proportional increase in atmospheric density. High densities of hot plasma gas are indeed observed in CTTSs, but at temperatures consistent with an origin in an accretion shock \citep{sacco2008, argiroffi2009}. As is the case for \heii, the increased X-ray flux due to accretion may result in a larger \civ\ population, and larger observed doublet flux. 

In summary, the relative contribution of the stellar transition region to the total \heii\ or \civ\ remains uncertain. While unlikely, we cannot rule out with these observations that at least some fraction of the NC in \heii\ or \civ\ originates in the stellar atmosphere of CTTSs. 




\subsection{Blueshifted profiles and outflows}
\label{outflow}

We have argued that if the post-shock radiation ionizes material far away from the accretion spot, we may end up with broad components having small or even negative velocities. On the other hand, the objects for which we observe a negative velocity in the NC of \civ, or in the overall profile, present considerable challenges. These are AK Sco, DK Tau, HN Tau A, T Tau N, and RW Aur A. The \heii\ line matches the \civ\ line for RW Aur A and HN Tau A (Figure \ref{all_civ_heii}) but  is centered at velocities closer to zero than \civ\ for the other stars.

If we observe a CTTS for which the accretion stream is moving towards us, the velocity components would be negative. This would be the case, for example, if we observe the stream below the disk through the inner truncation hole of the accretion disk. The only case in which this is a possibility is T Tau N as this is the only target for which v$_{NC}<0$, v$_{BC}<0$, and $\mid$v$_{NC}\mid<\mid$v$_{BC}\mid$. This requires an inclination close to face-on, which the system has, and perhaps a large difference between the stellar rotation axis and the magnetic field axis. 

Outflows present a more likely explanation for the blueshifted profiles. Outflow phenomena are common in CTTSs and high speed shocks between the jet material and the ISM may result in \civ\ emission. The \civ\ blueshifted emission in DG Tau \citep{ard02} is clearly related to the beautiful outflow imaged with HST/NICMOS \citep{padgett1999} and the \civ\ emission is likely produced by shocks in the jet \citep{schneider2013}.  In order to generate high temperature ($>10^5$ K) gas via an outflow shock, velocities larger than 100 \kms\ in the strong shock limit are necessary  \citep{gunther2008}, resulting in post-shock (observed) \civ\ velocities $>$25 \kms. The five objects we are considering have absolute NC velocities or velocities at maximum flux larger than this value. Therefore, at least energetically, it is possible for the emission to be produced by a shock in the outflow. Both HN Tau A and RW Aur A are known to have outflows \citep{hirth1994, hartigan1995}, and velocities in the approaching and receding jets that are comparable to the ones we observe in \civ\ and \htwo \citep{melnikov2009, coffey2012}.

For objects such as AK Sco and DK Tau the accretion shock and the outflow regions may both be contributing to the emission. For HN Tau A and RW Aur, if we accept that the observed profiles originate primarily in an outflow, the lack of accretion shock emission becomes puzzling. These are high-accretion rate objects, and perhaps in these conditions the low-density region in the periphery of the accretion column is not present, and so the shock is truly buried. Or maybe we are observing the objects in a rotational phase such that the accretion spot is away from us. The GHRS observations of RW Aur (red trace, Figure \ref{one_panel_multiepoch}) show the presence of  additional emission components to the red of the nominal \civ\ lines, which may be due to the accretion spot. 

Overall, the relationship between \htwo\ asymmetries and hot line shapes remains to be fully explored. The analysis of DAO data by \cite{france2012} shows that some \htwo\ emission lines in DK Tau, ET Cha (RECX 15), HN Tau, IP Tau, RU Lup, RW Aur, and V1079 Tau (LkCa 15) are asymmetric, presenting redshifted peaks and low-level emission to the blue of the profiles. In general, the systematic errors in the \COS\ wavelength scale make it difficult to determine whether the line peaks are truly shifted in velocity. For HN Tau A and RW Aur, the peak of the \htwo\ line R(6) (1-8), at 1556.87 \AA\ is shifted +19 \kms\ and +88 \kms\ respectively, from the stellar rest frame, larger than would be expected from pointing errors alone and suggests that outflows that are fast enough in \civ\ are accompanied by \htwo\ flows away from the observer. For RW Aur A, we observe that the (redshifted) \htwo\ emission covers the (blueshifted) \civ\ emission (Figure \ref{rw_aur}). 


If the gas heating occurs very close to the star, or if the wind is launched hot, one would expect to observe P-Cygni-like profiles in the hot gas lines. \cite{dupree2005} argued that the asymmetric shape of the \ovi\ profiles of TW Hya is the result of a hot wind in the star. Because the two \civ\ lines are close to each other, blueshifted wind absorption in the 1550 \AA\ line will decrease emission in the red wing of the 1548 \AA\ line.  \cite{cmj2007} compared both \civ\ lines and concluded that their similarity, as well as the absence of absorption below the local continuum (as seen in many neutral and singly ionized lines), suggest that a hot wind is not present in the case of TW Hya.

Within our sample, DX Cha is the only object that may have a high temperature wind, as it shows a deficit in the red wing of the 1548 \AA\  line for \civ, compared to the red wing of the 1550 \AA\  line and a very sharp blue cutoff in the \civ\ and the \siiv\ lines. A blueshifted absorption is seen in the 1548 \AA\ \civ\ line, suggesting outflow speeds of up to $400$ \kms. A blueshifted absorption is also seen in the 1403 \AA\ line of \siiv\ (the red doublet member). However, note that there is an O {\scshape iv} emission line at --340 \kms\ of the line, and its presence may give the illusion of a wind absorption. The narrow \htwo\ lines observed in the \siiv\ profiles suggest that the putative wind is collimated or inhomogeneous, as the \htwo\ lines are not absorbed by the wind. The difference between both members of \nv\ is not due to a wind but to absorption of the 1243 \AA\ line by circumstellar absorbers. 

DX Cha also shows absorption features in the region near \heii, although it is not clear that they are related to the outflows. Lower temperature outflows, like those observed in other CTTSs \citep{her05} are also observed in the DAO spectra of DX Cha: the \siii\ $\lambda$ 1526.71 \AA\ and 1533.43 \AA\ (not shown here) present very clear P-Cygni profile with wind absorption up to 600 \kms\ from the star.  Further examination of the outflows will appear in a future paper.

\subsection{Accretion in Herbig Ae stars}

In this analysis, we have considered DX Cha as one more member of the overall sample, in order to contrast the characteristics of Herbig Ae stars with those of CTTSs. It has the largest mass ($\sim$2.2 M$_{\odot}$, \citealt{bohm2004}), and earliest spectral type (A7.5) in the sample.

DX Cha is a spectroscopic system with a K3 secondary and an average separation of $\sim$0.15 AU between components \citep{bohm2004}. Even at these small separations, small circumstellar disks may be present, in addition to the circumbinary disk \citep{devalborro2011}.  To add to the complexity of the system, observations by \cite{tatulli2007} are consistent with the presence of a wind launched in the 0.5 AU region of the disk. \cite{testa2008} show that two different temperature plasmas are responsible for the X-ray emission. The overall flux is emitted from a relatively high-density region and dominated by the primary star. They argue that the hot component is created in the companion's corona, while the low-temperature component originates in an accretion shock. 

In the observations presented here, DX Cha does appear peculiar when compared to the CTTSs. The system has the largest \siiv\ luminosity of the sample, and the second largest \civ\ luminosity. The 1550 \AA\ \civ\ line has the second largest FWHM of the sample, and as we have noted, the \civ\ lines are not alike in shape, suggesting extra emission or absorption in one of the members. The \siiv\ lines are unlike any other \siiv\ or \civ\ lines in shape, in that they show a very sharp, blue cutoff. As indicated above, DX Cha is also the only clear candidate in the sample for the presence of a hot wind. 

For DX Cha we observe a strong continuum in the \heii\ region and no clear emission line. If in Figure \ref{panel_1} we identify the depression at +60 \kms\ in the \heii\ panel of DX Cha as \heii\ in absorption, this would be the largest redshift of any \heii\ line in our sample, and larger than the 11.5 \kms\ observed for the peak of \civ. The absence of the \heii\ line in emission is put in context by \cite{calvet2004}, who present low-resolution UV spectra of accreting objects with masses comparable to DX Cha, all of which show \heii\ in emission. For all of them L$_{He II}$/L$_{\odot} \sim 10^{-5}$ to $10^{-4.5}$, comparable to the luminosities of other stars observed here. This makes the absence of an \heii\ line a mystery. 

Are these characteristics the continuation of the standard accretion rate phenomena to larger masses, the consequence of the close companion and/or an outflow, or new phenomena related to the weak magnetic fields associated with Herbig Ae objects? A larger sample of high-resolution UV spectra of higher mass objects is required to answer these questions.

\section{Conclusions}
\label{conclusions}

The goal of this paper is to describe the hot gas lines of CTTSs and to provide measurements that will contribute to understand their origin. We describe the resonance doublets of \nv \ ($\lambda \lambda$ 1238.82, 1242.80 \AA), \siiv\  ($\lambda \lambda$ 1393.76, 1402.77 \AA), and \civ\ ($\lambda \lambda$ 1548.19, 1550.77 \AA), as well as the \heii\ ($\lambda$ 1640.47 \AA) line. If produced by collisional excitation in a low-density medium, these UV lines suggest the presence of a plasma with temperatures $\sim10^5$ K. We focus primarily on the \civ\ doublet lines, with the other emission lines playing a supporting role.

 We combine high resolution \COS\ and \STIS\ data from the Cycle 17  Hubble Space Telescope ({\it HST}) proposal ``The Disks, Accretion, and Outflows (DAO) of T Tau stars'' (PI G. Herczeg) with archive and literature data for 35 stars: one Herbig Ae star, 28 CTTSs, and 6 WTTSs. The sample includes 7 stars with transition disks. This is the largest single study of the UV hot gas lines in CTTSs and WTTSs, with high resolution and high sensitivity. 
 
 We use the centroids of the \htwo\ lines to argue that the systematic wavelength errors in these \COS\ observations are $\sim$7 \kms. We do not perform any systematic velocity correction to the spectra to account for these errors.
 
 The WTTSs establish the baseline characteristics of the hot line emission, and help to separate the effect of accretion from purely atmospheric effects. In particular, they provide the line luminosities and shapes that would be emitted by the young stars in the absence of the accretion process.
 
 The observations were analyzed using non-parametric shape measurements such as the integrated flux, the velocity at maximum flux (V$_{Max}$), the FWHM, and the line skewness. We also decomposed each \heii\ and \civ\ line into narrow and broad Gaussian components. We obtained accretion rate measurements from the literature (see Table \ref{TableAncil3}).

\subsection{The shape of the lines}

\begin{itemize}

\item The most common (50 \%) \civ\ line morphology is that of a strong, narrow emission component together with a weaker, redshifted, broad component (see for example BP Tau. Figure \ref{ctts_wtts_comparison}). In general, the \civ\ CTTSs lines are broad, with significant emission within $\pm400$ \kms\ of the lines. When compared to the WTTSs lines, the \civ\ lines in CTTSs are skewed to the red, broader (FWHM$\sim$200 \kms\ for CTTSs, FWHM $\sim$ 100 \kms\ for WTTSs), and more redshifted (20 \kms\ for CTTSs, 10 \kms\ for WTTSs). See Table \ref{velocities}. The 1548 \AA\  member of the \civ\ doublet is sometimes contaminated by the \htwo\ R(3)1-8 line and by \sii, \cii, and \feii\ emission lines. 

\item Overall, the \civ, \siiv, and \nv\ lines in CTTSs all have similar shapes. The 1243 \AA\ \nv\ line is strongly absorbed by circumstellar \niti\ and both lines of the \siiv\ doublet are strongly affected by \htwo\ emission (Figures \ref{panel_1} to \ref{panel_5}). We do no detect \htwo\ emission within $\pm$400 \kms\ of the hot gas lines for WTTSs. 

\item In general, the \heii\ CTTSs lines are symmetric and narrow, with FWHM$\sim$100 \kms. The FWHM and redshifts are  comparable to the same values in WTTSs. The \heii\ lines also have the same FWHM as the narrow component of \civ\ in CTTSs. They are less redshifted than the CTTSs \civ\ lines, by $\sim$10 \kms, but have the same redshift as the WTTSs. A comparison of the Gaussian parameters for \civ\ and \heii\ is shown in Figure \ref{histogram_vel}. 

\end{itemize}

\subsection{Correlations}

\begin{itemize}

\item We confirm that the \civ\ line luminosities are correlated with accretion luminosities and accretion rates, although the exact correlation depends on the set of extinctions that is adopted (Figure \ref{lacc_vs_lciv}).  \civ\ and \heii\ luminosities are also correlated with each other, and the \civ/\heii\ luminosity ratio is a factor of three to four times larger in CTTSs than in WTTS. This confirms that the ratio depends crucially on accretion phenomena and it is not an intrinsic property of the stellar atmosphere. 

\item We do not find any significant correlation between inclination and \civ\ $V_{Max}$, FWHM, or skewness, or between inclination and any of the Gaussian parameters for the \civ\ decomposition (Figure \ref{vel_max}).  We conclude that the dataset is not large enough for inclination to be a strong descriptor in the sample. The variability observed in the \civ\ lines of RU Lup and DR Tau (Figure \ref{one_panel_multiepoch}) may be consistent with the \civ\ accretion spot coming in and out of view.

\item We do not find correlations among $V_{Max}$, FWHM, or skewness, with each other or with accretion rate or line luminosity. On the same token, we do not find correlations between the width of the Gaussian components and accretion rate or line luminosity. 

\item The ratio between the flux in the 1548 \AA\ \civ\ line to the flux in the 1550 \AA\  \civ\ line is a measure of the opacity of the emitting region. For CTTSs, 70\% (19/29) of the stars have blue-to-red \civ\ line ratios consistent with optically thin or effectively thin emitting regions. DF Tau is anomalous, with a blue-to-red \civ\ line ratio close to 3, indicating either emission in the 1548 \AA\  line or absorption on the 1550 \AA\  one. 

\item We find that the contribution fraction of the NC to the \civ\ line flux in CTTSs increases with accretion rate, from $\sim$20\% up to $\sim$80\%, for the range of accretion rates considered here (Section \ref{shape}, Figure \ref{opacity}).  This suggests that for some stars the region responsible for the BC becomes optically thick with accretion rate. As a response to the accretion process, the \civ\ lines develop a BC first.  

\item We show resolved multi-epoch observations in \civ\ for six CTTSs: BP Tau, DF Tau, DR Tau, RU Lup, RW Aur A, and T Tau N.  The kind of variability observed is different for each star, but we confirm that the NC of BP Tau, DF Tau, and RU Lup increase with an increase in the line flux (Section \ref{variability}, Figure \ref{one_panel_multiepoch}).

\end{itemize}

\subsection{Abundance Anomalies?}

\begin{itemize}

\item The \siiv\ and \nv\ line luminosities are correlated with the \civ\ line luminosities (Figure \ref{siiv_civ_correlation}). The relationship between  \siiv\ and \civ\ shows large scatter with respect to a linear relationship.  If we model the emitting region as an optically thin plasma in coronal ionization equilibrium, we conclude that TW Hya and V4046 Sgr show evidence of silicon depletion with respect to carbon, as has already been noted in previous papers  \citep{her02, kastner2002, stelzer2004, gunther2006}. Other stars (AA Tau, DF Tau, GM Aur, and V1190 Sco) may also be silicon-poor, while CV Cha, DX Cha, RU Lup, and RW Aur may be silicon-rich.  The relationship between  \nv\ and \civ\ shows significantly less scatter. 
\item For WTTSs, we observe silicon to be generally more abundant than expected within our simple model. This is reminiscent of the first ionization potential (FIP) effect. However, an inverse FIP has been observed in the Fe/Ne abundance ratios of active, low-mass, non-accreting stars, which should behave similarly to WTTSs \citep{gudel2007b}. It is unclear whether the discrepancy is due to the simplicity of the assumed model or if the Si/C ratios in WTTSs behave differently than the Fe/Ne ratios in active stars. This points to the need for a more sophisticated model of the relationship between \siiv, \civ, and \nv\ both in WTTSs and CTTSs.

\end{itemize}

\subsection{Age effects}

\begin{itemize}

\item Our sample covers a range of ages from $\sim$2 Myr to $\sim$10 Myr, and a wide range in disk evolutionary state. We do no detect changes in any of the measured quantities (line luminosities, FWHM, velocity centroids, etc) with age. We do not find systematic differences in any quantity considered here between the CTTS subsample with transition disks and the whole CTTS sample.  
\item We do not find any significant difference in the accretion rates of the \siiv-rich objects compared to those \siiv-poor, nor in their disk characteristics. 

\end{itemize}

\subsection{The origin of the hot gas lines}

\begin{itemize}
\item  We find no evidence for a decrease in any of the Gaussian components of \civ\ that could be interpreted as burying of the accretion column in the stellar photosphere due to the ram pressure of the accretion flow \citep{drake2005, sacco2010}. In particular, we do not observe a decrease in the strength of the NC as a function of accretion, but exactly the opposite. In the context of the magnetospheric accretion model, we interpret this as an argument in favor of a line origin in either (a) an accretion spot with an aspect ratio such that the post-shock photons can escape the columns (f$>$0.001), and/or (b) the low-density, slow-moving periphery of an inhomogeneous accretion column \citep{romanova2004}, and/or (c) multiple, uncorrelated accretion columns of different densities \citep{sacco2010, orlando2010, ingleby2013}, which appear as the result of increasing accretion rate.
\item The accretion shock model predicts that the velocity of the post-shock emission should be 4 times smaller than the velocity of the pre-shock gas emission. If we identify the broad Gaussian component with the pre-shock emission and the NC with the post-shock emission, we find that V$_{BC}\gtrsim4$ V$_{NC}$ is the case in only 2 out of 22 CTTSs. Accounting for possible pointing errors, 10 more objects could be within the accretion shock predictions (see Figure \ref{vbc_vs_vnc}). For six systems the Gaussian decomposition is impossible (DX Cha, RW Aur) or requires only one Gaussian component (CV Cha, ET Cha, DM Tau, UX Tau A). 
\item When compared to the predicted infall velocities, the measured V$_{BC}$ values of these objects imply large ($\sim$90\degs) inclinations of the flow with respect to the line of sight. Because this is unlikely in general, we suggest that the accretion flow responsible for the emission may not be at the free-fall velocity (for example if the emission comes from the slow-region at the edge of the column). Another alternative is that we are observing emission from multiple columns with different lines of sight, in which case the ratio of pre- to post-shock emission is not the same. Finally, radiation from the post-shock gas may be ionizing regions far from it, which, because of the shape of the magnetic field channeling the flow, may have line-of-sight velocities unrelated to the post-shock ones.  Our models of the pre-shock gas show that pre-shock sizes can range from a few tens of kilometers to $\sim$1 R$_\odot$ for typical parameters. These concepts may also explain cases for which the NC velocity is positive, but the BC velocity is small or negative (DF Tau, DR Tau, GM Aur, IP Tau, RU Lup, SU Aur, and V1190 Sco).   
\item The accretion shock model predicts that, because the formation temperature of \heii\ is lower than that of \civ, the post-shock line emission from the former should have a lower velocity than the latter. We confirm this, by identifying the NC emission with the post-shock gas emission. However, we note that gas flows in the stellar transition region may result in similar velocity offsets between \civ\ and \heii.
\item Observationally, the amount of flux in the BC of the \heii\ line is small compared to the \civ\ line. We model the pre-shock column and conclude that for typical parameters, it can produce line ratios between \civ\ and \heii\ within the observed range. In other words, the weakness of the BC component in \heii\ is consistent with its origin in the pre-shock gas. 
\item Overall, we favor the origin of the emission lines in an accretion shock, but we cannot rule out that the true transition region is contributing part of the NC flux. 
\end{itemize}
\subsection{Outflow shocks and hot winds}

\begin{figure}
\epsscale{1}
\includegraphics[width=3.4in]{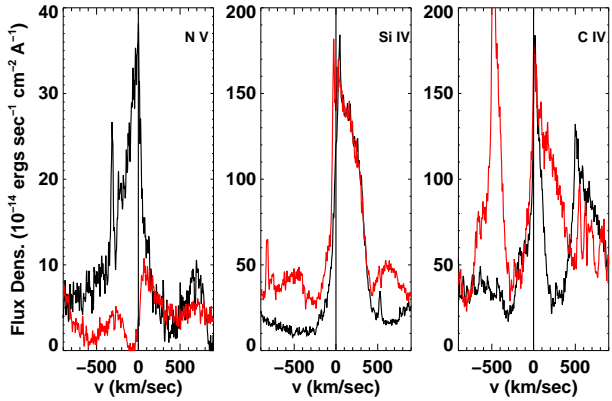}
\caption{The \nv, \siiv, and \civ\ doublets for DX Cha. The red line is the redder member of the doublet, scaled to match the blue member. The red member of the \nv\ lines shows strong \niti\ absorption. The red member of \siiv\ show \oiv\ emission at -341 \kms. Notice the very sharp blue cutoff in both \siiv\ lines. In the case of \civ, we believe that red wing of the bluer line (shown in black) is being absorbed by a hot wind. Blueshifted absorption is seen in the blue wing of the blue member.  \label{winds}}
\end{figure}

We find three different types of profiles that show evidence of outflows (Section \ref{outflow}):
\begin{itemize}
\item For HN Tau A and RW Aur A, most of the \civ\ and \heii\ line fluxes are blueshifted and the peaks of the \htwo\ lines are redshifted \citep{france2012}. The \civ\ and \heii\ emission in this case should be produced by shocks within outflow jets.  For these stars we may be observing the opposite sides of an outflow simultaneously (Section \ref{comparison}, Figure \ref{rw_aur}).

\item For the CTTSs AK Sco, DK Tau, T Tau N, we find that the NC velocity is negative. These represent less extreme examples of objects such as HN Tau A and RW Aur, although the observed velocities are high enough that internal shocks in the outflow can result in hot gas emission.  For these stars we do not observe any redshifted NC emission that would correspond to the postshock.
\item Within our sample, DX Cha is the only candidate for a high temperature wind, as it shows a deficit in the red wing of the 1548 \AA\  line for \civ, compared to the red wing of the 1550 \AA\  line, and blue absorption in 1548 \AA\  \civ\  line (Figure \ref{winds}). Lower temperature outflows in \siii\ are also observed in DX Cha.
\end{itemize}
\subsection{Peculiar Objects}
Certain well-known objects are tagged as peculiar in this work. In addition to DX Cha, HN Tau, and RW Aur, mentioned before, the most peculiar are:
\begin{itemize}
\item AK Sco: Even when accounting for extra emission lines, the \civ\ lines are different form each other, perhaps as a result of it being a spectroscopic binary. It may be Si-rich.
\item DF Tau: The 1548-to-1550 ratio is too large. The BC velocity is small compared to expectations of the accretion model.
\item DK Tau: The BC velocity is negative, the \civ\ profile is double-peaked, may be Si-rich.
\item DR Tau: For \civ, the BC velocity is small compared to expectations, the thermalization depth for the NC is small, but the BC flux ratio is anomalously high. The \civ\ lines are double-peaked. Multi-epoch observations show either absorption within 150 \kms\ of the line center, or evidence for rotational modulation in the profile. It may be Si-rich.
\item RU Lup: The red wings of the \civ\ lines are different from each other, the BC velocity,  the thermalization depth for the NC is small, and it may be Si-rich. In addition, some of the \htwo\ lines present a low level blushifted emission. As in DR Tau, we may be observing rotational modulation of the \civ\ profile in multi-epoch observations.
\end{itemize}

High-spectral resolution UV observations provide a crucial piece of the puzzle posed by young stellar evolution. They sample a much hotter plasma than optical observations and complement and enhance X-ray data. However the data considered here has two important limitations. One is the lack of time-domain information, which makes it difficult to understand what is the average behavior of a given target. The other is the small range of accretion rates well-covered by objects. Future observational work should focus on resolving these limitations. Even with these limitations, it is clear that with this work we have only scratched the surface of this magnificent dataset.


\clearpage

\clearpage

\clearpage

\clearpage

\clearpage

\clearpage

\clearpage



\clearpage
\appendix
\renewcommand{\thefigure}{A.\arabic{figure}} 
\setcounter{figure}{0}

\section{Multi-panel Figures}
\clearpage
\begin{figure}
\epsscale{1}
\plotone{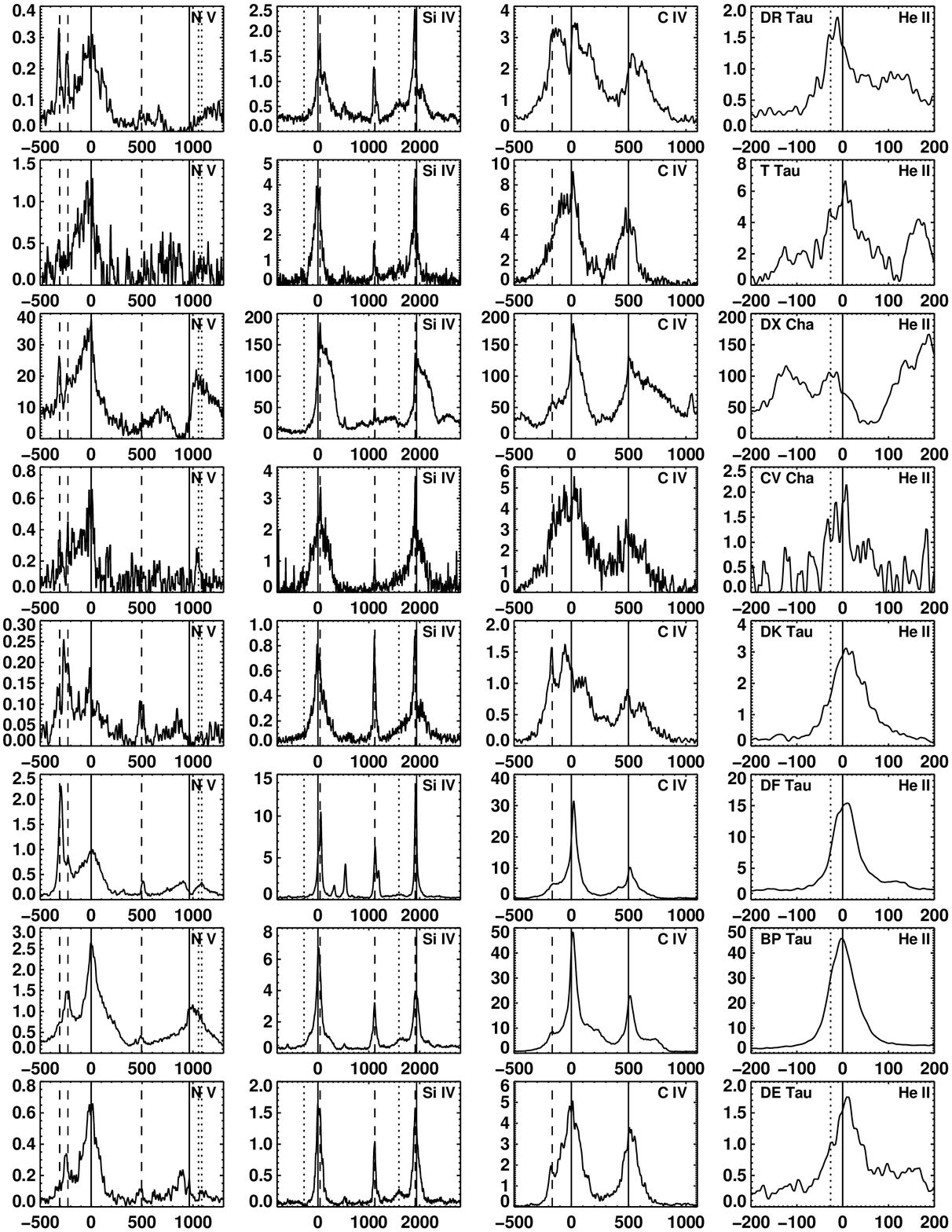} 
\caption{Hot lines for CTTSs ordered by decreasing ${\dot {M}}$. Left to right: \nv, \siiv, \civ, \heii. The plots show flux density ($10^{-14}$ erg sec$^{-1}$  cm$^{-2}$ \AA$^{-1}$) versus velocities (\kms) in the stellar rest frame. The spectra are not extinction-corrected nor continuum-subtracted. Transition disks are indicated with ``TD" after the name. {\bf Solid lines}: Nominal line positions;  {\bf Dashed lines}: Nominal locations of the strongest \htwo \ lines. {\bf \nv, dotted lines}: \niti\ (1243.18 \AA, +1055 \kms; 1243.31 \AA, +1085 \kms). {\bf \siiv, dotted lines}: CO A-X (5-0) bandhead (1392.5 \AA, -271 \kms), {O {\scshape iv}} (1401.17 \AA, +1633 \kms). {\bf  {He {\scshape ii}}, dotted line}: Location of the secondary \heii \ line (Section \ref{heii}). \label{panel_1}}
\end{figure}

\begin{figure}
\epsscale{1}
\plotone{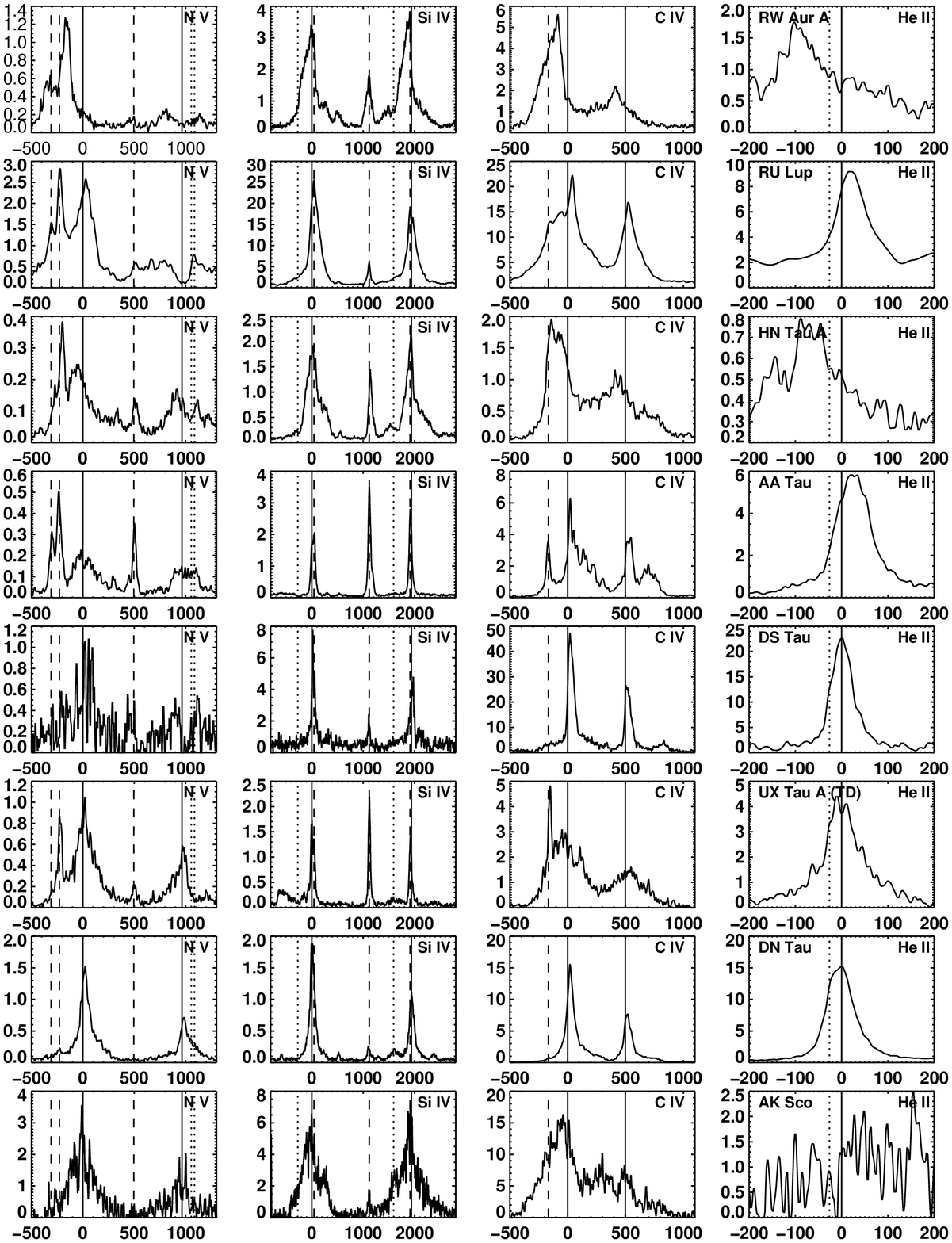} 
\caption{Same as Figure \ref{panel_1} \label{panel_2}}
\end{figure}

\begin{figure}
\epsscale{1}
\plotone{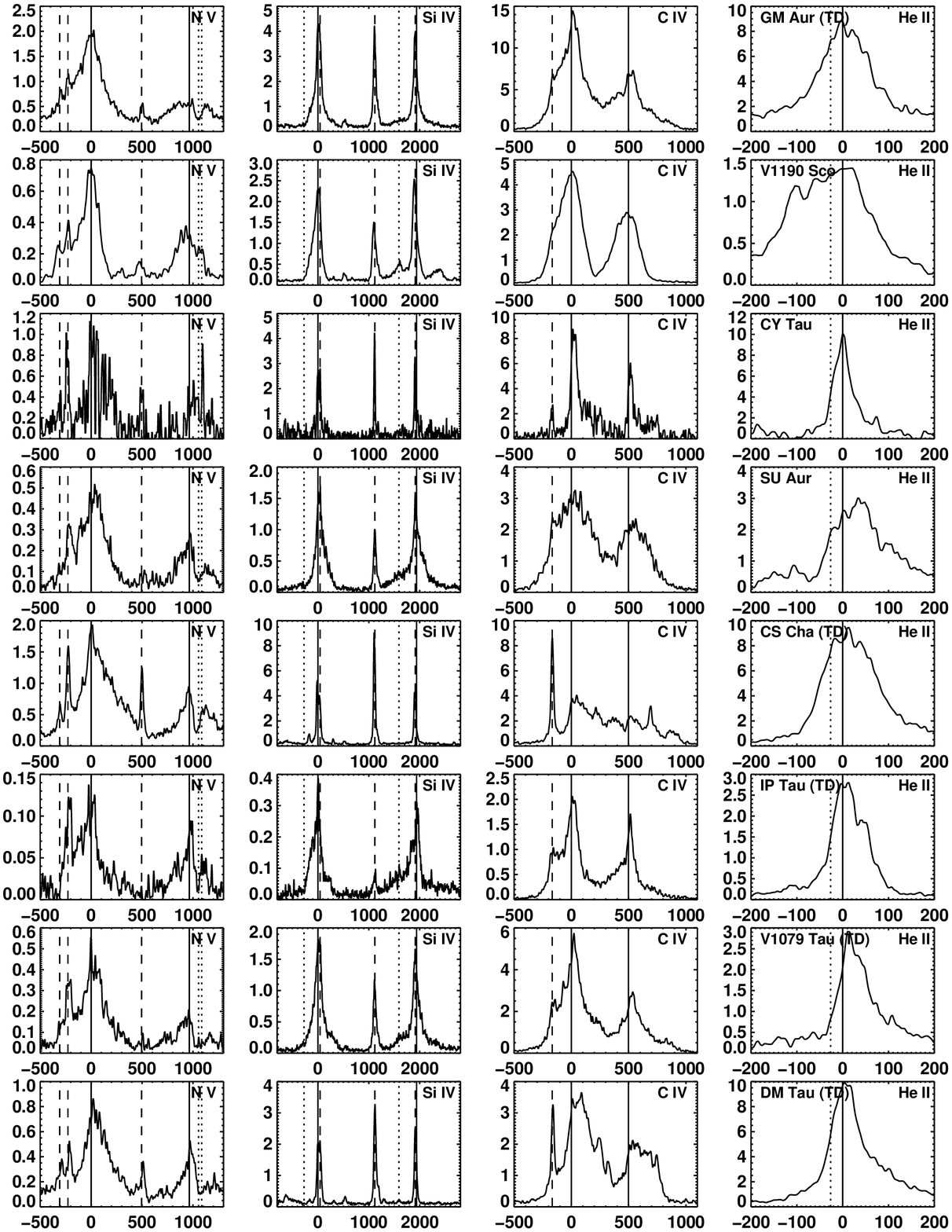} 
\caption{Same as Figure \ref{panel_1} \label{panel_3}}
\end{figure}

\begin{figure}
\epsscale{1}
\plotone{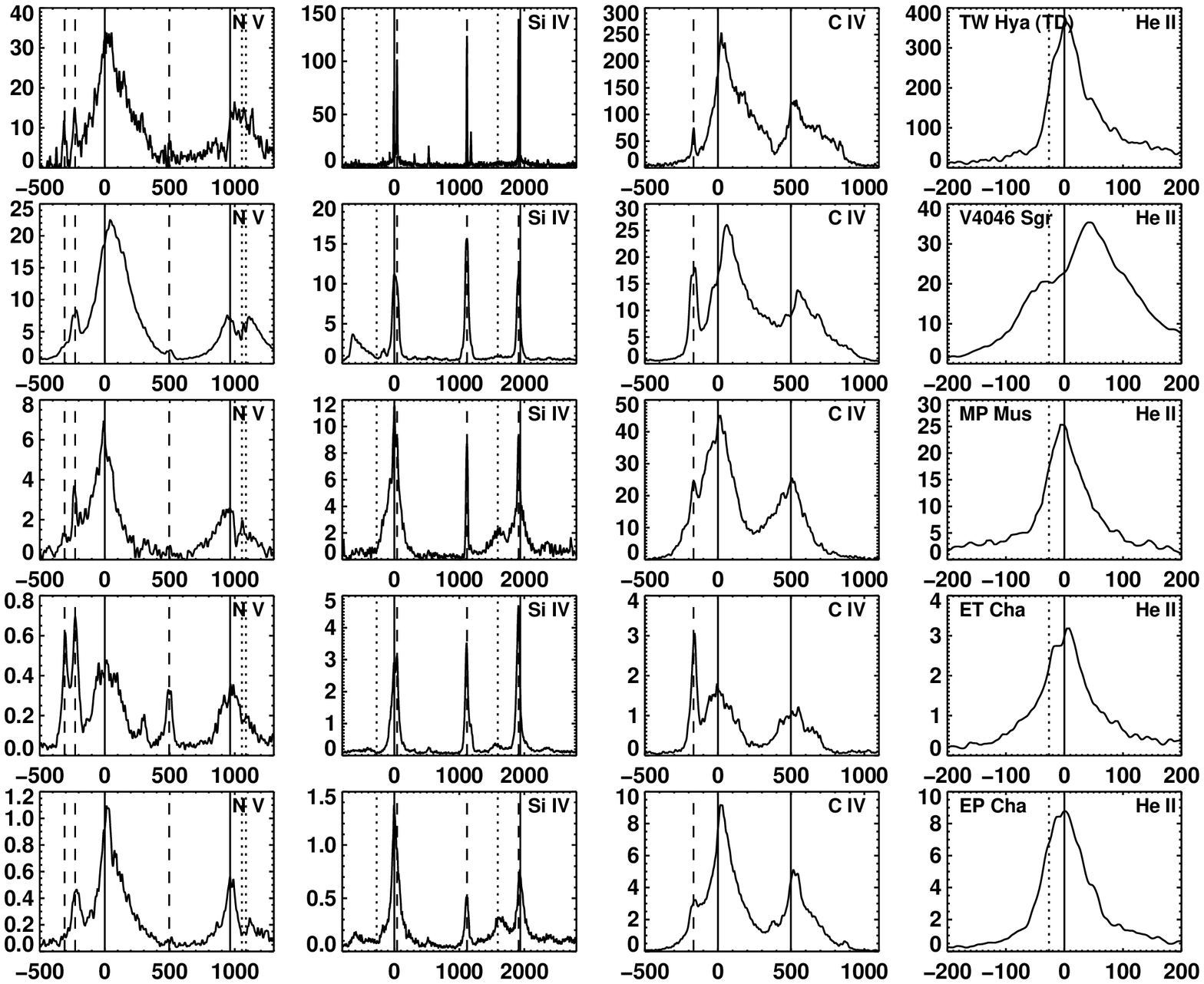} 
\caption{Same as Figure \ref{panel_1} \label{panel_4}}
\end{figure}

\begin{figure}
\epsscale{1}
\plotone{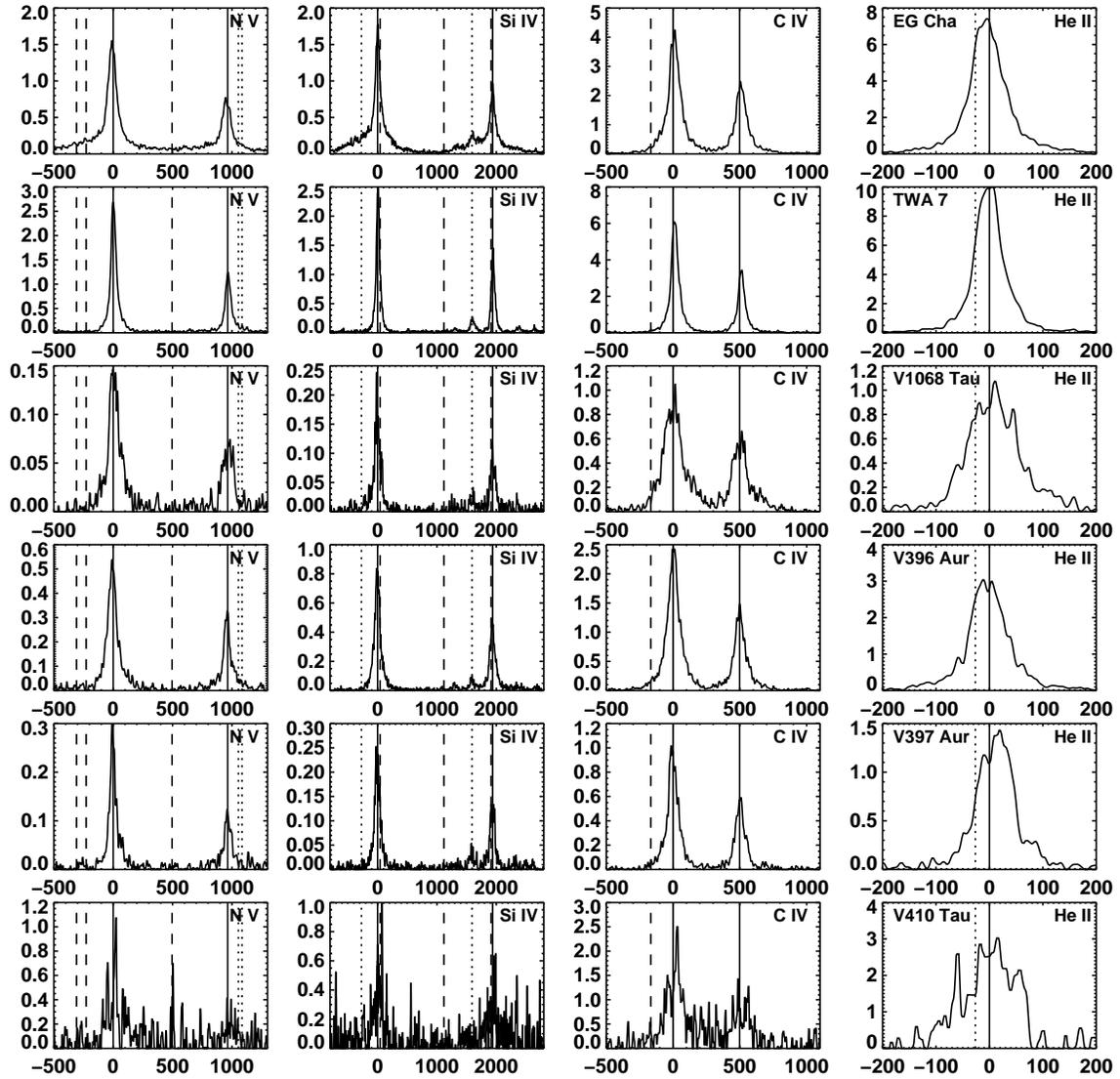} 
\caption{The ``hot" lines for WTTS. The stars are listed in alphabetical order. Axes and markings are the same as in Figure \ref{panel_1} \label{panel_5}}
\end{figure}

\begin{figure}
\epsscale{1}
\plotone{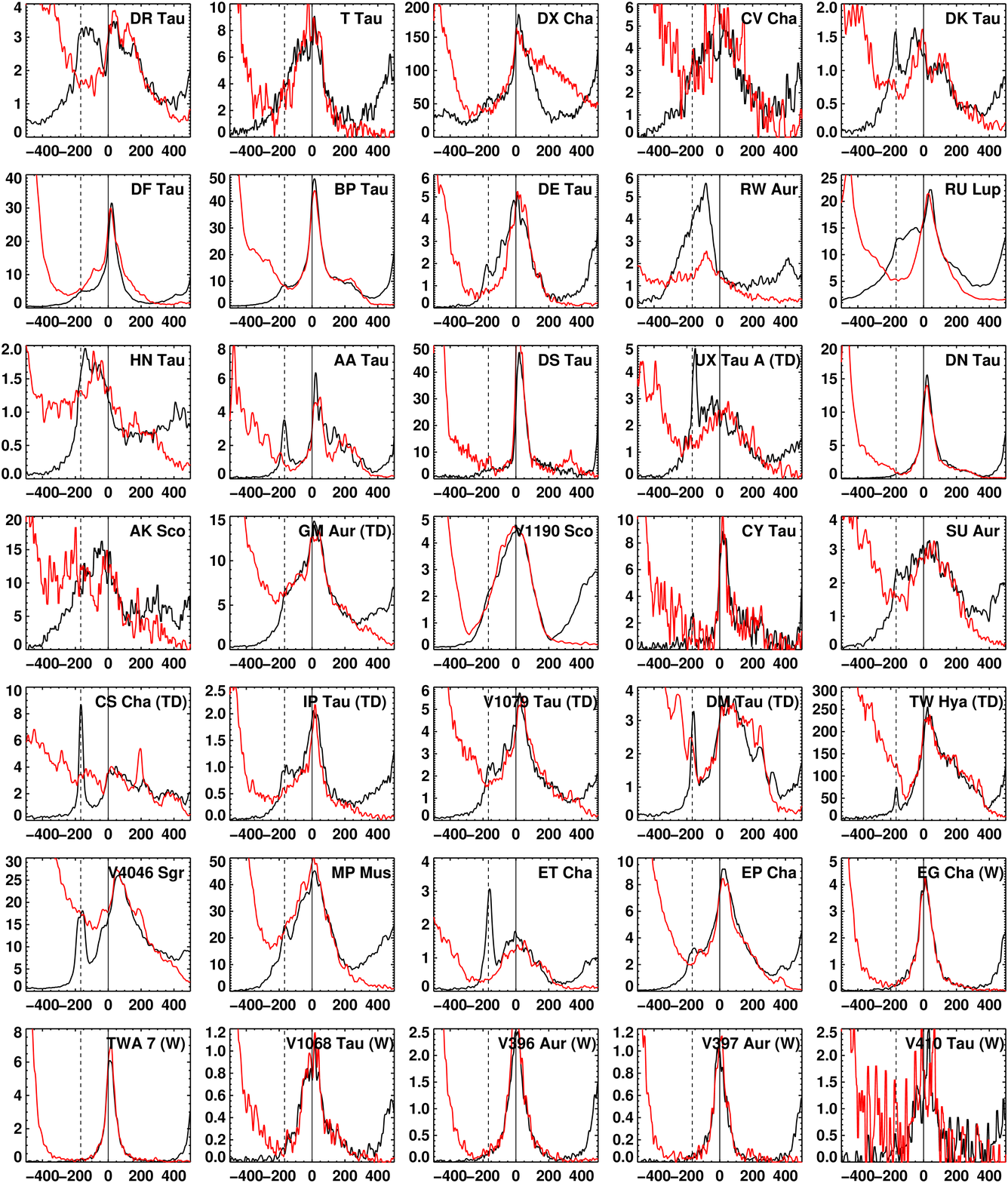}
\caption{Both members of the \civ \ doublet, scaled and overplotted. The 1550 \AA\ line (the red member of the doublet) is shown in red, and it has been scaled to match the red wing of the 1548 \AA\ line. Axes and units are the same as in figure \ref{panel_1}. \label{all_civ}}
\end{figure}

\begin{figure}
\epsscale{1}
\plotone{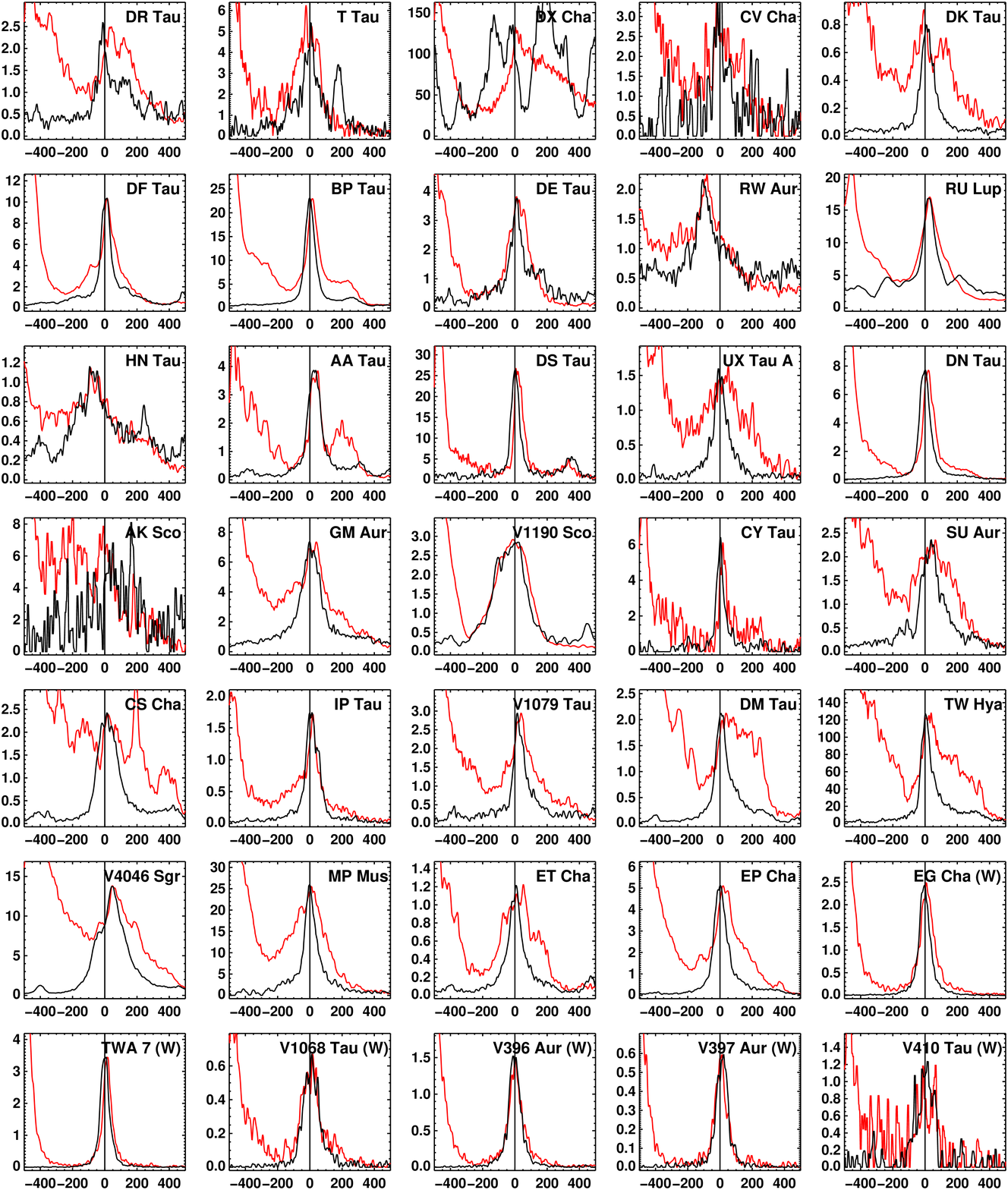}
\caption{The black trace shows the \heii\ line overploted to the 1550 \AA\ \civ\ and scaled to the same maximum value. Axes and units are the same as in figure \ref{panel_1}.  \label{all_civ_heii}}
\end{figure}

\clearpage
\acknowledgments

Based on observations made with the NASA/ESA Hubble Space Telescope. Support for this paper was provided by NASA through grant numbers HST-GO-11616.10 and HST-GO-12161.01 from the Space Telescope Science Institute (STScI), which is operated by Association of Universities for Research in Astronomy, Inc (AURA) under NASA contract NAS 5-26555. SGG acknowledges support from the Science $\&$ Technology Facilities Council (STFC) via an Ernest Rutherford Fellowship [ST/J003255/1]. RDA acknowledges support from the UK's Science \& Technology Facilities Council (STFC) through an Advanced Fellowship (ST/G00711X/1).

This research has made use of NASA's Astrophysics Data System Bibliographic Services and CHIANTI,  a collaborative project involving George Mason University, the University of Michigan (USA) and the University of Cambridge (UK).  

We thank the team from {\it HST} GTO programs 11533 and 12036 (PI J. Green) for allowing us early access to their data.

{\it Facilities:} \facility{HST (\COS, \STIS)}.

\bibliographystyle{/Users/ardila/bibtex/apj}


\clearpage
\LongTables 
\begin{landscape}
\begin{deluxetable}{lcccclcl}
\setlength{\tabcolsep}{2pt}
\tablecolumns{8}
\tabletypesize{\tiny}
\tablewidth{0pt}
\tablecaption{Stars analyzed in this paper\label{TableAncil1}}
\tablehead{\colhead{Name (a)} & \colhead{Alternate name} & \colhead{Region} &  \colhead{d} & \colhead{SpT (b)} & \colhead{Ref. SpT} & \colhead{Av} & \colhead{Ref. Av} \\
\colhead{} & \colhead{} & \colhead{} &  \colhead{(pc)} &\colhead{} & \colhead{} & \colhead{(mag)} & \colhead{} } 
\startdata
\cutinhead{Classical T Tauri Stars}
AA Tau & HBC 63 & Taurus-Auriga & 140 & K7 & \cite{donati2010} & 0.74 & \cite{gull98} \\
AK Sco & HBC 271  & Upper-Scorpius & 145 & F5+F5 & \cite{andersen1989} & 0.5 & \cite{alencar2003} \\
BP Tau & HBC 32 & Taurus-Auriga & 140 & K7 & \cite{cmj1999} & 0.5 & \cite{gull98} \\
CV Cha & HBC 247 & Chamaeleon I & 160 & G8 & \cite{guenther2007} & 1.5 & \cite{furlan2009} \\
CY Tau & HBC 28 & Taurus-Auriga & 140 & M1 & \cite{hartmann1998} & 0.32 & \cite{gull98} \\
DE Tau & HBC 33 & Taurus-Auriga & 140 & M0 & \cite{furlan2009} & 0.62 & \cite{gull98} \\
DF Tau & HBC 36  & Taurus-Auriga & 140 & M2+? & \cite{her06} & 0.6 & \cite{her06} \\
DK Tau & HBC 45 & Taurus-Auriga & 140 & K7 & \cite{furlan2009} & 1.42 & \cite{gull98} \\
DN Tau & HBC 65 & Taurus-Auriga & 140 & M0 & \cite{furlan2009} & 0.25 & \cite{gull98} \\
DR Tau & HBC 74 & Taurus-Auriga & 140 & K7 & \cite{petrov2011} & 1.2 & \cite{gullbring2000} \\
DS Tau & HBC 75 & Taurus-Auriga & 140 & K5 & \cite{muzerolle1998} & 0.34 & \cite{gull98} \\
DX Cha & HD 104237 & $\epsilon$ Chamaeleontis & 114 & A7.5+K3 & \cite{bohm2004} & 0.56 &  \cite{sartori2003} \\
EP Cha & RECX 11 & $\eta$ Chamaeleontis & 97 & K4 &  \cite{mamajek1999} & 0 & \cite{ingleby2013} \\
ET Cha & RECX 15 & $\eta$ Chamaeleontis & 97 & M2 & \cite{lawson2002} & 0 & \cite{ingleby2013} \\
HN Tau A & HBC 60 & Taurus-Auriga & 140 & K5 & \cite{furlan2009} & 0.65 & \cite{gull98} \\
MP Mus & PDS 66 & Isolated? & 100 & K1 & \cite{torres2006} & 0.17 & \cite{mamajek2002} \\
RU Lup & HBC 251  & Lupus I & 150 & K7 & \cite{her05} & 0.1 & \cite{her06} \\
RW Aur A & HBC 80  & Taurus-Auriga & 140 & K4+? & \cite{cmj1999} & 1.2 & \cite{valenti1993} \\
SU Aur & HBC 79 & Taurus-Auriga & 140 & G1 & \cite{furlan2009} & 0.9 & \cite{gullbring2000} \\
T Tau N & HBC 35 & Taurus-Auriga & 140 & K1 & \cite{walter2003} & 0.3 & \cite{walter2003} \\
V1190 Sco & HBC 617; Sz 102 & Upper Scorpius & 145 & K0 & \cite{hughes1994} & 0.32 & \cite{sartori2003}  \\
V4046 Sgr & HBC 662 & $\beta$ Pictoris & 72 & K5+K7 & \cite{quast2000} & 0 & \cite{curran2011} \\
 \cutinhead{Transition Disks}
CS Cha & HBC 569 & Chamaeleon I & 160 & K6+? & \cite{luhman2004} & 0.3 & \cite{furlan2009} \\
DM Tau & HBC 62 & Taurus-Auriga & 140 & M1.5 & \cite{espaillat2010} & 0.6 & \cite{ingleby2009} \\
GM Aur & HBC 77 & Taurus-Auriga & 140 & K5.5 & \cite{espaillat2010} & 0.31 & \cite{gull98} \\
IP Tau & HBC 385 & Taurus-Auriga & 140 & M0 & \cite{furlan2009} & 0.32 & \cite{gull98} \\
TW Hya & HBC 568 & TW Hydrae Association & 55 & K6 & \cite{torres2006} & 0.0 & \cite{her04} \\
UX Tau A & HBC 43 & Taurus-Auriga & 140 & K2 & Herbig (1977) & 0.7 & \cite{furlan2009} \\
V1079 Tau  & LkCa 15; HBC 419  & Taurus-Auriga & 140 & K3 & \cite{espaillat2010} & 1.00 & \cite{white2001} \\
 \cutinhead{Weak T Tauri Stars}
EG Cha & RECX 1 & $\eta$ Chamaeleontis & 97 & K4+? & \cite{mamajek1999} & 0 & \cite{ingleby2013} \\
TWA 7 & 2MASS J1042-3340  & TW Hydrae Association & 55 & M2 & \cite{torres2006} & 0 & \cite{ingleby2009} \\
V1068 Tau & LkCa 4; HBC 370 & Taurus-Auriga & 140 & K7 & \cite{herbig1988} & 0.69 & \cite{bertout2007} \\
V396 Aur & LkCa 19; HD 282630 & Taurus-Auriga & 140 & K0 & \cite{herbig1988} & 0 & \cite{walter1988} \\
V397 Aur & HBC 427 & Taurus-Auriga & 140 & K7+M2 & \cite{steffen2001} & 0.0 & \cite{kenyon1995} \\
V410 Tau & HBC 29 & Taurus-Auriga & 140 & K2 (A) +M? (B) & \cite{stelzer2003} & 0.03 & \cite{bertout2007} \\
 \enddata
\tablecomments{\\
(a) In this paper we refer to the stars with their variable star name, if it exists. The exception is TWA 7, which does not have variable star name.\\
(b) For binary stars, the spectral type refers to the primary component, unless indicated.}
\end{deluxetable}

\clearpage
\end{landscape}

\clearpage
\LongTables 
\begin{landscape}
\begin{deluxetable}{lclclcl}
\setlength{\tabcolsep}{1.1pt}
\tablecolumns{7}
\tabletypesize{\tiny}
\tablewidth{0pt}
\tablecaption{Ancillary Data \label{TableAncil2}}
\tablehead{\multicolumn{1}{l}{Name} & \colhead{Radial Velocity} & \multicolumn{1}{l}{Ref. Rad. Vel.} & \colhead{Inc. (a)} & \multicolumn{1}{l}{Ref. Inc.} & \colhead{Binarity (b)} &\multicolumn{1}{l}{Ref. Binarity} \\
\colhead{} & \colhead{(km s$^{-1}$)} & \colhead{} & \colhead{(deg)} & \colhead{} & \colhead{} &\colhead{}
 } 
\startdata
\cutinhead{Classical T Tauri Stars}
AA Tau & 16.5 & \cite{bouvier1999} & 70$\pm$10 & \cite{donati2010}  & S & \cite{najita2007} \\
AK Sco & -1.15 & \cite{andersen1989} & 63 & \cite{andersen1989}  & 2 (SB) & \cite{alencar2003} \\
BP Tau & 15.8 & \cite{hartmann1986} & $\sim$45 & \cite{donati2008}  & S & \cite{leinert1993} \\
CV Cha & 16.1 & \cite{guenther2007} & 35$\pm$10 & \cite{hussain2009}  & S & \cite{melo2003} \\
CY Tau & 19.1 & \cite{hartmann1986} & 47$\pm$8 & \cite{simon2000}  & S & \cite{najita2007} \\
DE Tau & 14.9 & \cite{hartmann1986} & 57 & \cite{appenzeller2005}  & S & \cite{najita2007} \\
DF Tau & 11.0 & Edwards et al. (1987) & 50 & \cite{appenzeller2005}  & 2  (0.09") & \cite{ghez1993} \\
DK Tau & 15.3 & \cite{hartmann1986} & 44 & \cite{appenzeller2005}  & 2 (2.304") & \cite{correia2006}  \\
DN Tau & 16.1 & \cite{hartmann1986} & 30 & \cite{appenzeller2005}  & S & \cite{najita2007} \\
DR Tau & 27.6 & \cite{alencar2000} & 20$\pm$4 & \cite{petrov2011}  & S & \cite{leinert1993} \\
DS Tau & 16.3 & \cite{hartmann1986}  & $\sim$90 & Derived from & S & \cite{najita2007} \\
 & &  & & \cite{kundurthy2006} &  & \\
 & &  & & \cite{glebocki2005} &  & \\
DX Cha & 13 & \cite{kharchenko2007}; & 18 & \cite{grady2004}  & 2 (SB) & \cite{bohm2004} \\
EP Cha (RECX 11)& 18 & Assumed to be as EG Cha & N/A & $\ldots$  & S & \cite{brandeker2006} \\
ET Cha (RECX 15) & 15.9 & \cite{barbier2000} & 40 & \cite{woitke2011}   & S & \cite{woitke2011} \\
HN Tau A & 4.6 & \cite{nguyen2012} & N/A & $\ldots$  & 2 (A-B: 3.109") & \cite{correia2006} \\
MP Mus & 11.6 & \cite{torres2006} &32 &\cite{curran2011}  & S & \cite{curran2011} \\
RU Lup & -0.9 & \cite{melo2003} & 24 & \cite{her05}  & S & \cite{lamzin1995}\\
RW Aur A & 14.0 & \cite{hartmann1986} & 37 & \cite{appenzeller2005}  & M  & \cite{gahm1999};\\
   &  &  &  &   & (Aa-Ab: 0.01"; AaAb-BC: 1.41")  &\cite{correia2006} \\
SU Aur & 9.7 & \cite{hartmann1986} & 52$\pm$10 & \cite{akeson2005}  & S & \cite{nguyen2012} \\
T Tau N & 19 & \cite{hartmann1986} & 19 & \cite{herbst1997}  & 2 (0.7") & \cite{dyck1982} \\
V1190 Sco & 5.0 & \cite{graham1988} & $<$5 & \cite{comeron2010}  & N/A & $\ldots$ \\
V4046 Sgr & -6.94 & \cite{quast2000} & 35 & \cite{quast2000}  & 2 (SB) & \cite{quast2000} \\
 \cutinhead{Transition Disks}
CS Cha & 15.0 & \cite{reipurth1995} & 60 & \cite{espaillat2011}  & 2 (SB)? (c) & \cite{guenther2007} \\
DM Tau & 16.9 & \cite{hartmann1986} & 45$\pm$5 & \cite{simon2000}  & S & \cite{najita2007} \\
GM Aur & 14.8 & \cite{simon2000} & 54$\pm$5 & \cite{simon2000}  & S & \cite{najita2007} \\
IP Tau & 14.8 & \cite{hartmann1987} & 30 & Derived from & S & \cite{najita2007} \\
  & &  & & \cite{glebocki2005} &  & \\
    & &  & & \cite{glebocki2005} &  & \\
TW Hya & 13.5 & \cite{sterzik1999}; & 18$\pm$10 & \cite{her02}  & S & \cite{curran2011} \\
   &  & measured by \cite{her02} & &   & &  \\
UX Tau A & 22.9 & \cite{kharchenko2007}; & 49 & \cite{andrews2011}  & S & \cite{najita2007} \\
V1079 Tau  (LkCa 15)& 17.0 & \cite{hartmann1987} & 42$\pm$5 & \cite{simon2000}  & S & \cite{najita2007} \\
\cutinhead{Weak T Tauri Stars}
EG Cha (RECX 1)& 18 & \cite{malaroda2000}; & N/A & $\ldots$  & M  & \cite{kohler2002} \\
 &  &  &  &   & (AB 0.18"; AB-C: 8.6") &  \\
TWA 7 & 10.6 & \cite{kharchenko2007}; & N/A & $\ldots$  & S & \cite{lowrance2005} \\
V1068 Tau (LkCa 4)& 16.9 & \cite{hartmann1987} & N/A & $\ldots$  & S & \cite{najita2007} \\
V396 Aur (LkCa 19)& 14.3 & \cite{hartmann1987} & N/A & $\ldots$  & S & \cite{najita2007} \\
V397 Aur & 15.0 & \cite{walter1988} & N/A & $\ldots$  & 2 (0.0328") & \cite{steffen2001} \\
V410 Tau & 17.8 & \cite{hartmann1986} & N/A & $\ldots$  & M & \cite{correia2006} \\
 &  &  &  &   & (AB: 0.07"; AB-C: 0.290")   \\
\enddata
\tablecomments{\\
(a) Inclination from face-on ($=0^{\circ}$). When inclination is listed as "Derived from" we have derived it from the reported $v * \sin i$, the period, and the stellar radius.\\
(b) Binarity: S=Single; M=Multiple; SB=Spectroscopic Binary. The number in parenthesis indicates the separation of the components.\\
(c) Possible spectroscopic long-period ($>$2482 days) binary. Unknown companion characteristics \citep{guenther2007}.
}
\end{deluxetable}

\clearpage
\end{landscape}

\clearpage
\begin{deluxetable}{lclc}
\setlength{\tabcolsep}{2pt}
\tablecolumns{4}
\tabletypesize{\tiny}
\tablewidth{0pt}
\tablecaption{Ancillary Data (cont.)\label{TableAncil3}}
\tablehead{\colhead{Name} & \colhead{$\dot{M}$} &  \colhead{Ref. $\dot{M}$} & \colhead{Simult. $\dot{M}$? (a)} \\
\colhead{} &\colhead{$10^{-8}$ M$_\odot$/yr} &  \colhead{} &  \colhead{Y/N}} 
\startdata
\cutinhead{Classical T Tauri Stars}
AA Tau &   1.3 & \cite{ingleby2013} & Y  \\
AK Sco &   1 & Assumed (b) & N \\
BP Tau &  2.9 & \cite{ingleby2013} & N \\
CV Cha &3.5 & \cite{ingleby2013} & N \\
CY Tau &  0.75 & \cite{gull98} & N \\
DE Tau &   2.6 & \cite{ingleby2013} & Y \\
DF Tau &   3.45 & \cite{her06} - Average & N \\
DK Tau & 3.5 & \cite{ingleby2013} & Y \\
DN Tau &   1.0 & \cite{ingleby2013} & Y \\
DR Tau & 5.2 & \cite{ingleby2013} & Y \\
DS Tau &   1.3 & \cite{gull98} & N \\
DX Cha &  3.55 & \cite{garcia2006} & N \\
EP Cha  (RECX 11)&   0.017 & \cite{ingleby2013} & Y \\
ET Cha  (RECX 15)&   0.042 & \cite{ingleby2013} & Y \\
HN Tau A &  1.4 & \cite{ingleby2013} & Y \\
MP Mus &   0.045 & \cite{curran2011} &  N \\
RU Lup &   1.8 & \cite{her06} & N \\
RW Aur A &   2 & \cite{ingleby2013} & Y \\
SU Aur & 0.55 & \cite{cal04} & N \\
T Tau N & 4.45 & \cite{cal04} & N \\
V1190 Sco &   0.79 & \cite{gudel2010} & N \\
V4046 Sgr &   0.060 & \cite{curran2011} & N \\
 \cutinhead{Transition Disks}
CS Cha &  0.53 & \cite{ingleby2013} & Y \\
DM Tau &   0.22 & \cite{ingleby2013} & Y \\
GM Aur &   0.96 & \cite{ingleby2013} & Y \\
IP Tau &  0.48 & \cite{ingleby2013} & Y \\
TW Hya &   0.15 & \cite{her04} - Average & N \\
UX Tau A &   1.1 & \cite{espaillat2010} & N \\
V1079 Tau  (LkCa 15)&   0.31 & \cite{ingleby2013} & Y \\
 \cutinhead{Weak T Tauri Stars}
EG Cha  (RECX 1)&   $\ldots$ & $\ldots$ & $\ldots$  \\
TWA 7 &  $\ldots$ & $\ldots$ & $\ldots$  \\
V1068 Tau (LkCa 4)& $\ldots$ & $\ldots$ & $\ldots$  \\
V396 Aur (LkCa 19)&   $\ldots$ & $\ldots$ & $\ldots$  \\
V397 Aur &   $\ldots$ & $\ldots$ & $\ldots$  \\
V410 Tau &  $\ldots$ & $\ldots$ & $\ldots$  \\
\enddata
\tablecomments{\\
(a) Simultaneous (within $\sim$ 10 hrs) near and far ultraviolet observations are available for some of the DAO targets \citep{herczeg2013}. The last column indicates whether simultaneous NUV observations were used to calculate the accretion rate, as described in \cite{ingleby2013}.
(b) We assume $\dot{M}$=1 $\times 10^{-8}$ M$_\odot$/yr for AK Sco \citep{gdc2009}.\\
}
\end{deluxetable}

\clearpage
\begin{deluxetable}{cccc}
\tablecolumns{4}
\tabletypesize{\tiny}
\tablewidth{0pt}
\tablecaption{Data Sources\label{TableData}}
\tablehead{\colhead{Name} & \colhead{FUV (a)} & \colhead{FUV Dataset (b)} & \colhead{FUV Slit (c)} } 
\startdata
\cutinhead{Classical T Tauri Stars}
AA Tau & DAO: COS G130M; G160M & LB6B070XX & PSA \\
AK Sco & DAO: STIS E140M & OB6B21040  & 0.2"x0.2" \\
BP Tau & GTO 12036: COS G130M; G160M & LBGJ010XX & PSA \\
CV Cha & DAO: STIS E140M & OB6B180XX & 0.2"x0.2" \\
CY Tau & GO 8206: STIS E140M & O5CF030XX & 0.2"x0.2" \\
DE Tau & DAO: COS G130M; G160M & LB6B080XX & PSA \\
DF Tau & GTO 11533: COS G130M; G160M & LB3Q020XX & PSA \\
DK Tau & DAO: COS G130M; G160M & LB6B120XX & PSA \\
DN Tau & DAO: COS G130M; G160M & LB6B040XX & PSA \\
DR Tau & DAO: COS G130M; G160M & LB6B140XX & PSA \\
DS Tau & GO 8206: STIS E140M & O5CF010XX & 0.2"x0.2" \\
DX Cha & DAO: STIS E140M & OB6B25050 & 0.2"x0.2" \\
EP Cha & DAO: COS G130M; G160M & LB6B240XX & PSA \\
ET Cha & DAO: COS G130M; G160M & LB6B170XX & PSA \\
HN Tau A & DAO: COS G130M; G160M & LB6B090XX & PSA \\
MP Mus & DAO: STIS E140M & OB6B230XX & 0.2"x0.2" \\
RU Lup & GTO 12036: COS G130M; G160M & LBGJ020XX & PSA \\
RW Aur & DAO: COS G130M; G160M & LB6B150XX & PSA \\
SU Aur & DAO: COS G130M; G160M & LB6B110XX & PSA \\
T Tau & GO 8157: STIS E140M & O5E3040XX & 0.2X0.06 \\
V1190 Sco & DAO: COS G130M; G160M & LB6B590XX & PSA \\
V4046 Sgr & GTO 11533: COS G130M; G160M & LB3Q010XX & PSA \\
\cutinhead{Transition Disks}
CS Cha & DAO: COS G130M; G160M & LB6B160XX & PSA \\
DM Tau & DAO: COS G130M; G160M & LB6B020XX & PSA \\
GM Aur & DAO: COS G130M; G160M & LB6B010XX & PSA \\
IP Tau & DAO: COS G130M; G160M & LB6B050XX & PSA \\
TW Hya & GO 11608: STIS E140M & OB3R070XX & 0.2"x0.2" \\
UX Tau A & DAO: COS G130M; G160M & LB6B530XX & PSA \\
V1079 Tau  & DAO: COS G130M; G160M & LB6B030XX & PSA \\
 \cutinhead{Weak T Tauri Stars}
EG Cha & DAO: COS G130M; G160M & LB6B320XX & PSA \\
TWA 7 & DAO: COS G130M; G160M & LB6B300XX & PSA \\
V1068 Tau & DAO: COS G130M; G160M & LB6B270XX & PSA \\
V396 Aur & DAO: COS G130M; G160M & LB6B280XX & PSA \\
V397 Aur & DAO: COS G130M; G160M & LB6B260XX & PSA \\
V410 Tau & GO 8157: STIS E140M & O5E3080XX & 0.2"x0.06" \\
\enddata
\tablecomments{
(a) In addition to the DAO data; we have made use of data from the following HST proposals:	\\	
\ \ GO 8157: Molecular Hydrogen in the Circumstellar Environments of T Tauri Stars; PI Walter	\\	
\ \ GO 8206: The Structure of the Accretion Flow on pre-main-sequence stars; PI: Calvet\\			
\ \ GTO 11533: Accretion Flows and Winds of Pre-Main Sequence Stars; PI: Green		\\	
\ \ GO 11608: How Far Does H2 Go: Constraining FUV Variability in the Gaseous Inner Holes of Protoplanetary Disks; PI: Calvet			\\
\ \ GTO 12036: Accretion Flows and Winds of Pre-Main Sequence Stars Part 2; PI: Green	\\		
(b ) The data set columns indicate the suffix or the full name (if only one) of the HST dataset used.\\
(c) PSA: Primary Science Aperture (for COS). 2.5" diameter. For the STIS data the slit size used is indicated. \\
}
\end{deluxetable}

\clearpage
\LongTables 
\begin{landscape}
\begin{deluxetable}{lcccccccc}
\setlength{\tabcolsep}{2pt}
\tablecolumns{9}
\tabletypesize{\tiny}
\tablewidth{0pt}
\tablecaption{Non-parametric line measurements \label{ForPaper_NonParametric}}

\tablehead{
 \colhead{} &  \multicolumn{4}{c}{\civ} & \colhead{} & \multicolumn{3}{c}{\heii} \\ 
 \cline{2-5}   \cline{7-9} \\
\colhead{Name} &  \colhead{Ratio Blue/Red (a)}&  \colhead{FWHM (b)} &  \colhead{Vel. at max. flux (c)}  &  \colhead{Skewness (d)} &   \colhead{} & \colhead{FWHM (e)} &  \colhead{Vel. at max. flux}  &  \colhead{Skewness (f)}\\
\colhead{} &    \colhead{} &  \colhead{km s$^{-1}$}  &  \colhead{km s$^{-1}$} &  \colhead{} &   \colhead{} & \colhead{km s$^{-1}$}  &   \colhead{km s$^{-1}$} & \colhead{}}
\startdata
\cutinhead{Classical T Tauri Stars}
AA Tau &  1.28$\pm$0.1 & 207.4 & 32.2$\pm$1 & 0.09 & &87.6 & 23.7$\pm$1 & 0.01 \\
AK Sco &  1.89$\pm$0.2 & 116.4 & -27.2$\pm$2 & -0.06 && \ldots & \ldots & \ldots \\
BP Tau &  1.92$\pm$0.1 & 82.9 & 14.1$\pm$1 & 0.04 & &65.7 & -1.2$\pm$1 & 0.01 \\
CV Cha &  2.15$\pm$0.2 & 196.7 & -3.9$\pm$6 & 0.01 & &53.7 & 5.7$\pm$1 & 0.00 \\
CY Tau &  1.69$\pm$0.3 & 45.5 & 19.6$\pm$2 & 0.12 & &43.8 & -0.6$\pm$2 & 0.02 \\
DE Tau &  1.38$\pm$0.1 & 141.8 & 6.6$\pm$2 & 0.00 & &81.6 & 9.0$\pm$1 & 0.02 \\
DF Tau &  2.89$\pm$0.1 & 96.3 & 17.5$\pm$1 & -0.03 & &64.7 & 6.8$\pm$1 & 0.01 \\
DK Tau &  1.79$\pm$0.1 & 262.2 & -31.5$\pm$1 & 0.05 && 83.6 & 8.3$\pm$1 & 0.02 \\
DN Tau  & 1.82$\pm$0.1 & 74.9 & 19.0$\pm$1 & 0.05 & &69.7 & -3.3$\pm$1 & 0.01 \\
DR Tau  & 1.52$\pm$0.1 & 236.8 & 33.7$\pm$1 & 0.07 && 66.7 & -12.4$\pm$1 & 0.05 \\
DS Tau & 2.11$\pm$0.1 & 61.5 & 15.5$\pm$1 & 0.06 & &55.7 & 0.1$\pm$2 & 0.01 \\
DX Cha  & 1.24$\pm$0.1 & 374.6 & 11.5$\pm$1 & 0.07 & &\ldots & \ldots & \ldots \\
EP Cha  & 1.66$\pm$0.1 & 153.8 & 21.2$\pm$1 & 0.03 && 79.6 & -0.2$\pm$1 & 0.01 \\
ET Cha  & 1.24$\pm$0.1 & 260.9 & -5.0$\pm$1 & 0.03 & &80.6 & 6.1$\pm$1 & -0.01 \\
HN Tau A  &1.67$\pm$0.1 & 310.4 & -84.5$\pm$1 & 0.05 & &226.9 & -82.9$\pm$3 & 0.01 \\
MP Mus  & 1.97$\pm$0.2 & 267.6 & 9.8$\pm$1 & -0.04 & &78.6 & -4.6$\pm$2 & 0.02 \\
RU Lup  & 1.26$\pm$0.2  & 148.5 & 31.6$\pm$2 & -0.05 & &90.5 & 19.6$\pm$2 & 0.01 \\
RW Aur A  &  1.45$\pm$0.1 & 387.0 & -86.4$\pm$1 & -0.021 && 131.7 & -99.2$\pm$2 & 0.02 \\
SU Aur      &1.38$\pm$0.1 & 335.8 & 44.3$\pm$2 & -0.02 & &120.4 & 35.1$\pm$1 & 0.00 \\
T Tau N  & 1.71$\pm$0.2 & 169.9 & -4.2$\pm$1 & -0.07 & &79.5 & 4.9$\pm$1 & -0.01 \\
V1190 Sco & 1.58$\pm$0.2 & 251.5 & -2.7$\pm$2 & -0.05 & &210.9 & 7.6$\pm$2 & -0.05 \\
V4046 Sgr  &  2.00$\pm$0.1 & 301.0 & 55.3$\pm$1 & 0.06 & &180.1 & 42.4$\pm$1 & 0.00 \\
\cutinhead{Transition Disks}
CS Cha  & 1.69$\pm$0.1 & 235.8 & 32.0$\pm$1 & 0.09 & &134.5 & 11.3$\pm$1 & 0.01 \\
DM Tau & 1.72$\pm$0.1 & 297.0 & 62.4$\pm$1 & 0.06 && 75.6 & 7.5$\pm$1 & 0.02 \\
GM Aur & 1.83$\pm$0.1 & 270.2 & 26.0$\pm$1 & -0.05 & &127.4 & -3.3$\pm$1 & 0.02 \\
IP Tau & 1.27$\pm$0.1 & 66.9 & 12.9$\pm$1 & -0.06 & &87.6 & 8.0$\pm$2 & 0.02 \\
TW Hya & 1.85$\pm$0.1 & 247.5 & 26.4$\pm$2 & 0.13 && 64.7 & 3.5$\pm$1 & 0.04 \\
UX Tau A & 1.78$\pm$0.1 & 290.8 & 1.3$\pm$3 & -0.02 && 83.4 & -10.5$\pm$1 & 0.03 \\
V1079 Tau  &1.89$\pm$0.1 & 171.2 & 30.1$\pm$1 & -0.01 & &68.7 & 14.9$\pm$1 & 0.04 \\
\cutinhead{Weak T Tauri Stars}
EG Cha & 1.72$\pm$0.1 & 88.3 & 9.5$\pm$1 & 0.01 & &67.7 & -5.7$\pm$1 & -0.01 \\
TWA 7 & 2.03$\pm$0.1 & 59.4 & 12.3$\pm$1 & 0.01 & &58.4 & -0.3$\pm$1 & 0.00 \\
V1068 Tau & 1.74$\pm$0.1 & 115.1 & 15.9$\pm$1 & -0.02 & &96.5 & 11.6$\pm$1 & -0.02 \\
V396 Aur & 2.14$\pm$0.1 & 77.6 & 3.1$\pm$1 & 0.00 && 74.6 & -12.6$\pm$1 & 0.04 \\
V397 Aur &1.91$\pm$0.1 & 69.6 & -2.4$\pm$1 & 0.01 & &77.6 & 18.0$\pm$1 & -0.02 \\
V410 Tau &  2.15$\pm$0.7 & 145.5 & 25.3$\pm$7 & -0.01 && 136.3 & 13.7$\pm$1 & -0.05 \\

\enddata
\tablecomments{All velocities are calculated on the stellar rest frame.\\
(a) The ratio of the two \civ\  lines is calculated by matching the red wings of both lines between 0 and 150 km s$^{-1}$. For RW Aur A, the scaling is calculated from 0 to 100 \kms\ only.\\	
(b) This is the FWHM of the red \civ\  line. The error is $\pm$5 km s$^{-1}$. \\
(c) Velocity at maximum flux for the \civ\  profile. Except for RW Aur A, this is the average of the two doublet lines. For RW Aur A it is only the velocity of the red line.\\
(d) Skewness of the \civ\  profile, averaged over the two lines. For each line, skewness is defined as $(V_{Max}-\overline{V})/\Delta V$, where $\overline{V}$ is the flux-weighted mean velocity over an interval $\Delta V$. For the blue and red lines, the intervals are $\pm$250 km s$^{-1}$ and $\pm$150 km s$^{-1}$ from the maximum, respectively. To calculate the average, we normalize the red \civ\  skewness to the blue one, by dividing by 250/150. To calculate the skewness, we have subtracted the continuum and interpolated over the \htwo\ lines.  Values of the skewness within $\sim$[-0.02,0.02] indicate a symmetric line.\\
(e) FWHM of the \heii\  line. The error is $\pm$5 km s$^{-1}$. \\
(f) Skewness of the \heii\  profile. Measured over $\pm$100 km s$^{-1}$, but normalized to the \civ\  interval.
}
\end{deluxetable}

\clearpage
\end{landscape}

\clearpage
\LongTables 
\begin{landscape}

\begin{deluxetable}{llccccccccccccccccccc}
\setlength{\tabcolsep}{1pt}
\tablecolumns{21}
\tabletypesize{\tiny}
\tablewidth{0pt}
\tablecaption{Gaussian fits to the \civ\ lines.\label{ForPaper_TableFits_CIV}}

\tablehead{
\colhead{} &  \colhead{} &
\multicolumn{13}{c}{Blue Line}  &  \colhead{} &
 \multicolumn{5}{c}{Red Line}  \\
\cline{3-15} \cline{17-21} \\ 
\colhead{} &  \colhead{} &
\multicolumn{6}{c}{Narrow Comp.}  &  \colhead{} &
\multicolumn{6}{c}{Broad Comp.}  & \colhead{} &
\multicolumn{2}{c}{Narrow Comp.}  & \colhead{} &
 \multicolumn{2}{c}{Broad Comp.}  \\
\cline{3-8} \cline{10-15} \cline{17-18} \cline{20-21}  \\ 
\multicolumn{1}{l}{Name (a)} &   \colhead{} &
\colhead{A0} &  \colhead{$\Delta$A0}  &  \colhead{V0} &  \colhead{$\Delta$V0} &  \colhead{$\sigma$0}  &  \colhead{$\Delta\sigma$0} &  \colhead{} &
 \colhead{A1} &  \colhead{$\Delta$A1}  &  \colhead{V1} &  \colhead{$\Delta$V1} &  \colhead{$\sigma$1}  &  \colhead{$\Delta\sigma$1} &    \colhead{} &
 \colhead{A2} &  \colhead{$\Delta$A2}  &   \colhead{} & \colhead{A3} &  \colhead{$\Delta$A3}  \\
\colhead{} &  \colhead{} &
 \colhead{flux} &  \colhead{flux}  &  \colhead{\kms} &  \colhead{\kms} &  \colhead{\kms}  &  \colhead{\kms} &  \colhead{} &
 \colhead{flux} &  \colhead{flux}  &  \colhead{\kms} &  \colhead{\kms} &  \colhead{\kms}  &  \colhead{\kms} &   \colhead{} &
 \colhead{flux} &  \colhead{flux}  &  \colhead{} &\colhead{flux} &  \colhead{flux} \\
  }
\startdata
\cutinhead{Classical T Tauri Stars}
AA Tau (COS) &  & 2.81 & 0.1 & 23.65 & 0.3 & 17.50 & 0.2 &  & 2.70 & 0.1 & 82.86 & 0.4 & 120.94 & 0.5 &  & 2.78 & 0.1 &  & 1.37 & 0.1 \\
AK Sco (STIS) &  & 9.56 & 0.1 & -78.18 & 0.2 & 115.08 & 0.3 &  & 3.69 & 0.1 & -11.05 & 0.1 & 257.8 & 1 &  & 2.67 & 0.1 &  & 2.46 & 0.1 \\
BP Tau (COS) &  & 32.28 & 0.2 & 14.49 & 0.1 & 24.91 & 0.1 &  & 13.99 & 0.1 & 45.11 & 0.1 & 151.00 & 0.2 &  & 16.12 & 0.2 &  & 6.19 & 0.1 \\
BP Tau (GHRS) &  & 10.2 & 2 & 9.3 & 1 & 33.5 & 3 &  & 6.85 & 0.4 & -12.5 & 1 & 123.6 & 2 &  & 4.4 & 1 &  & 2.94 & 0.2 \\
CV Cha (STIS) &  & \ldots & \ldots & \ldots & \ldots & \ldots & \ldots &  & 4.43 & 0.2 & -4.77 & 0.1 & 140.04 & 0.3 &  & \ldots & \ldots &  & 2.16 & 0.1 \\
CY Tau (STIS) &  & 7.35 & 0.5 & 20.85 & 0.7 & 20.43 & 0.8 &  & 2.24 & 0.1 & 104.48 & 3.3 & 126.46 & 3.4 &  & 4.19 & 0.4 &  & 1.10 & 0.1 \\
DE Tau (COS) &  & 3.12&	0.4	&21.0&	1 &	54.7	&2& &	1.37	&0.2&	23.6&2	&129.7&	5	& &2.69	&0.3& &	0.76	& 0.2\\
DF Tau (COS) &  & 20.45 & 0.2 & 23.00 & 0.1 & 23.81 & 0.1 &  & 7.74 & 0.1 & -3.96 & 0.1 & 112.93 & 0.2 &  & 5.06 & 0.1 &  & 4.18 & 0.1 \\
DF Tau (GHRS) &  &  48.7 & 2 &   34.0 & 1 & 36.1 & 1 & & 5.2 & 1   & -7.96 & 0.4   & 133.6 & 1 & & 17.2 & 1 & & 2.56 & 0.1\\
DF Tau (STIS) &  &  49.0 & 2   & 26.54 & 0.4 & 19.65 & 0.3 & & 7.43 & 0.3 &   26.0 & 1   & 104.3 & 2 & &14.5&1& &3.52&0.2 \\
DK Tau (COS) &  & 0.45	&0.1&	-20.7 & 	8&	25.3	& 8	& &0.93	&0.1&	36.4&	3	&142.5&	6& &	0.23	&0.1	& &0.49&	0.1 \\
DN Tau (COS) &  & 10.86 & 0.1 & 21.33 & 0.1 & 28.87 & 0.2 &  & 2.92 & 0.1 & 63.76 & 0.4 & 120.79 & 0.5 &  & 6.22 & 0.1 &  & 1.26 & 0.1 \\
DR Tau (COS) &  &1.55	&0.3&	80.7&4&	80.2&	4	& &1.02&	0.2&	98.4&	6	&197.2&	9	& &1.38	&0.2& &	0.35&	0.1 \\
DR Tau (STIS) &  & \ldots & \ldots & \ldots & \ldots & \ldots & \ldots &  & 1.28 & 0.2 & 226.3 & 7 & 58.3 & 4 &  & \ldots & \ldots &  & 0.70 & 0.1\\
DR Tau (GHRS 1995) &  & 1.7 & 1 & 112.2 & 7 & 50.8 & 4 &  & 1.54 & 0.4 & 17.1 & 2 & 76.4 & 5 &  & 2.5 & 1 &  & $<$0.2 & \ldots \\
DS Tau (STIS) &  & 38.9 & 1 & 27.61 & 0.2 & 23.34 & 0.3 &  & 5.71 & 0.2 & 57.3 & 1 & 147.4 & 2 &  & 23.5 & 1 &  & 2.32 & 0.1 \\
DX Cha (STIS) &  & \ldots & \ldots & \ldots & \ldots & \ldots & \ldots &  & \ldots & \ldots & \ldots & \ldots & \ldots & \ldots &  & \ldots & \ldots &  & \ldots & \ldots \\
EP Cha (COS) &  & 3.84 & 0.1 & 28.94 & 0.3 & 30.25 & 0.3 &  & 5.20 & 0.1 & 40.17 & 0.2 & 136.23 & 0.3 &  & 2.31 & 0.1 &  & 2.74 & 0.1 \\
ET Cha (COS) &  & \ldots & \ldots & \ldots & \ldots & \ldots & \ldots &  & 1.58 & 0.1 & 9.76 & 0.2 & 99.80 & 0.4 &  & \ldots & \ldots &  & 0.98 & 0.1\\
HN Tau A (COS) &  & 1.60 &0.1&-80.42&	0.5&	84.48&	0.6&	&0.51&	0.1&	179.0&	2&	104.0	&2&	&0.84&	0.1& &	0.38&	0.1 \\
MP Mus (STIS) &  & 18.49 & 0.1 & -4.47 & 0.1 & 74.92 & 0.1 &  & 22.63 & 0.1 & 6.06 & 0.1 & 153.34 & 0.1 &  & 12.96 & 0.1 &  & 10.16 & 0.1 \\
RU Lup (COS) &  & 12.09 & 0.1 & 36.63 & 0.1 & 43.36 & 0.1 &  & 6.77 & 0.1 & 0.03 & 0.1 & 192.95 & 0.1 &  & 11.07 & 0.1 &  & 3.85 & 0.1 \\
RU Lup (STIS) &  & 20 & 0.5 & 10.22 & 0.2 & 35.00 & 0.4 &  & 7.98 & 0.2 & -0.08 & 0 & 184.5 & 2 &  & 14.85 & 0.4 &  & 3.86 & 0.1 \\
RU Lup (GHRS) &  & \ldots & \ldots & \ldots & \ldots & \ldots & \ldots &  & 5.5 & 1 & -15 & 15 & 100.5 & 5 &  & \ldots & \ldots &  & 4.5 & 1 \\
RW Aur (COS) &  & \ldots & \ldots & \ldots & \ldots & \ldots & \ldots &  & \ldots & \ldots & \ldots & \ldots & \ldots & \ldots &  & \ldots & \ldots &  & \ldots & \ldots \\
SU Aur (COS) &  & 1.00	&0.3&	85.2&	7	&113.2&	9	& &1.88&	0.3&	17.0&	1	&160.3&5&	&1.03&	0.3& &	1.02	&0.3 \\
T Tau N (STIS)    &  & 3.31&0.5&-22.2 &	1	&75.2	&2& &	3.91&	0.3&	-56.3	&2&	144.1&	3	& &4.71&	0.3& &	$<$0.1&	\ldots \\
T Tau N (GHRS) &  & 6 	& 0.2 & 0 & 0.2 & 79 & 0.5 &  & 3.5 & 0.1 & -52 & 1.2 & 152 & 0.7 &  & 5.3 & 0.2 &  & 0.25 & 0.1 \\
V1190 Sco (COS) &  & 3.55 & 0.1 & 9.44 & 0.1 & 83.97 & 0.1 &  & 1.50 & 0.1 & -103.54 & 0.1 & 104.13 & 0.1 &  & 2.24 & 0.1 &  & 1.00 & 0.1 \\
V4046 Sgr (COS) &  & 11.2&	1	&60.4&	2	&54.2&	2& &	13.69&	0.5&	117.6&	1	&164.9&	2& &	4.45&	0.7& &	7.47&	0.4 \\
 \cutinhead{Transition Disks}
CS Cha (COS) &  & 1.86 &0.3& 60&4& 78.8 &5 & &1.83& 0.2&208.7&6& 183.9 &  7 &&  0.68 & 0.2 & & 1.02 &0.2\\

DM Tau (COS) &  & \ldots & \ldots & \ldots & \ldots & \ldots & \ldots &  & 3.11 & 0.1 & 81.54 & 0.1 & 131.73 & 0.1 &  & \ldots & \ldots &  & 1.78 & 0.1 \\
GM Aur (COS) &  & 4.93 & 0.2 & 23.34 & 0.5 & 30.23 & 0.7 &  & 8.86 & 0.1 & 1.17 & 0.1 & 141.68 & 0.4 &  & 1.97 & 0.2 &  & 4.84 & 0.1 \\
IP Tau (COS) &  & 1.20	&0.1&	19.75&	0.4&	28.28	&0.6& &	0.93&	0.1&	-20.62&	0.3&	129.67&	0.8&	&0.72&	0.1& &	0.63&	0.1\\
TW Hya (STIS) &   &122.93 & 2.2 & 27.74 & 0.3 & 35.32 & 0.3 &  & 133.68 & 0.9 & 116.58 & 0.3 & 142.43 & 0.3 &  & 55.02 & 1.5 &  & 73.64 & 0.6 \\
UX Tau A (COS) &  & \ldots & \ldots & \ldots & \ldots & \ldots & \ldots &  & 2.44 & 0.1 & 18.5 & 0.3 & 135.5 & 1 &  & \ldots & \ldots &  & 1.22 & 0.1 \\
V1079 Tau (COS) &  & 2.35	&0.5&	21.3&	2	&40.6&	3&&	2.67&	0.2&	50.9&	2&	149.9&	3	&&1.13&	0.3&&	1.34&	0.1\\
\cutinhead{Weak T Tauri Stars}
 EG Cha (COS) &  & 2.91 & 0.1 & 4.16 & 0.1 & 33.41 & 0.1 &  & 1.03 & 0.1 & 1.42 & 0.2 & 104.19 & 0.1 &  & 1.75 & 0.1 &  & 0.53 & 0.1 \\
TWA 7 (COS) &  & 4.05 & 0.2 & 16.44 & 0.3 & 19.25 & 0.4 &  & 2.17 & 0.1 & 17.03 & 0.3 & 53.40 & 0.5 &  & 2.37 & 0.1 &  & 0.98 & 0.1 \\
V1068 Tau (COS) &  &0.7	&0.2&	-4.28&	1	&54.9&	5	& &0.22&	0.1	&68.7&	10	&130.8&	12&	&0.41	&0.1& &	0.111	&0.05 \\
V396 Aur (COS) &  & 2.06 & 0.7 & -4.03 & 1.5 & 29.10 & 4.2 &  & 0.44 & 0.1 & 7.57 & 0.4 & 101.82 & 1.2 &  & 1.07 & 0.5 &  & 0.24 & 0.1 \\
V397 Aur (COS) &  & 0.54 & 0.1 & -1.41 & 0.1 & 23.28 & 0.1 &  & 0.44 & 0.1 & -3.07 & 0.1 & 64.10 & 0.1 &  & 0.35 & 0.1 &  & 0.22 & 0.1 \\
V410 Tau (STIS) &  & 1.42 & 0.1 & -0.04 & 1.0 & 66.51 & 2.7 &  & \ldots & \ldots & \ldots & \ldots & \ldots & \ldots &  & 0.76 & 0.1 &  & \ldots & \ldots \\
\enddata
\tablecomments{
The flux has been fit in units of $10^{-14} ergs\ sec^{-1}\ cm^{-2}$. For each complex of two \civ\ lines we fit either two or four gaussians. In the case of two gaussians: $F=A0\  exp(-(v-\mu_0)^2/2\sigma_0^2)+A2 \ exp(-(v-500.96)^2/2\sigma_0^2)$. In the case of four gaussians: $F=A0\ exp(-(v-\mu_0)^2/2\sigma_0^2)+A1\ exp(-(v-\mu_1)^2/2\sigma_1^2)+A2\ exp(-(v-500.96)^2/2\sigma_0^2)+A3\ exp(-(v-500.96)^2/2\sigma_3^2)$. \\								
(a)	The instrument used to obtain the data is indicated in parenthesis.\\								
(b)	Velocity difference between the BC and the NC as a fraction of the velocity of the BC. The error in the ratio is ~5\%. \\
(c)    For DX Cha and RW Aur the red wing of the red line is different than the red wing of the blue line and neither a four nor an eight parameter fit is possible.  
}
\end{deluxetable}

\clearpage
\end{landscape}

\clearpage
\LongTables 
\begin{landscape}
\begin{deluxetable}{lcccccc}

\tablecolumns{7}
\tabletypesize{\tiny}
\tablewidth{0pt}
\tablecaption{\heii\ Gaussian fits\label{ForPaper_HeII}}
\tablehead{\colhead{Name (a)} &   \colhead{A0} &  \colhead{V0}  &  \colhead{$\sigma$0} &  \colhead{A1} &  \colhead{V1}  &  \colhead{$\sigma$1} \\
\colhead{} & \colhead{10$^{-14}$ ergs sec$^{-1}$ cm$^{-2}$} &  \colhead{\kms}  &  \colhead{\kms} &  \colhead{10$^{-14}$ ergs sec$^{-1}$ cm$^{-2}$} &  \colhead{\kms}  &  \colhead{\kms}
}
\startdata
\cutinhead{Classical T Tauri Stars}
AA Tau (COS) & 5.20$\pm$0.1 & 25.54$\pm$0.2 & 36.17$\pm$0.3 & 0.47$\pm$0.1 & 127.0$\pm$3 & 219.2$\pm$3 \\
AK Sco (STIS) & \ldots & \ldots & \ldots & \ldots & \ldots & \ldots \\
BP Tau (COS) & 38.75$\pm$0.2 & 2.89$\pm$0.1 & 29.76$\pm$0.1 & 2.31$\pm$0.1 & -2.13$\pm$0.2 & 195.2$\pm$1 \\
CV Cha (STIS) & 1.24$\pm$0.2 & 10.5$\pm$2.0 & 44.28$\pm$3.8 & \ldots & \ldots & \ldots \\
CY Tau (STIS) & 6.72$\pm$0.8 & 3.36$\pm$0.6 & 21.37$\pm$1.0 & 1.01$\pm$0.4 & 15.0$\pm$4 & 54.1$\pm$8 \\
DE Tau (COS) & 1.24$\pm$0.1 & 5.92$\pm$0.1 & 25.28$\pm$0.1 & 0.31$\pm$0.1 & 21.18$\pm$0.6 & 170.4$\pm$2 \\
DF Tau (COS) & 12.62$\pm$0.1 & 6.88$\pm$0.1 & 24.71$\pm$0.2 & 2.21$\pm$0.1 & 36.14$\pm$0.4 & 141.49$\pm$0.8 \\
DK Tau (COS) & 2.92$\pm$0.1 & 11.38$\pm$0.4 & 32.43$\pm$0.5 & 0.15$\pm$0.1 & 44.20$\pm$0.1 & 191.23$\pm$0.4 \\
DN Tau (COS) & 13.18$\pm$0.2 & -0.04$\pm$0.1 & 31.38$\pm$0.2 & 0.98$\pm$0.1 & 72.5$\pm$1 & 129.3$\pm$2 \\
DR Tau (COS) & 1.22$\pm$0.1 & -14.42$\pm$0.5 & 28.66$\pm$0.7 & 0.62$\pm$0.1 & 104.9$\pm$3 & 48.4$\pm$2 \\
DS Tau (STIS) & 20.43$\pm$0.7 & -1.53$\pm$0.4 & 25.64$\pm$0.4 & 3.83$\pm$0.3 & 325.2$\pm$3 & 33.6$\pm$1 \\
DX Cha (STIS) & \ldots & \ldots & \ldots & \ldots & \ldots & \ldots \\
EP Cha (COS) & 6.92$\pm$0.1 & -5.79$\pm$0.2 & 28.79$\pm$0.2 & 1.86$\pm$0.1 & 25.74$\pm$0.2 & 98.38$\pm$0.4 \\
ET Cha (COS) & 1.51$\pm$0.1 & -2.38$\pm$0.1 & 20.47$\pm$0.9 & 1.51$\pm$0.1 & 1.15$\pm$0.1 & 68.22$\pm$0.7 \\
HN Tau A (COS) & 0.33$\pm$0.1 & -89.34$\pm$1.0 & 57.80$\pm$0.8 & 0.23$\pm$0.1 & 0.16$\pm$0.1 & 204.1$\pm$2 \\
MP Mus (STIS) & 17.80$\pm$0.1 & -0.57$\pm$0.1 & 25.15$\pm$0.1 & 6.49$\pm$0.1 & 6.83$\pm$0.2 & 107.86$\pm$0.4 \\
RU Lup (COS) & 7.27$\pm$0.1 & 20.58$\pm$0.1 & 35.78$\pm$0.1 & \ldots & \ldots & \ldots \\
RW Aur (COS) & 0.78$\pm$0.1 & -96.6$\pm$3 & 27.2$\pm$2 & 0.48$\pm$0.1 & -48.5$\pm$4 & 91.8$\pm$5 \\
SU Aur (COS) & 2.26$\pm$0.1 & 28.44$\pm$0.1 & 43.37$\pm$0.1 & 0.44$\pm$0.1 & -0.15$\pm$0.1 & 197.74$\pm$0.9 \\
T Tau N (STIS) & 3.53$\pm$0.3 & 0.14$\pm$0.1 & 47.1$\pm$2 & 0.79$\pm$0.2 & 0.49$\pm$0.1 & 146.5$\pm$5 \\
V1190 Sco (COS) & 1.28$\pm$0.1 & -23.37$\pm$0.1 & 88.85$\pm$0.1 & \ldots & \ldots & \ldots \\
V4046 Sgr (COS) & 19.24$\pm$0.2 & 35.25$\pm$0.2 & 66.85$\pm$0.2 & 11.01$\pm$0.2 & 58.01$\pm$0.2 & 120.43$\pm$0.4 \\
 \cutinhead{Transition Disks}
CS Cha (COS) & 8.12$\pm$0.1 & 17.53$\pm$0.2 & 53.03$\pm$0.3 & 1.03$\pm$0.1 & 170.9$\pm$3 & 191.6$\pm$3 \\
DM Tau (COS) & 8.00$\pm$0.1 & 7.74$\pm$0.1 & 31.09$\pm$0.2 & 1.65$\pm$0.1 & 76.10$\pm$0.9 & 132.79$\pm$0.5 \\
GM Aur (COS) & 6.29$\pm$0.1 & -5.24$\pm$0.3 & 42.39$\pm$0.5 & 1.78$\pm$0.1 & 8.95$\pm$0.8 & 209.0$\pm$1 \\
IP Tau (COS) & 2.73$\pm$0.1 & 11.46$\pm$0.3 & 31.97$\pm$0.3 & 0.15$\pm$0.1 & 1.46$\pm$0.1 & 163.50$\pm$0.3 \\
TW Hya (STIS) & 292.00$\pm$4.0 & -0.16$\pm$0.1 & 21.99$\pm$0.1 & 78.00$\pm$2.1 & 49.18$\pm$0.6 & 34.25$\pm$0.5 \\
UX Tau A (COS) & 3.55$\pm$0.2 & -0.25$\pm$0.1 & 35.5$\pm$1 & 0.65$\pm$0.1 & 15.5$\pm$2 & 150.4$\pm$5 \\
V1079 Tau (COS) & 2.12$\pm$0.1 & 16.00$\pm$0.1 & 24.37$\pm$0.1 & 0.67$\pm$0.1 & 73.24$\pm$0.4 & 44.27$\pm$0.5 \\
 \cutinhead{Weak T Tauri Stars}
EG Cha (W) (COS) & 6.31$\pm$0.1 & -7.75$\pm$0.1 & 27.70$\pm$0.1 & 0.90$\pm$0.1 & -4.98$\pm$0.1 & 88.27$\pm$0.1 \\
TWA 7 (COS) & 8.30$\pm$0.3 & 0.15$\pm$0.1 & 21.66$\pm$0.3 & 1.85$\pm$0.1 & 0.15$\pm$0.1 & 55.1$\pm$1 \\
V1068 Tau (W) (COS) & 0.94$\pm$0.1 & 6.30$\pm$0.1 & 48.15$\pm$0.1 & \ldots & \ldots & \ldots \\
V396 Aur (W) (COS) & 2.56$\pm$0.1 & -0.09$\pm$0.1 & 33.88$\pm$0.2 & 0.32$\pm$0.1 & -1.75$\pm$0.1 & 83.73$\pm$0.2 \\
V397 Aur (W) (COS) & 1.37$\pm$0.1 & 13.42$\pm$0.1 & 33.80$\pm$0.1 & \ldots & \ldots & \ldots \\
V410 Tau (W) (STIS) & 2.64$\pm$0.2 & 11.3$\pm$1 & 46.22$\pm$2.0 & \ldots & \ldots & \ldots \\
\enddata
\tablecomments{The \heii\ region of AK Sco is too noisy to allow the gaussian fit. DX Cha does not show a single clear emission line in the spectral region.\\
(a) WTTSs are indicated with a (W) after their name; The instrument used to obtain the data is indicated in parenthesis.\\				   
}
\end{deluxetable}

\clearpage
\end{landscape}

\clearpage
\LongTables 
\begin{landscape}
\begin{deluxetable}{lcccccc}
\setlength{\tabcolsep}{2pt}
\tablecolumns{7}
\tabletypesize{\tiny}
\tablewidth{0pt}
\tablecaption{Hot line fluxes, not corrected for extinction\label{ForPaper_TableLineFluxes}}
\tablehead{\colhead{Name} &  \colhead{N V - 1238.8 \AA} &  \colhead{N V - 1242.8 \AA}  &  \colhead{Si IV - 1393.8 \AA (a)} &  \colhead{Si IV - 1402.8 \AA (a)} &  \colhead{C IV - 1548.2 + 1550.8 \AA (b)}  &  \colhead{HeII - 1640.5 \AA}\\
\colhead{} &  \colhead{10$^{-14}$ erg sec$^{-1}$ cm$^{-2}$} &  \colhead{10$^{-14}$ erg sec$^{-1}$ cm$^{-2}$}  &  \colhead{10$^{-14}$ erg sec$^{-1}$ cm$^{-2}$} &  \colhead{10$^{-14}$ erg sec$^{-1}$ cm$^{-2}$} &  \colhead{10$^{-14}$ erg sec$^{-1}$ cm$^{-2}$}  &   \colhead{10$^{-14}$ erg sec$^{-1}$ cm$^{-2}$}
}
\startdata
\cutinhead{Classical T Tauri Stars}
AA Tau & 0.221$\pm$0.01 & 0.150$\pm$0.01 & $<$0.3 & $<$0.3 & 8.66$\pm$0.1 & 3.94$\pm$0.1 \\ 
AK Sco & 2.43$\pm$0.2 & 1.37$\pm$0.2 & 10.2$\pm$3 & 11.0$\pm$3 & 37.6$\pm$1 & $<$1 \\ 
BP Tau & 2.186$\pm$0.02 & 0.984$\pm$0.02 & 4.76$\pm$0.3 & 3.34$\pm$0.3 & 55.47$\pm$0.1 & 23.20$\pm$0.2 \\ 
CV Cha & 0.63$\pm$0.1 & 0.09$\pm$0.1 & 4.0$\pm$1 & 3.6$\pm$1 & 12.28$\pm$0.1 & 1.78$\pm$0.6 \\ 
CY Tau & 1.20$\pm$0.1 & $<$0.3 & $<$4 & $<$4 & 9.10$\pm$0.3 & 4.12$\pm$0.9 \\ 
DE Tau & 0.475$\pm$0.01 & 0.084$\pm$0.01 & $<$0.5 & $<$0.5 & 8.68$\pm$0.1 & 1.31$\pm$0.2 \\ 
DF Tau & 1.080$\pm$0.02 & 0.314$\pm$0.02 & $<$0.9 & $<$0.8 & 25.69$\pm$0.1 & 9.26$\pm$0.2 \\ 
DK Tau & 0.145$\pm$0.01 & 0.015$\pm$0.01 & 0.68$\pm$0.1 & 0.55$\pm$0.1 & 3.503$\pm$0.05 & 1.78$\pm$0.2 \\ 
DN Tau & 0.901$\pm$0.01 & 0.400$\pm$0.01 & 1.20$\pm$0.2 & 0.072$\pm$0.01 & 13.07$\pm$0.1 & 7.51$\pm$0.2 \\ 
DR Tau & 0.287$\pm$0.01 & $<$0.03 & 1.46$\pm$0.3 & 1.27$\pm$0.3 & 8.71$\pm$0.1 & 1.31$\pm$0.3 \\ 
DS Tau & 0.84$\pm$0.1 & 0.23$\pm$0.1 & 3.0$\pm$1 & 2.7$\pm$1 & 35.93$\pm$0.5 & 11.8$\pm$1 \\ 
DX Cha & 35.92$\pm$0.6 & 5.90$\pm$0.6 & 218$\pm$10 & 169$\pm$10  & 147.1$\pm$2 & N/A \\ 
EP Cha & 0.983$\pm$0.01 & 0.399$\pm$0.01 & 0.98$\pm$0.1 & 0.62$\pm$0.1 & 16.57$\pm$0.7 & 5.62$\pm$0.1 \\ 
ET Cha & 0.411$\pm$0.01 & 0.254$\pm$0.01 & $<$0.5 & $<$0.5 & 3.619$\pm$0.03 & 2.10$\pm$0.1 \\ 
HN Tau A & 0.2837$\pm$0.007 & 0.1335$\pm$0.007 & 1.90$\pm$0.2  & 1.67$\pm$0.2 & 4.132$\pm$0.04 & 0.97$\pm$0.1 \\ 
MP Mus & 5.68$\pm$0.1 & 2.49$\pm$0.1 & 8.6$\pm$1 & 4.9$\pm$1 & 94.3$\pm$1 & 17.30$\pm$1.0 \\ 
RU Lup & 2.574$\pm$0.05 & 0.308$\pm$0.05 & 25.9$\pm$1 & 18.8$\pm$1 & 44.8$\pm$1 & 5.79$\pm$0.5 \\ 
RW Aur A& 0.812$\pm$0.02 & 0.038$\pm$0.01 & 4.68$\pm$0.4 & 4.66$\pm$0.4 & 6.00$\pm$0.1 (c)& 1.85$\pm$0.3 \\ 
SU Aur & 0.463$\pm$0.01 & 0.202$\pm$0.01 & 1.26$\pm$0.2 & 1.54$\pm$0.2 & 9.53$\pm$0.1 & 2.59$\pm$0.2 \\ 
T Tau N & 1.14$\pm$0.1 & 0.55$\pm$0.1 & 3.1$\pm$1 & 2.8$\pm$1 & 16.15$\pm$0.1 & 5.89$\pm$1.0 \\ 
V1190 Sco & 0.700$\pm$0.01 & 0.354$\pm$0.01 & $<$0.4 & $<$0.4 & 9.65$\pm$0.1 & 1.63$\pm$0.1 \\ 
V4046 Sgr & 29.66$\pm$0.1 & 10.43$\pm$0.1 & $<$1 & $<$1 & 57.10$\pm$0.1 & 38.05$\pm$0.2 \\ 
\cutinhead{Transition Disks}
CS Cha & 2.491$\pm$0.03 & 0.848$\pm$0.02 & $<$0.8 & $<$0.7 & 9.93$\pm$0.1 & 9.01$\pm$0.2 \\ 
DM Tau & 0.853$\pm$0.02 & 0.281$\pm$0.02 & $<$0.4 & $<$0.4 & 8.57$\pm$0.1 & 7.11$\pm$0.1 \\ 
GM Aur & 2.383$\pm$0.03 & 0.662$\pm$0.03 & $<$1 & $<$1 & 28.23$\pm$0.2 & 8.50$\pm$0.2 \\ 
IP Tau & 0.1211$\pm$0.006 & 0.0491$\pm$0.006 & 0.34$\pm$0.1 & 0.31$\pm$0.1 & 3.560$\pm$0.04 & 1.64$\pm$0.1 \\ 
TW Hya & 38.30$\pm$0.6 & 17.13$\pm$0.4 & 8.8$\pm$3 & 7.6$\pm$3 & 473.4$\pm$2 & 228.9$\pm$4 \\ 
UX Tau A & 0.949$\pm$0.03 & 0.391$\pm$0.01 & 1.08$\pm$0.2 & 0.73$\pm$0.2 & 6.9$\pm$0.1 & 2.80$\pm$0.3 \\ 
V1079 Tau  & 0.485$\pm$0.01 & 0.136$\pm$0.01 & 1.32$\pm$0.2 & 1.11$\pm$0.2 & 9.80$\pm$0.1 & 1.78$\pm$0.1 \\ 
\cutinhead{Weak T Tauri Stars}
EG Cha & 0.948$\pm$0.01 & 0.460$\pm$0.01 & 1.76$\pm$0.1 & 1.08$\pm$0.1 & 4.111$\pm$0.05 & 3.50$\pm$0.1 \\ 
TWA 7 & 0.759$\pm$0.02 & 0.399$\pm$0.01 & 1.03$\pm$0.2 & 0.68$\pm$0.2 & 3.925$\pm$0.04 & 4.03$\pm$0.1 \\ 
V1068 Tau  & 0.0813$\pm$0.004 & 0.0428$\pm$0.004 & 0.116$\pm$0.03 & 0.0064$\pm$0.001 & 1.412$\pm$0.03 & 0.69$\pm$0.1 \\ 
V396 Aur  & 0.2335$\pm$0.008 & 0.1118$\pm$0.008 & 0.52$\pm$0.1 & 0.298$\pm$0.07 & 2.357$\pm$0.04 & 1.47$\pm$0.1 \\ 
V397 Aur  & 0.0895$\pm$0.004 & 0.0455$\pm$0.004 & 0.142$\pm$0.04 & 0.082$\pm$0.02 & 0.855$\pm$0.02 & 0.75$\pm$0.1 \\ 
V410 Tau  & 0.30$\pm$0.1 & $<$0.3 & $<$3 & $<$3 & 2.37$\pm$0.2 & 2.51$\pm$0.8 \\ 
\enddata
\tablecomments{\\
(a)	Measured fluxes not corrected for extinction. The fluxes were obtained by integrating between -400 \kms and 400 \kms, except for EG Cha. For this star, the \siiv\ flux was was obtained by integrating from -800 to 600 \kms. \\		
(b )	Measured fluxes not corrected for extinction. The flux has been measured from -400 to 900 \kms, interpolating over the R(3)1-8 \htwo\ line.	\\
(c) The \civ\ line flux for RW Aur A was obtained by creating a synthetic blue line, as a copy of the red \civ\ line but scaled by 1.4. \\
}
\end{deluxetable}

\clearpage
\end{landscape}

\clearpage
\LongTables 


\begin{deluxetable}{ccccc}
\tablecolumns{5}
\tabletypesize{\tiny}
\tablewidth{0pt}
\tablecaption{Average Line Kinematic Parameters\label{velocities}}
\tablehead{\colhead{Parameter} & \colhead{WTTSs C IV} & \colhead{CTTSs C IV} & \colhead{WTTSs He II} & \colhead{CTTSs He II} \\
\colhead{} & \colhead{km/sec} & \colhead{km/sec} & \colhead{km/sec} & \colhead{km/sec}} 
\startdata
\cutinhead{Non-parametric averages (Section \ref{non_par})}
V$_{Max}$ & 10.6$\pm$4 & 18.3$\pm$4 (a) & 4.1$\pm$5 & 7.1$\pm$3 (a) \\
FWHM & 92$\pm$10 & 210$\pm$18 & 85$\pm$11 & 96$\pm$9 \\
Skewness & -0.005$\pm$0.01 & 0.018$\pm$0.01 & -0.007$\pm$0.01& 0.0126$\pm$0.004 \\
\cutinhead{Parametric averages (Section \ref{fits})}
V$_{NC}$& 1.6$\pm$2 & 26.6$\pm$	6 (a) & 3.9$\pm$3&6.0$\pm$3 (a) \\
V$_{BC}$&17.2$\pm$5 & 39$\pm$10 (a) & -2.2$\pm$1 & 53$\pm$20 (a) \\
FWHM$_{NC}$& 91$\pm$20 &118$\pm$10 & 83$\pm$10 &  85$\pm$10 \\
FWHM$_{BC}$& 223$\pm$40&334$\pm$30 & 178$\pm$20 & 335$\pm$	30 \\
\cutinhead{Velocity differences (b)}
V$_{CTTS\ Max\ CIV}$-V$_{Max}$& 7.7$\pm$6 (a) & 0 & 14.2$\pm$6 (a) & 11.0$\pm$4 \\
V$_{WTTS\ Max\ CIV}$-V$_{Max}$ & 0 & -7.7$\pm$6 (a) &6.5$\pm$6 & 3.4$\pm$6 (a)     \\
V$_{CTTS\ NC\ CIV}$-V$_{NC}$  & 24.8$\pm$7 (a)& 0 & 22.7$\pm$7 (a)& 19.7$\pm$6 \\
V$_{WTTS\ NC\ CIV}$-V$_{NC}$ &  0 & -24.8$\pm$7  (a) & -2.2$\pm$5 & -4.4$\pm$3 (a)\\
V$_{CTTS\ BC\ CIV}$-V$_{BC}$ & 21$\pm$20 (a)    & 0 &39.2$\pm$10 (a)& -4$\pm$10 (a)\\
V$_{CTTS\ BC\ CIV}$-V$_{NC}$  & 38$\pm$10 (a) &26$\pm$17 & 39$\pm$10 (a)&42$\pm$10 \\
V$_{CTTS\ NC\ HeII}$-V$_{NC}$ & 4.4$\pm$3  (a)& -19.7$\pm$6 & 2.1$\pm$4(a)& 0 \\
V$_{WTTS\ NC\ HeII}$-V$_{NC}$ &2.2$\pm$5 &  -39.2$\pm$10 (a) &0 &-2.1$\pm$4 (a)\
\enddata
\tablecomments{All calculations include the \htwo\ velocity correction (Section \ref{datared}).\\
(a) Does not include HN Tau, RW Aur, AK Sco.\\
(b) Velocity differences measured in the same spectrum (for example V$_{C IV}$-V$_{He II}$) are not subject to errors due to pointing. \\
}
\end{deluxetable}

\end{document}